\documentclass[12pt,preprint]{aastex}

\usepackage{amssymb}
\usepackage[caption=false]{subfig} 
\usepackage{color}

\slugcomment{}

\shorttitle{The Y-Type Brown Dwarfs: Estimates of Mass and Age}
\shortauthors{Leggett et al.}

\begin{document}


%%\title{Near-Infrared Photometric Followup of WISE Y Dwarfs}

\title{
The Y-Type Brown Dwarfs: Estimates of Mass and Age from New Astrometry, Homogenized Photometry and \\
Near-Infrared Spectroscopy}


\author{S. K. Leggett\altaffilmark{1}}
\email{sleggett@gemini.edu}
\author{P. Tremblin\altaffilmark{2}}
\author{T. L. Esplin \altaffilmark{3}}
\author{K. L. Luhman\altaffilmark{3,4}}
\author{Caroline V. Morley\altaffilmark{5}}

\altaffiltext{1}{Gemini Observatory, Northern Operations Center, 670
  N. A'ohoku Place, Hilo, HI 96720, USA} 
\altaffiltext{2}{Maison de la Simulation, CEA-CNRS-INRIA-UPS-UVSQ, USR 3441, Centre d'\'etude de Saclay, F-91191 Gif-Sur-Yvette, France}
\altaffiltext{3}{Department of Astronomy and Astrophysics, The Pennsylvania State University, University Park, PA 16802, USA}
\altaffiltext{4}{Center for Exoplanets and Habitable Worlds, The Pennsylvania State University, University Park, PA 16802, USA}
\altaffiltext{5}{Harvard-Smithsonian Center for Astrophysics, Harvard University,Cambridge, MA 02138, USA}

\begin{abstract}

The survey of the mid-infrared sky by the {\it Wide-field Infrared Survey Explorer} ({\it WISE}) led to the discovery of  extremely cold low-mass brown dwarfs, classified as Y dwarfs, which extend the T class to lower temperatures. Twenty-four  Y dwarfs are known at the time of writing.  Here we present improved parallaxes for four of these, determined using {\it Spitzer} images. We give new photometry for four late-type T and three Y dwarfs, and new spectra of three Y dwarfs, obtained at Gemini Observatory. We also present previously unpublished photometry taken from  $HST$, $ESO$, {\it Spitzer} and {\it WISE} archives of 11 late-type T and 9 Y dwarfs.  The near-infrared data are put on to the same photometric system, forming a homogeneous data set for the coolest brown dwarfs. We compare recent models to our  photometric and spectroscopic data set.
We confirm that non-equilibrium atmospheric chemistry is important for these objects. Non-equilibrium cloud-free models reproduce well the near-infrared spectra and mid-infrared photometry  for the warmer Y dwarfs with  $425 \leq  T_{\rm eff}$~K $\leq 450$. A small amount of cloud cover may improve the model fits in the near-infrared for the Y dwarfs with $325 \leq  T_{\rm eff}$~K $\leq 375$. Neither cloudy nor cloud-free models reproduce the near-infrared photometry for the $T_{\rm eff} = 250$~K Y dwarf W0855. We use the mid-infrared region, where most of the flux originates, to constrain our models of W0855. We find that W0855 likely has a mass of 1.5 -- 8 Jupiter masses and an age of 0.3 -- 6 Gyr. The Y dwarfs with measured parallaxes are within 20~pc of the Sun and have tangential velocities typical of the thin disk. The metallicities and ages we derive for the sample  are generally solar-like. We estimate that the known Y dwarfs are 3 to 20 Jupiter-mass objects with ages of 0.6 -- 8.5 Gyr.

\end{abstract}


\section{Introduction}

Brown dwarfs are stellar-like objects with a mass too low for stable nuclear fusion. During the first Gyr of a brown dwarf's life, the luminosity decreases by a factor of $\sim$100, and  1 -- 73 Jupiter-mass brown dwarfs cool to 
effective temperatures ($T_{\rm eff}$) of $\sim$200 -- 2000~K respectively (Baraffe et al. 2003, Saumon \& Marley 2008). As photometric sky surveys are executed at longer wavelengths and with larger mirrors,  fainter and cooler brown dwarfs are identified. Most recently, the {\it Wide-field Infrared Survey Explorer} ({\it WISE}; Wright et al. 2010) revealed a population with $250 \lesssim T_{\rm eff}$~K $\lesssim 500$, and these have been classified as Y dwarfs (Cushing et al. 2011, Kirkpatrick et al. 2012). 
The Y dwarfs are an extension of the T-type brown dwarfs which typically have  $500 \lesssim T_{\rm eff}$~K $\lesssim 1300$ (e.g. Golimowski et al. 2004; Leggett et al. 2009, 2012).

Significantly, even for the predominantly isolated brown dwarfs in the solar neighborhood, the fundamental parameters mass and age can be estimated if models can be fit to observations and $T_{\rm eff}$ and surface gravity $g$ constrained. Evolutionary models show that $g$ constrains mass, because the radii of brown dwarfs do not change significantly after about 200 Myr and are within 25\% of a Jupiter radius (Burrows et al. 1997). Also, the cooling curves as a function of mass are well understood, so that $T_{\rm eff}$ combined with $g$ constrains age (Saumon \& Marley 2008). 

Models of brown dwarf atmospheres have advanced greatly in recent years. Opacities have been updated for CH$_4$, H$_2$ and NH$_3$  (Yurchenko, Barber \& Tennyson 2011, Saumon et al. 2012, Yurchenko \& Tennyson 2014). Models which include non-equilibrium chemistry driven by vertical gas transport are available (Tremblin et al. 2015, hereafter T15), as are models which include sedimentation by various species i.e. clouds (Morley et al. 2012, 2014, hereafter M12 and M14). 
The models are accurate enough that the physical parameters of the brown dwarf atmospheres can be constrained by comparing the observed output energy of the brown dwarf, in the form of a flux-calibrated spectral energy distribution (SED), to synthetic colors and spectra. This paper enhances the number and quality of Y dwarf SEDs in order to improve our understanding of this cold population. We do this by presenting new photometry and spectra, and improved trigonometric parallaxes. 

We present new near-infrared spectra for three Y dwarfs obtained with the Gemini near-infrared spectrograph (GNIRS; Elias et al. 2006) and the Gemini imager and spectrometer FLAMINGOS-2 (Eikenberry et al. 2004). 
We also present new infrared photometry of late-type T and Y dwarfs, obtained with the Gemini observatory  near infrared imager  (NIRI; Hodapp et al. 2003), and previously unpublished near- and mid-infrared photometry of late-type T and Y dwarfs taken from data archives. The near-infrared archive photometry is either on the 
Mauna Kea Observatories (MKO) system (Tokunaga \& Vacca 2005) or on the {\em Hubble Space Telescope (HST)} WFC3 system.
We derive transformations between these two systems and use these to produce a sample of late-T and Y-type dwarfs with single-system photometry. We also present improved parallaxes for  four  Y dwarfs using   {\em Spitzer} images. 

Our new  data set is large enough that trends and outliers can be identified. We compare color-color and color-magnitude plots, and near-infrared spectra, to available models. The Y dwarf atmospheric parameters  $T_{\rm eff}$, $g$ and metallicity are constrained, and mass and age estimated.   We also compare models to the photometric SED of the coolest known brown dwarf, 
WISE J085510.83$-$071442.5 (Luhman 2014, hereafter W0855) and constrain the properties of this extreme example of the known Y class.

In \S 2 we set the context of this work by illustrating how the shape of synthetic Y dwarf SEDs  
vary as model parameters are changed. We show the regions of the spectrum sampled by the filters used in this work, and demonstrate the connection between luminosity, temperature, mass and age as given by evolutionary models.  In \S 3 we describe the model atmospheres used in this work. \S 4 presents the new GNIRS and FLAMINGOS-2 spectra and the new NIRI photometry, and \S 5 presents the previously unpublished photometry extracted from data archives; \S 6 gives transformations between the MKO and WFC3 photometric systems. New parallaxes and proper motions are given in \S 7. \S 8 compares models to the photometric data set, which allows us to estimate some of the Y dwarf properties, and also allows us to select a preferred model type for a comparison to near-infrared spectra, which we present in \S 9. \S 10 combines our results to give final estimates of atmospheric and evolutionary parameters for the sample. Our conclusions are given in \S 11.  

\section{Spectral Energy Distributions and Filter Bandpasses}

This work focusses on observations and models of Y dwarfs. Figure 1 shows synthetic spectra for a Y dwarf with $T_{\rm eff} = 400$~K, generated from T15 models.  Flux emerges through windows between strong absorption bands of primarily CH$_4$, H$_2$O and NH$_3$ (e.g. M14, their Figure 7).  The four panels demonstrate the effect of varying the atmospheric parameters. Near- and mid-infrared filter bandpasses used in this work are also shown. 
 
Figure 1 shows that for a Y dwarf with $T_{\rm eff} \sim 400$~K changes in   $T_{\rm eff}$ of 25~K have large, factor of  $\sim$2, affects on the absolute brightness of the near-infrared spectrum at all of $YJH$; the flux in the [4.5] bandpass changes by 15\%. 
An increase in metallicity  or a decrease in surface gravity $g$ changes the slope of the near-infrared spectrum,
brightening the $Y$ and $J$ flux while having only a small effect on $H$. An increase in metallicity [m/H] of 0.2 dex increases the $YJ$ flux by 20 -- 30\%  and decreases the [4.5] flux by 40\%.
An increase in gravity $g$ cm$\,$s$^{-2}$  of 0.5 dex decreases the $YJ$ and [4.5] flux by about 15\%.
Finally, an increase in the eddy diffusion coefficient $K_{\rm zz}\,$cm$^2\,$s$^{-1}$ (the chemical mixing parameter, see \S 3) from $10^6$ to $10^8$ increases the $YJK$ flux by 15\%  while decreasing the [4.5] flux by 50\%. The parameter $\gamma$ is discussed in \S 3.

The near-infrared spectra of Y dwarfs are therefore expected to be sensitive to all the atmospheric parameters, and especially sensitive to $T_{\rm eff}$. The shape and brightness of the near-infrared spectrum combined with the [4.5] flux can usefully contrain a Y dwarf's atmospheric parameters. We test this later, in \S 9.

The shape of the $Y$-band flux peak appears sensitive to gravity in Figure 1. Not shown in Figure 1 (but demonstrated later in the fits of synthetic to observed spectra), a decrease in metallicity has a similar effect. 
Leggett et al. (2015, their Figure 5) show that the 1~$\mu$m flux from a 400~K Y dwarf emerges between H$_2$O and CH$_4$ absorption bands, in a region where NH$_3$ and pressure-induced H$_2$ opacity is important. H$_2$ opacity is sensitive to both gravity and metallicity (e.g. Liu, Leggett \& Chiu 2007) and the change in shape of the $Y$-band flux peak is likely due to changes in the H$_2$ opacity.

Figure 1 shows that much of the flux from a 400~K Y dwarf is emitted in the  {\em Spitzer} [4.5] bandpass, which is similar to the {\it WISE} W2 bandpass. In fact the T15 models show that, for $300 \leq T_{\rm eff}$~K $\leq 500$ and $4.0 \leq \log g \leq 4.5$, 45--54\% of the total flux is emitted though this bandpass. The percentage of the total flux emitted at $\lambda < 2.5~\mu$m decreases from 20\% to $< 1$\% as $T_{\rm eff}$  decreases from 500~K to 300~K, with the remaining 30--50\% emitted at $\lambda > 5~\mu$m.

Because half the energy of Y dwarfs is emitted in the [4.5] bandpass, the value of [4.5] is an important constraint on  bolometric luminosity and therefore $T_{\rm eff}$. Figure 2  shows $T_{\rm eff}$ as a function of $M_{[4.5]}$ in the left panel, and $T_{\rm eff}$ as a function of $\log g$ in the right panel. The sequences in the left panel are from the various atmospheric models used in the work, which are described below. The sequences in the right panel are taken from the evolutionary models of Saumon \& Marley (2008). 

The cold Y dwarfs are intrinsically faint as they have a radius similar to that of Jupiter's. This low luminosity limits
detection to nearby sources only, and all the known Y dwarfs with measured parallaxes are within 20~pc of the Sun (see \S 7). We assume therefore that the ages of the Y dwarfs should be typical of the solar neighborhood and we limit the evolutionary sequences in Figure 2 to ages of 0.4 -- 10 Gyr. For reference, the Galactic thin disk is estimated to have an age of $4.3 \pm 2.6$ Gyr (e.g. Bensby et al. 2005). The right panel of Figure 2 shows that for our sample we expect a range in $\log g$ of 3.8--4.8, and a range in mass of 3 -- 23 Jupiter masses.

\section{Model Atmospheres}

In this work we use cloud-free model atmospheres from Saumon et al. (2012, hereafter S12) and T15. We also use models which include  homogeneous layers of  chloride and sulphide  clouds from M12, and patchy water cloud models from M14. We do not use PHOENIX models which have not been validated for $T_{\rm eff} < 400$~K 
\footnote{http://www.perso.ens-lyon.fr/france.allard/} or the Hubeny \& Burrows (2007) models which do not include the recent improvements to the CH$_4$, H$_2$ and NH$_3$ line lists.

Models for surface gravities given by $\log g = 4.0$, 4.5 and 4.8 were used, with a small number of  $\log g = 3.8$ models for the lowest temperatures, as appropriate for this sample (see Figure 2). The  T15 models include non-solar metallicities of [m/H]$ = -0.5$ and [m/H]$ = +0.2$, and a few models were also generated with  [m/H]$ = -0.2$. This range in metallicities covers the expected range for stars in the Galactic thin disk (e.g. Bensby et al. 2005). The T15 models include an updated CH$_4$ line list (Yurchenko \& Tennyson 2014) however they do not include opacities of PH$_3$ or rain-out processes for condensates, as do the  S12, M12 and M14  models. For this work, a small number of T15 models with an adjusted adiabat were generated as described below. 

The  T15 models include  non-equilibrium chemistry driven by vertical gas transport  and parameterized with an eddy diffusion coefficient $K_{\rm zz}\,$cm$^2\,$s$^{-1}$. The S12, M12 and M14 models are in chemical equilibrium. Vertical gas transport brings long-lived molecular species such as CO and N$_2$ up into the brown dwarf photosphere.  Mixing occurs in the convective zones of the atmosphere and may occur in the  nominally quiescent radiative zone via processes such as gravity waves (Freytag et al. 2010) or fingering instability (T15). If mixing occurs faster than local chemical reactions can return the species to local equilibrium, then abundances can be different by orders of magnitude from those expected for a gas in equilibrium  (e.g. Noll, Geballe \& Marley 1997, Saumon et al. 2000, Golimowski et al. 2004, Leggett et al. 2007, Visscher \& Moses 2011, Zahnle \& Marley 2014). The left panel of Figure 2 shows that, for a given $T_{\rm eff}$ and for $T_{\rm eff} \gtrsim 450$~K, the introduction of mixing increases $M_{[4.5]}$. This is due to the dredge up of CO which absorbs at 4.4--5.0~$\mu$m (e.g. M14, their Figure 7). For the coldest objects the CO lies very deep in the atmosphere and is not expected to significantly impact the 4.5~$\mu$m flux. While CO absorption is enhanced by mixing, NH$_3$ absorption is diminished because of the dredge up of N$_2$. In Figure 1 the black lines in the top and bottom panels are model spectra calculated for the same temperature, gravity and metallicity, but with different values of $K_{\rm zz}$. The increased mixing in the bottom panel results in stronger CO absorption at 4.5~$\mu$m and weaker NH$_3$ absorption in the near-infrared and at $\lambda \sim 10~\mu$m.

Various species condense in these cold atmospheres, forming cloud decks. For T dwarfs with $500 \lesssim T_{\rm eff}$~K $\lesssim 1300$ the condensates consist of  chlorides and sulphides (e.g. Tsuji et al. 1996, Ackerman \& Marley 2001, Helling et al. 2001, Burrows et al. 2003, Knapp et al. 2004, Saumon \& Marley 2008, Stephens et al. 2009, Marley et al. 2012, M12, Radigan et al. 2012, Faherty et al. 2014). As 
$T_{\rm eff}$ decreases further, the next species to condense are calculated to be H$_2$O for $T_{\rm eff} \approx$ 350~K and  NH$_3$ for  $T_{\rm eff} \approx$ 200~K (Burrows et al. 2003, M14). Comparison of the cloudy and cloud-free sequences in Figure 2 shows that the clouds are not expected to impact the  4.5~$\mu$m flux until temperatures are low enough for water clouds to form. These water clouds are expected to scatter light in the near-infrared and absorb at $\lambda \gtrsim 3~\mu$m (e.g. M14, their Figure 2). For the warmer Y dwarfs with $T_{\rm eff} \approx$ 400~K,
the chloride and sulphide clouds lie deep in the atmosphere but they may nevertheless impact light emitted in particularly clear opacity windows, such as the $Y$ and $J$ bands. Such clouds may be the cause of the (tentative) variability seen at $Y$ and $J$ for the Y0 WISEA J173835.52$+$273258.8 (hereafter W1738), which also exhibits low-level variability at [4.5] (Leggett et al. 2016b). 

Tremblin et al. (2016) show that brown dwarf atmospheres can be subject to thermo-chemical instabilities which could induce turbulent energy transport. This can change the temperature gradient in the atmosphere which in turn can produce the observed brightening at $J$ across the L- to T-type spectral boundary, without the need for cloud disruption (e.g. Marley, Saumon \& Goldblatt 2010). Tremblin et al. model the L to T transition by increasing the adiabatic index 
which leads to warmer temperatures in the deep atmosphere and cooler temperatures in the upper regions. We have similarly experimented with modified adiabats for this work, i.e. using pressure-temperature profiles not described by adiabatic cooling of an ideal gas.  For an ideal gas, adiabatic cooling is described by $P^{(1-\gamma)}T^{\gamma} =$ constant. $\gamma$ is the ratio of specific heats at constant pressure and volume. For hydrogen gas  $\gamma = 1.4$. Model spectra were generated with $\gamma =$ 1.2, 1.3 and 1.35. We found that $\gamma =$ 1.35 produced spectra indistinguishable from adiabatic cooling, and the observations presented later do not support a $\gamma$ value as low as 1.2.  Hence we only explore models with   $\gamma =$ 1.3 here.

Brown dwarf atmospheres are turbulent. It is likely that vertical mixing, cloud formation, thermal variations and non-adiabatic energy transport are all important. Full three-dimensional hydrodynamic models are needed. In the mean time, we compare available models to new data we present in the next two sections. Although no model is perfect, we do find that the models which include vertical mixing can be used to estimate the properties of Y dwarfs.

\section{New Gemini Observations}

\subsection{GNIRS Near-Infrared Spectrum for WISEA J041022.75$+$150247.9}

WISEA J041022.75$+$150247.9 (hereafter W0410)  is a Y0 brown dwarf that was discovered in the {\it WISE} database by Cushing et al. (2011). 
Cushing et al. (2014) present a spectrum of W0410 which covers the wavelength range 1.07--1.70~$\mu$m at a resolution $R \approx 130$. 
 The shape of the spectrum at 0.98--1.07~$\mu$m is sensitive to gravity and metallicity (\S 2), and for this reason we obtained a 
$0.95 \leq \lambda~\mu$m $\leq 2.5$ spectrum using GNIRS at Gemini North on 2016 December 24 and 25, via program GN-2016B-Q-46. GNIRS was used in cross-dispersed mode  with the 32 l/mm grating, the short camera and the 0$\farcs$675 slit, giving   $R \approx 700$. A central wavelength of 1.65~$\mu$m resulted in wavelength  coverage for orders 3 to 7 of 
1.87--2.53~$\mu$m, 1.40--1.90~$\mu$m, 1.12--1.52~$\mu$m, 0.94--1.27~$\mu$m, 0.80--1.08~$\mu$m. 
Flatfield and arc images were obtained using lamps on the telescope, and pinhole images were obtained to trace the location of the cross-dispersed spectra. 
A total of 18  300$\,$s frames were obtained on W0410 on December 24 and 10 300$\,$s frames on December 25. Both nights were clear, with seeing 
around 0$\farcs$8 on the first night and around 1$\farcs$0 on the second. GNIRS suffered from electronic noise on the second night, and we used the data from December 24 only. An ``ABBA'' offset pattern was used with offsets of $3\arcsec$ along the slit. Bright stars were observed before and after W0410 on December 24 to remove telluric absorption features and produce an instrument response function;
the F2V HD 19208 was observed before and the F3V HD 33140 was observed after. Template spectra for these spectral types were obtained from the spectral library of Rayner et al. (2009).
The data were reduced in the standard way using routines supplied in the IRAF Gemini package.
The final flux calibration of the W0410 spectrum was achieved using the observed $YJHK$ photometry. Figure 3 shows the new spectrum, and the lower resolution Cushing et al. (2014) spectrum for reference. We compare the spectrum to models later, in \S 9.

\subsection{FLAMINGOS-2 Near-Infrared Spectra for \\ WISE J071322.55$-$291751.9 and WISEA  J114156.67$-$332635.5}

WISE J071322.55$-$291751.9 (hereafter W0713)  is a Y0 brown dwarf that was discovered in the {\it WISE} database by Kirkpatrick et al. (2012). Kirkpatrick et al. present a spectrum of W0713 which covers the $J$-band only. WISEA  J114156.67$-$332635.5 (hereafter W1141) was discovered in the {\it WISE} database and presented by Tinney et al. (2014). No near-infrared spectrum has been published for this object, but a type of Y0 was estimated by Tinney et al. based on absolute magnitudes and colors. We obtained near-infrared spectra for these two objects using FLAMINGOS-2 at Gemini South on 2017 February 3 and 7, via program GS-2017A-FT-2. The $JH$ grism was used with the 4-pixel (0$\farcs$72) slit, giving  $R \approx 600$. The wavelength coverage was 0.98--1.80~$\mu$m. We compare the spectra to models later, in \S 9.

W1141 was observed on 2017 February 3 in thin cirrus with seeing 0$\farcs$8.  An ``ABBA'' offset pattern was used with offsets of $10\arcsec$ along the slit. A total of 18  300$\,$s frames were obtained, as well as flat field and arc images using lamps on the telescope. The F5V star HD 110285 was observed immediately following W1141. A template spectrum for F5V was obtained from the spectral library of Rayner et al. (2009). The bright star was used to remove telluric features and provide an instrument response function,
the final flux calibration was achieved using the $J$ photometry given by Tinney et al. (2014). The data were reduced in the standard way using routines supplied in the IRAF Gemini package. Figure 3 shows our spectrum for this object, compared to the low-resolution Cushing et al. (2011) $JH$ spectrum of W1738, which currently defines the Y0 spectral standard. The spectral shapes are almost identical, and we confirm the Y0 spectral type estimated photometrically by Tinney et al. (2014). 

W0713 was observed on 2017 February 7 in clear skies with seeing 0$\farcs$7.  An ``ABBA'' offset pattern was used with offsets of $10\arcsec$ along the slit. 12 300$\,$s frames were obtained, as well as flat field and arc images using lamps on the telescope. The A3V star HD 43119 was observed immediately before W0713. A template spectrum for A3V was obtained from the Pickles (1998) spectral atlas. The bright star was used to remove telluric features and provide an instrument response function, the final flux calibration was achieved using the $J$ photometry given by Leggett et al. (2015). This was consistent with the more uncertain $Y$ magnitude, but inconsistent with $H$ by 5~$\sigma$. Although variability cannot be excluded, the  model fit shown later does a reasonable job of reproducing the entire spectrum, and we believe the discrepancy is 
due to the lower signal to noise in the $H$ spectral region. Figure 3 shows our W0713 spectrum together with the Kirkpatrick et al. (2012) $J$-band spectrum of this object, which has been flux-calibrated by the $J$ photometry. The Kirkpatrick et al. spectrum appears noisier, but is consistent with our data.

\subsection{NIRI $Y$, $CH_4$(short) and $M^{\prime}$ for WISE J085510.83$-$071442.5}

W0855 was discovered as a high proper motion object by Luhman (2014) in {\it WISE} images. W0855 is the intrinsically faintest and coolest object known outside of the solar system at the time of writing, with effective temperature $T_{\rm eff} \approx$ 250~K and  $L/L{\odot} \approx 5e-8$ (based on Stefan’s law and radii given by evolutionary models). W0855 is 2.2~pc away and has a high proper motion of $-8\farcs10$ yr$^{-1}$ in Right Ascension and   $+0\farcs70$ yr$^{-1}$ in Declination (Luhman \& Esplin 2016, hereafter LE16). 

We obtained  photometry for W0855 on Gemini North using NIRI at $Y$ and $CH_4$(short) via program GN-2016A-Q-50, and  at $M^{\prime}$  via program GN-2016A-FT-10.  The photometry is on the MKO system however there is some variation in the $Y$ filter bandpass between the cameras used on Maunakea, and $Y_{\rm NIRI} - Y_{\rm MKO} = 0.17 \pm 0.03$ magnitudes for late-type T and Y dwarfs (Liu et al. 2012). At the time of our observations (2015 December to 2016 March), the only published near-infrared detection of W0855 was a $J$-band measurement (Faherty et al. 2014). The $Y$ and $CH_4$(short) observations were obtained in order to provide a near-infrared SED for this source. The  $M^{\prime}$ observation was obtained to probe the degree of mixing in the atmosphere, as described in \S 8.1.

All nights were photometric, and the seeing was $0\farcs5$ -- $0\farcs8$. Photometric standards  FS 14, FS 19 and FS 126 were used for the  $Y$ and $CH_4$(short) observations, and HD 77281 and LHS 292 were used for the $M^{\prime}$ observations (Leggett et al. 2003, 2006; UKIRT
online catalogs\footnote{http://www.ukirt.hawaii.edu/astronomy/calib/phot\_cal/}).  The photometric standard FS 20 with a type of DA3 was also observed in the    $CH_4$(short) filter. This standard has $J - H = -0.03$  magnitudes and $H - K = -0.05$  magnitudes, i.e. very close to a Vega energy distribution across the $H$ bandpass. FS 20 confirmed that NIRI $CH_4$(short) zeropoints could be determined by adopting $CH_4 = H$ for all the standards observed, and we found the zeropoint to be 22.95 $\pm$ 0.03 magnitudes. W0855 and the calibrators were offset slightly between exposures using a 5- or 9-position telescope dither pattern. Atmospheric extinction corrections between W0855 and the nearby calibrators were not applied as these are much smaller than the measurement uncertainty (Leggett et al. 2003, 2006). The measurement uncertainties were estimated from the sky variance and the variation in the aperture corrections. 

The  $Y$ and $CH_4$(short) data were obtained using the NIRI f/6 mode, with a pixel size of $0\farcs12$ and a field of view (FOV) of 120''. Individual exposures were 120~s at $Y$ and 60~s at  $CH_4$(short). $Y$ data were obtained at airmasses ranging from 1.1 to 1.9 on 2016 February 16, 17, 18 and 23. $CH_4$(short) data were obtained at airmasses ranging from 1.1 to 1.5 on 2015 December 25 and 26, 2016 January 19, 2016 February 1, 2016 March 12 and 13. The total on-source integration time was 7.1 hours at $Y$ and 14.7 hours at  $CH_4$(short). Calibration lamps on the telescope were used for  flat fielding and the data were reduced in the standard way using routines supplied in the IRAF Gemini package. Images from different nights were combined after shifting the coordinates to allow for the high proper motion of the target. The shift per day was $-0.191$ pixels in $x$ and $+0.017$  pixels in $y$. Aperture photometry with annular skies was carried out, using an aperture diameter of  $1\farcs2$ and using point sources in the image to determine the aperture corrections. 

The $M^{\prime}$ data  were obtained using the NIRI f/32 mode,  with a pixel size of $0\farcs02$ and a FOV of 22''; individual exposures were 24~s composed of 40 coadded 0.6~s frames. Data were obtained 
at an airmass of 1.1 to 1.4 on 2016 March 11. The total on-source integration time was  1.6 hours at $M^{\prime}$. Flat fields were generated from sky images created by masking sources in the science data. Although the exposure time was short, the background signal through this $5~\mu$m filter is high and can vary quickly. Because of this, after flat fielding the data we subtracted adjacent frames and then shifted the subtracted frames to align the calibrator or W0855 before combining the images. As the data were taken on one night no correction had to be made for W0855's proper motion. Aperture photometry with annular skies was carried out, using an aperture diameter of  $0\farcs8$. Aperture corrections were determined from the photometric standards.

W0855 was not detected in the $Y$ filter, but was detected in $CH_4$(short) and  $M^{\prime}$. Figure 4 shows two $CH_4$(short) images. One uses data taken in December 2015 and January 2016,  and the other uses data taken in March 2016. The North-West motion of W0855 is apparent. Figure 4 also shows 
the stacked $M^{\prime}$ and $Y$ image. The measured magnitudes or detection limits are given in Table 1.
Our measurement of $Y > 24.5$ magnitudes is consistent with the Beamin et al. (2014) measurement of  $Y > 24.4$ magnitudes.
Our measurement of  $CH_4$(short) $= 23.38 \pm 0.20$ magnitudes is consistent with the 23.2 $\pm$ 0.2 magnitudes measured by LE16 and the 23.22 $\pm$ 0.35 magnitudes determined by Zapatero Osorio et al. (2016) from analysis of the LE16 data.

\subsection{NIRI $M^{\prime}$ for CFBDS J005910.90$-$011401.3, 2MASSI J0415195$-$093506, UGPS J072227.51$-$054031.2, 2MASSI J0727182$+$171001 and \\ WISEPC J205628.90$+$145953.3}

$M^{\prime}$ data for a sample of T and Y dwarfs was obtained  via program GN-2016B-Q-46 using NIRI on Gemini North in the same configuration as described in \S 4.3. The  $M^{\prime}$ observations were obtained to probe the degree of mixing in brown dwarf atmospheres, as described in \S 8.1.
All nights were photometric with seeing varying night to night from  $0\farcs4$ to  $1\farcs1$.  The data were reduced in the same way as the  W0855 $M^{\prime}$ data. The results are given in Table 1.

CFBDS J005910.90$-$011401.3 is a T8.5 dwarf discovered by Delorme et al. (2008).  The T dwarf was observed on 2016 July 18 and  2016 October 11. The second data set was taken at a lower airmass, and we used the data from October 11 only. 207 24-second dithered images were obtained for an on-source time of 1.4 hours. The airmass range was 1.07--1.26. Also observed on 2016 October 11 was UGPS J072227.51-054031.2, a T9 dwarf discovered by Lucas et al. (2010).  Thirty-six 24-second dithered images were obtained for an on-source time of 14 minutes, at an airmass of 1.2. The photometric standards HD 1160, HD 22686 and HD 40335 were used as $M^{\prime}$ calibrators on 2016 October 11.

2MASSI J0415195$-$093506 is a T8 dwarf discovered by Burgasser et al. (2002). The T dwarf was observed on 2016 October 22. 181 24-second dithered images were obtained for an on-source time of 1.2 hours. The airmass range was 1.18--1.30. The photometric standard HD 22686 was used for calibration.

2MASSI J0727182$+$171001 is a T7  dwarf discovered by Burgasser et al. (2002). The T dwarf was observed on 2017 January 9. 153  24-second dithered images were obtained for an on-source time of 1.0 hours. The airmass range was 1.0--1.1. The photometric standards HD 40335 and HD 44612 were used for calibration.

WISEPC J205628.90$+$145953.3 (hereafter W2056) is a Y0 brown dwarf that was discovered in the {\it WISE} database by Cushing et al. (2011). W2056 was observed on 
2016 July 13.  244 24-second dithered images were obtained for an on-source time of 1.6 hours.  The airmass range was 1.0--1.5. The photometric standards  G 22-18 and HD 201941 were used as calibrators. The offsets were such that one corner of the stacked image contained the 2MASS star 20562847$+$1500092. This star has 2MASS magnitudes $J=13.45\pm0.03$,  $H=13.18\pm0.04$ and $K_s=13.20\pm0.03$ magnitudes. We measure  $M^{\prime}=13.21\pm0.15$ magnitudes for this star. The near-infrared colors suggest a spectral type of G0 (Covey et al. 2007), and the measured $K_s - M^{\prime}=-0.01\pm0.15$ magnitudes is consistent with the color expected for the spectral type (e.g. Davenport et al. 2014).

\subsection{Revised FLAMINGOS-2 $H$ for WISEA J064723.24$–$623235.4 }

We obtained $H$ data for the Y1 WISEA J064723.24$-$623235.4 (Kirkpatrick et a. 2013, hereafter W0647) using FLAMINGOS-2 on Gemini South, which were presented in Leggett et al. (2015). Leggett et al. (2015) give a lower limit for $H$ only. We have examined in more detail the reduced image at the location of the source (provided by the contemporaneous $J$ detection) and obtained a 3.5 $\sigma$ measurement, which is given in Table 1.

\section{Photometry from Image Archives}

We searched various archives for late-type T and Y dwarf images in order to determine transformations between photometric systems and complement our data set. The archived images were downloaded in calibrated form, and we carried out aperture photometry using annular sky regions. Aperture corrections were derived using bright sources in the field of the target. This section gives the resulting, previously unpublished, photometry. 

We have also updated our near-infrared photometry for the T8 dwarf  ULAS J123828.51$+$095351.3 using data release 10 of the UKIRT Infrared Deep Sky Survey (UKIDSS), processed by the Cambridge Astronomy Survey Unit (CASU) and available via the WFCAM Science Archive WSA\footnote{http://www.wsa.roe.ac.uk}. We added two late-type T dwarfs which have UKIDSS and WISE data and which were identified by Skrzypek, Warren \& Faherty (2016): J232035.29$+$144829.8 (T7) and J025409.58$+$022358.7 (T8).

\subsection{{\em HST} WFC3}

We used the {\em HST} Mikulski Archive for Space Telescopes (MAST) to search for Wide Field Camera 3 (WFC3) data for late-type T and Y dwarfs taken with the F105W, F125W, F127M or F160W filters. These filters were selected as they more closely map onto the ground-based $Y$, $J$ and $H$ bandpasses (Figure 1), compared to for example the F110W and F140W which have also been used for brown dwarf studies. The ``drz'' files were used, which have been processed through the {\tt calwf3} pipeline and geometrically corrected using {\tt AstroDrizzle}. The photometric zeropoints for each filter were taken from the WFC3 handbook\footnote{http://www.stsci.edu/hst/wfc3/phot\_zp\_lbn}.  Previously unpublished WFC3 photometry for five T dwarfs and one Y dwarf was obtained, and is presented in Table 2.

\subsection{ESO VLT HAWK-I}

Images obtained with the European Southern Observatory's (ESO) High Acuity Wide field K-band Imager (HAWK-I), are published as reduced data via the ESO science archive facility. The data were processed by CASU which produced astrometrically and photometrically calibrated stacked and tiled images. The integration time was obtained from the ``DIT'' and ``NDIT'' entries in the FITS headers. Previously unpublished $J$ and $H$ photometry on the MKO system was obtained for four T dwarfs and is presented in Table 2.

\subsection{{\em Spitzer}}

The NASA/IPAC Infrared Science Archive (IRSA) was used to search the mid-infrared {\em Spitzer} 
[3.6], [4.5], [5.8] and [8.0] IRAC images. The post basic calibrated data (PBCD) were downloaded and photometry obtained using the Vega fluxes given in the IRAC instrument handbook\footnote{http://irsa.ipac.caltech.edu/data/SPITZER/docs/irac/iracinstrumenthandbook/17/}.  Previously unpublished IRAC photometry for four T dwarfs, two confirmed Y dwarfs and one unconfirmed Y dwarf   is presented in Table 3.
We also re-extracted [3.6] photometry for W2056 from six images taken in 2012, 2013 and 2014, in order to more accurately remove artefacts caused by a nearby bright star. This result is also given in Table 3. 

\subsection{{\em WISE}}

IRSA was used to examine the ALLWISE calibrated images taken in the W1 (3.4~$\mu$m) and W3 (12~$\mu$m) filters where the photometry was not listed in the WISE catalog. Zeropoints were provided in the data FITS headers. Previously unpublished W1 photometry is given for two Y dwarfs, and W3 for 
two T and three Y dwarfs, in Table 3.

In the process of examining the {\em WISE} image data for the known Y dwarfs we also determined that the W1 photometry for the Y dwarf WISE J154151.65$-$225024.9
was compromised by nearby sources and this measurement was removed from our photometric database.

\section{ Synthesized Photometry and Transformations between \\ {\em HST} F1.05W, F1.25W, F1.27M, F1.60W; $CH_4$(short); MKO $Y$, $J$, $H$ }

Schneider et al. (2015) present observed WFC3  F105W and/or F125W photometry for five late-T and eleven Y dwarfs. Beichman et al. (2014) present  F105W and F125W photometry for an additional Y dwarf. Schneider et al. also present grism spectroscopy from which they calculate synthetic  F105W and F125W photometry.  In order to transform {\em HST} photometry and our CH$_4$ photometry on to the MKO system, we calculated the following colors (or a subset) from available  near-infrared spectra:   $Y -$ F1.05W, $J -$ F1.25W, $J -$ F1.27M, $ H -$ F1.60W and $H - CH_4$(short). Table 4 lists these newly synthesized colors for five T dwarfs and six Y dwarfs, using spectra from  this work, Kirkpatrick et al. (2012), Knapp et al. (2004), Leggett et al. (2014, 2016a), Lucas et al. (2010), Schneider et al. (2015), and Warren et al. (2007). 

Table 4 also gives synthetic MKO-system colors for three T dwarfs and two Y dwarfs using spectra  from  this work, Kirkpatrick et al. (2011) and Pinfield et al. (2012, 2014). These five objects were selected as additions to our data set because they are either very late-type, or have been classified as peculiar and so potentially sample unusual regions of color-color space.  

We have used the synthesized colors and measured MKO and {\em HST} photometry where they both exist, to determine a set of transformations between the two systems as a function of type, for late-T and Y dwarfs. The  photometry is taken from Leggett et al. (2015) and references therein, Schneider et al. (2015) and references therein, and this work. We have included colors from T15 spectra for $T_{\rm eff} = 400$, 300, 250 and 200~K, with $\log g = 4.5$ and $\log K_{\rm zz} = 6$, to constrain the transformations at very late spectral types. For the purposes of the fit, we adopt spectral types of Y0.5, Y1.5, Y2 and Y2.5 for the colors generated by models with $T_{\rm eff} = 400$, 300, 250 and 200~K, respectively.   
We explored the sensitivity of the synthetic colors to the atmospheric parameters using $T_{\rm eff} = 300$~K models with $\log g = 4.0$ and 4.5, [m/H] $=$ 0.0 and $-0.5$ and   $\log K_{\rm zz} = 6$ and 8. We found a dispersion in $Y -$ F1.05W, $J -$ F1.25W, $J -$ F1.27M, $ H -$ F1.60W and $H - CH_4$(short) of 0.01 -- 0.09 magnitudes. We adopt a
$\pm 0.1$ magnitude uncertainty in these model colors.
 
We performed weighted least-squares quadratic fits to the data.  Figure 5 shows the data and the fits, and Table 5 gives the fit parameters for the transformations.  Based on the scatter seen in Figure 5 we estimate the uncertainty in the transformations for the Y dwarfs to be $\pm$0.10 magnitudes.
We have used the relationships given in Table 5 to estimate $Y$, $J$ and $H$ magnitudes (or a subset) on the MKO system  for seven Y dwarfs with {\em HST} and $CH_4$(short) photometry. The results are given in Table 6. We have expanded wavelength coverage for five of these Y dwarfs by adopting the synthetic colors derived by Schneider et al. (2015) from their spectra. These values are also given in Table 6. Table 6 also gives MKO photometry for W1141, determined using the synthesized colors given in Table 4.

WD0806$-$661B and W0855 have estimates of $J$, and $J$ and $H$, respectively, determined in two ways (Table 6). The two values of $J$ agree within the quoted uncertainties for both Y dwarfs (although only marginally so for W0855). The two values of $H$ for W0855 differ by $1.8 \sigma$. We use a weighted average of the two measurements in later analysis, and estimate the uncertainty in the average to be the larger of the uncertainty in the mean, or half the difference between the two values.

\section{New Astrometry, and the Luminosity of WISE J014656.66$+$423410.0AB}

LE16 refined the parallax and proper motion for W0855 using astrometry measured with multi-epoch
images from {\it Spitzer} and {\it HST}. They also presented new parallaxes for three Y dwarfs whose previous measurements had large uncertainties, consisting of WISE J035000.32$-$565830.2 (hereafter W0350), WISE J082507.35$+$280548.5 (hereafter W0825) and WISE J120604.38$+$840110.6 (hereafter W1206). Those measurements were based on the {\it Spitzer} IRAC images of these objects that were publicly available and the distortion corrections for IRAC
from Esplin \& Luhman (2016). We have measured new proper motions and parallaxes
in the same way for three additional Y dwarfs whose published measurements are uncertain: WISE J053516.80$-$750024.9 (hereafter W0535), WISEPC J121756.91$+$162640.2AB (hereafter W1217AB) and WISEPC J140518.40$+$553421.5 (hereafter W1405).
We have also determined an improved parallax for WISE J014656.66$+$423410.0AB (hereafter W0146AB) which was classified as a Y0 in the discovery paper (Kirkpatrick et al. 2012), and reclassified as a T9 when it was resolved into a binary with components of T9 and Y0 (Dupuy, Liu \& Leggett 2015). 

In Table 7, we have compiled the parallaxes for W0350, W0825 and W1206 from
LE16, the proper motions for those objects that were derived by LE16 but
were not presented, and our new parallaxes and proper motions for W0146AB, W0535, W1217AB and
W1405. 
The uncertainty in the new parallax measurements are significantly smaller than those of the previously published values --- 5 -- 12 mas compared to 14 -- 80 mas. The measurements for W1217AB and W1405 are consistent with previous measurements by Dupuy \& Kraus (2013), Marsh et al. (2013) and Tinney et al. (2014). The measurement for W0535 differs from the previous measurement by Marsh et al. by $2 \sigma$.  The measurement for W0146AB differs from the previous measurement by Beichman et al. (2014)  by $3 \sigma$. In the Appendix, Tables 11 -- 14 give the astrometric measurements  for W0146AB, W0535, W1217AB amd W1405.

We show in the next section that the revised parallax for W0146AB places the binary, and its components, in a region of the color-magnitude diagrams that is occupied by other T9/Y0 dwarfs. The previous parallax measurement implied an absolute magnitude 1.2 magnitudes fainter, suggesting an unusually low luminosity (Dupuy, Liu \& Leggett 2015). The upper panel of Figure 6 shows that the combined-light spectrum is very similar to what would be produced by a pair of Y0 dwarfs, i.e. the system is not unusual. We have deconvolved the spectrum using near-infrared spectra of late-T and early-Y dwarfs as templates (a larger number of spectra are available compared to when Dupuy, Liu \& Leggett deconvolved the spectrum).
The absolute brightness of each input spectrum has been ignored, but the relative brightness of each input pair has been made to match the $\delta J$ magnitudes measured for the resolved system. The lower panel of Figure 6 shows that T9 $+$ T9.5 and T9 $+$ Y0 composite spectra have slightly broader $J$ and $H$ flux peaks than observed for W0146AB, while a T9.5 primary with a Y0 secondary reproduces the spectrum quite well. We adopt a spectral type of T9.5 for W0146AB and W0146A, and a type of Y0 for W0146B.

\section{Photometry: The Sample and Comparison to Models}

Table 8 compiles the following observational data for the currently known sample of 24 Y dwarfs: parallax (in the form of a distance modulus), MKO-system $YJHK$, {\em Spitzer} [3.6] and [4.5], and {\em WISE} W1, W2 and W3 magnitudes.  The data sources are given in the Table.  In the Appendix, Table 15 gives an on-line data table with these values for the larger sample of late-T and Y dwarfs used in this work.  
We have compared these data  to calculations by the models described in \S 3 via a large number of color-color and color-magnitude plots. 

\subsection{Constraining the Eddy Diffusion Coefficient $K_{\rm zz}\,$cm$^2\,$s$^{-1}$}

The $M^{\prime}$ observations allow a direct measurement of the strength of the CO absorption at  $4.7~\mu$m, as shown in Figure 1. Figure 7 shows [4.5] $-M^{\prime}$  as a function of $M_{[4.5]}$. The reduction in $M^{\prime}$ flux for the T dwarfs  is evident in the Figure. The CO absorption does not appear to be a strong function of gravity, as indicated by the similarity between the T15 $\log g = 4.0$ and  $\log g = 4.5$ sequences. The absorption does appear to be a function of metallicity, and of the adiabat used for heat transport (see \S 3, note the $\gamma=1.3$ sequence in Figure 7 has   $\log K_{\rm zz}=8$). We make the assumption that the majority of the dwarfs shown in Figure 7 do not have metallicities as low as $-0.5$ dex, and we show below that  while the ad hoc change to the adiabat improves the model fits at some wavelengths, it is not preferred over the models with standard adiabatic cooling. With those assumptions Figure 7 indicates that $4 \lesssim \log K_{\rm zz} \lesssim 6$ for mid-T to early-Y type brown dwarfs. This is consistent with previous model fits to T6 -- T8 brown dwarfs, where the fits were well constrained by mid-infrared spectroscopy (Saumon et al. 2006, 2007; Geballe et al. 2009).  For the latest-T and early-Y dwarfs we adopt  $\log K_{\rm zz}=6$. Figure 7 suggests that the coolest object currently known, W0855, may have a larger diffusion coefficient and we explore this further in \S 10.5.

\subsection{Metallicity and Multiplicity}

The color-color plot best populated by the sample of Y dwarfs is $J -$ [4.5]:[3.6] $-$ [4.5]. This plot is shown in Figure 8. Figure 9 shows the color-magnitude plot  $J -$ [4.5]:$M_{[4.5]}$. The plots are divided into three panels. The top panel is data only, with a linear fit to sources with $J-[4.5] > 3.0$ magnitudes, excluding sources that deviated by $> 2\sigma$ from the fit. The average deviation from the linear fit along the $y$ axes is 0.09 and 0.14 magnitudes in Figures 8 and 9 respectively.  The fit parameters are given in the Figures. The middle panel of each Figure compares the data to non-equilibrium T15 models which differ in metallicity and adiabat gradient. The bottom panel compares the data to S12, M12 and M14 equilibrium models which differ in gravity and cloud cover. 

Figures 8 and 9 show that none of the models reproduce the observed [3.6] $-$ [4.5] color. The cloud-free non-equilibrium models are better than the cloud-free equilibrium models for the T dwarfs, which are more impacted by the dredge-up of CO than the Y dwarfs (Figure 7). The reduction in the adiabatic index and the introduction of clouds improves the fit for the T dwarfs because in both cases the $\lambda \sim 1~\mu$m light emerges from cooler regions of the atmosphere than in the adiabatic or cloud-free case (Morley et al. 2012 their Figure 5, Tremblin et al. 2016 their Figure 5). In the  $J -$ [4.5]:$M_{[4.5]}$ plot (Figure 9), however, the chemical equilibrium chloride and sulphide cloud model does not reproduce the observations of the T dwarfs and the modified adiabat model does not reproduce the observations as well as the adiabatic model does. If we assume that the model trends in the colors with gravity, cloud and metallicity are nevertheless correct, we can extract important information from Figures 8 and 9 for the Y dwarfs. 

Figure 8 suggests that the  $J -$ [4.5]:[3.6] $-$ [4.5] colors of Y dwarfs are insensitive to gravity and clouds, but are sensitive to metallicity. The model trends imply that the following objects are metal-rich: W0350 (Y1) and WISE J041358.14$-$475039.3 (W0413, T9). Similarly the following are metal-poor: WISEPA J075108.79$-$763449.6 (W0751, T9), WISE J035934.06$-$540154.6 (W0359, Y0), WD0806$-$661B (Y1) and W1828 ($>$Y1).

Figure 9 suggests that $J -$ [4.5]:$M_{[4.5]}$ is insensitive to gravity, but is sensitive to clouds and metallicity. Figure 9 supports a sub-solar metallicity for W0359 and W1828, and a super-solar metallicity for W0350. Note that 
SDSS J1416$+$1348B (S1416B, T7.5) is a known metal-poor and high-gravity T dwarf (e.g. Burgasser, Looper \& Rayner 2010).   The Y dwarfs, including W0855 in this plot, appear to be essentially cloud-free. 

In any color-magnitude plot multiplicity leads to over-luminosity. The Y 
dwarf sample size is now large enough, and the data precise enough, that 
we can identify W0535 and W1828 as likely multiple objects. We examined 
the drizzled WFC3 images of these two Y dwarfs for signs of elongation 
or ellipticity. Images of W0535 taken in 2013 September and December show 
no significant elongation or ellipticity, implying that if this is a 
binary system then the separation is $<$ 3 AU. A tighter limit was found by 
Opitz et al. (2016)
who used Gemini Observatory's multi-conjugate adaptive optics system to determine
that any similarly-bright companion must be within $\sim 1$ AU.
For W1828 five  WFC3
images taken in 2013 April, May, June and August show marginal 
elongation and ellipticity of $16\pm8$\%. For this (nearer) source the 
putative binary separation is $\lesssim$ 2 AU.

\subsection{Further Down-Selection of Models}

Figures 10 and 11 show near-infrared colors as a function of  $J -$ [4.5], and absolute $J$ as a function of near-infrared colors.  In both figures, the T15 non-equilibrium models reproduce the trends in $Y-J$ and $J-H$ quite well. The S12, M12 and M14 equilibrium models reproduce $Y-J$ but do poorly with $J-H$ except for the chloride and sulphide models which reproduce the T dwarfs' location.  Non-equilibrium effects are important in the near-infrared as gas transport leads to an enhancement of N$_2$ at the expense of NH$_3$. This increases the flux in the near-infrared, especially at $H$ (e.g. Leggett et al. 2016a), hence the better fit to $J - H$ by the T15 models. The only model that reproduces the $J-K$ colors of the Y dwarfs is the T15 model with the change to the adiabatic index, because of the large reduction in the $J$ flux (Figure 1).

Comparison of the S12 and M14 sequences in Figures 10 and 11 suggest that the near-infrared colors of Y dwarfs are insensitive to clouds, although clouds may be important at the $\sim 10$\% level for Y dwarfs (see also \S 3). Gravity appears to be an important parameter for $J-H$ and $J-K$, and metallicity appears to be important for $Y-J$, $J-H$ and $J-K$. The interpretation of $Y-J$ is not straightforward however as both W0350 and W1828 appear bluer in $Y-J$ than the other Y dwarfs, while Figures 8 and 9 implied that W0350 is metal-rich and W1828 is metal-poor. Figures 10 and 11 suggest that the W0146AB system may be metal-rich, while WISE J033515.01+431045.1 (W0335, T9) may be metal-poor. 
 
Figure 12 shows absolute [4.5] as a function of the mid-infrared colors [3.6]$-$[4.5] and [4.5]$-$W3 (see Figure 1 for filter bandpasses). In this mid-infrared color-color space the change of the adiabat in the T15 models does not significantly change the location of the model sequence. None of the models can reproduce the [3.6]$-$[4.5]:$M_{[4.5]}$ observations of the Y dwarfs, although the non-equilibrium models do a much better job of reproducing these colors for the T dwarfs. The models are mostly within $2\sigma$ of the observational error in the  [4.5]$-$W3:$M_{[4.5]}$ plot, although there is a suggestion that for the Y dwarfs the model [4.5] fluxes are too high and/or the W3 fluxes are too low (see also \S 10.5). 

Comparison of the S12 and M14 sequences in Figure 12 suggests that the mid-infrared colors of Y dwarfs with $M_{[4.5]}>16$ magnitudes, such as W0855, are sensitive to the presence of water clouds.  Gravity does not appear to play a large role in these mid-infrared color-magnitude diagrams, but metallicity does. The discrepancy with observations however makes it difficult to constrain parameters, or determine whether or not W0855 is cloudy, from this Figure.

In the next section we compare near-infrared spectra of Y dwarfs to synthetic spectra. The photometric comparisons have demonstrated that the late-T and Y dwarfs are mostly cloud-free and that non-equilibrium chemistry is important for interpretation of their energy distributions (see also Leggett et al. 2016a). We therefore compare the spectra to T15 models only. We use a single diffusion coefficient of $\log K_{\rm zz} = 6$, as indicated by our $M^{\prime}$ measurements (\S 8.1). We also use only non-modified adiabats. Although the modified adiabat produces redder colors which in some cases agree better with observations, it does so by reducing the $YJH$ flux (Figure 1). Based on the mid-infrared color-magnitude plot (Figure 12), the problem appears to be a shortfall of flux in the models at [3.6] (and $K$).  Note that the [3.6] (and W1) filter covers a region where the flux increases sharply to the red as the very strong absorption by CH$_4$ decreases  (Figure 1, bottom panel). A relatively small change in this slope may resolve the observed discrepancy.

\section{Spectroscopy: The Sample and Comparison to Models}

In this section we compare near-infrared spectra of Y dwarfs to T15 non-equilibrium cloud-free models. We analyse Y dwarfs that have trigonometric parallax measurements only, so that the model fluxes can be scaled to the distance of the Y dwarf. 
Given the problems at $K$ (\S 8.3), we only use the $YJH$ wavelength region (most of the observed spectra only cover this region). 

Figures 13 and 14 are color-magnitude plots for the known Y dwarfs  and latest T dwarfs. Sequences for the T15 solar, super-solar and sub-solar metallicity models are shown, as well as sequences for $\log g=$4.0, 4.5 and 4.8.  For this low--temperature solar--neighborhood sample $\log g$ almost directly correlates with mass, and   $T_{\rm eff}$ provides age once $\log g$ (mass) is known (Figure 2). Figures 2 and 14 suggest that $M_{[4.5]}$ is almost directly correlated with   $T_{\rm eff}$ for the Y dwarfs. This is consistent with the radii of the Y dwarfs being approximately constant, and the model calculation that half the total flux is emitted through the [4.5] bandpass for this range of $T_{\rm eff}$ (\S 2).  Figures 13 and 14 show that the T15 models 
indicate effectively the same value of  $T_{\rm eff}$ based on $M_J$ or $M_{[4.5]}$ for each Y dwarf. Excluding the very low-luminosity W0855, the Y dwarfs have $325 \lesssim T_{\rm eff}$~K $\lesssim 450$. 

Near-infrared spectra of 20 Y dwarfs or Y dwarf systems with trigonometric parallaxes are available from this work, Cushing et al. (2011),  Kirkpatrick et al. (2012), Tinney et al. (2012), Kirkpatrick et al. (2013), Leggett et al. (2014), Schneider et al. (2015), and Leggett et al. (2016a). We flux calibrated the spectra using the observed near-infrared photometry which has a typical uncertainty of 10 -- 20\% (Table 8).   For W0535 the flux calibration of the $H$ region of the spectrum is inconsistent with the $Y$ and $J$ region by a factor of four, and our fit suggests that the $H$ region of the spectrum is too bright.  For W1828 the flux calibration of the $J$ and $H$ spectral regions differ by a factor of two. We explored fits to the spectrum using both scaling factors, and the fits suggest there is a spurious flux contribution in the shorter wavelengths of the spectrum. The color-magnitude plots imply that W0535 and W1828 are multiple systems (\S 8.2), and contemporaneous near-infrared spectroscopy and photometry would be helpful in excluding variability in these sources, and enabling a more reliable spectral fit.

We compared the spectra to a set of T15 cloud-free, standard-adiabat models with $\log K_{\rm zz} = 6$. Solar metallicity models were computed with surface gravities given by  $\log g =$ 4.0, 4.5 and 4.8, for $T_{\rm eff}$ values of 300~K to 500~K in steps of 25~K. Solar metallicity $\log g=$3.8 models were calculated for $T_{\rm eff} = 325$, 350 and 375~K also, which evolutionary models show are plausible for a solar neighborhood sample (Figure 2).   A few metal-poor 
([m/H]$=-0.2$ and $-0.5$) and metal-rich  ([m/H]$=+0.2$, $+0.3$ and $+0.4$) models were calculated as needed, when exploring individual fits.
The model fluxes are converted from stellar surface flux to flux at the distance of the Y dwarf using the observed trigonometric parallax and the radius that corresponds to the $T_{\rm eff}$ and $\log g$ of the model as given by Saumon \& Marley (2008) evolutionary models. The typical uncertainty in the parallax-implied distance modulus is  
10 -- 20\% (Table 8). No other scaling was done to the models to improve agreement with the observations.

Due to the coarse nature of our model grid and the poor signal to noise of most of the spectra (due to the faintness of the sources), we fit the spectra by eye only.   We  determined the difference between the computed and observed [4.5] magnitude as a further check of the validity of the selected models. The models that are preferred for the near-infrared spectral fitting give [4.5] values that are within 0.35 magnitudes of the observed value, and on average they are within 0.15 magnitudes of the observed [4.5] magnitude. The spectroscopic fits and $\delta$[4.5] values support the photometrically identified 
non-solar metallicity values for W0350 (metal-rich), W0359 (metal-poor) and W1828 (metal-poor)  (\S 8.2).

Our selected fits for the sample of 20 Y dwarfs are shown in Figures 15 -- 18, where the spectra are grouped by  $T_{\rm eff}$.  For several of the Y dwarfs we show two fits which straddle the observations. Those multiple fits indicate that the uncertainty in the derived  $T_{\rm eff}$, $\log g$ and [m/H] is approximately half the model grid spacing: $\pm 15$~K, $\pm 0.25$ dex and $\pm 0.15$  dex respectively.  Better fits could be determined with a finer grid of models and a least-squares type of approach, but this would only be worthwhile when higher signal to noise spectra are available. 

Overall, the fits to the warmer half of the sample, with  $425 \leq T_{\rm eff}$~K $\leq 450$, are very good. For the cooler half of the sample, with $325 \leq T_{\rm eff}$~K $\leq 375$, the model spectra appear to be systematically too faint in the $Y$-band or, alternatively, too bright at $J$ and $H$. The discrepancy may be associated with the formation of water clouds, which are expected to become important at these temperatures, and which are not included in the T15 models. We discuss this further in \S 10.5.

\section{Properties of the Y Dwarfs}

Table 9 gives the estimated properties of the sample of   24 Y dwarfs, based on near-infrared spectra and photometry, or photometry only if there is no spectrum available, or in one case the photometry and the properties of its companion. Mass and age is estimated from  $T_{\rm eff}$ and $\log g$ using the evolutionary models of Saumon \& Marley (2008, Figure 2), allowing for the uncertainty in the temperature and gravity determinations. Table 9 also lists the tangential velocities ($v_{\rm tan}$) for the Y dwarfs with parallax measurements. Dupuy \& Liu (2012, their Figure 31) use a Galaxy model to show that low-mass dwarfs with $v_{\rm tan} < 80$~kms$^{-1}$ are likely to be thin disk members, and those with  $80 < v_{\rm tan}$~kms$^{-1} <100$ may be either thin or thick disk members.  
Twenty-one of the twenty-two Y dwarfs with $v_{\rm tan}$ measurements have $v_{\rm tan} < 80$~kms$^{-1}$, the remaining Y dwarf, the very low-temperature W0855, has   $v_{\rm tan} = 86$~kms$^{-1}$. There is significant overlap in the Galactic populations in kinematics, age and metallicity but generally the thin disk is considered to be younger than $\sim7$~Gyr  and have a metallicity  $-0.3 \lesssim$ [Fe/H] $\lesssim +0.3$, while the thick disk is older than  $\sim9$~Gyr and has  $-1.0 \lesssim$ [Fe/H] $\lesssim -0.3$ (e.g. Bensby, Feltzing \& Oey 2014). 

Leggett et al. (2016a) compare near-infrared spectra and photometry to T15 models for three Y dwarfs in common with this work: W0350, W1217B and W1738. The technique used is similar to that used here (although fewer models were available) and the derived  $T_{\rm eff}$ and $\log g$ are in agreement, given our new determination for the distance to W0350.  Schneider et al. (2015) compare {\it HST} near-infrared spectra and {\it Spitzer} mid-infrared photometry for a set of Y dwarfs to the S12, M12 and M14 solar-metallicity equilibrium-chemistry models. A goodness-of-fit parameter is used which incorporates the distance and evolutionary radius associated with the model parameters. As stated by Schneider et al., the  fits to the data are poor in many cases. This is likely due to a combination of the omission of chemical non-equilibrium in the models and poorly constrained parallaxes for some of the Y dwarfs. The range in the Schneider et al.  temperature and gravity values for each Y dwarf is about twice that derived here. There is generally good agreement between our values and those of Schneider et al.. Of the sample of 16 objects in common, only five have   $T_{\rm eff}$ or $\log g$ values that differ by more than the estimated uncertainty.  For two of these we use different values for the parallax (W0535 and W0825); for another pair (W0647 and WISEA J163940.84$-$684739.4 (W1639)) the  $T_{\rm eff}$ values are consistent but the Schneider et al. gravities are significantly higher; and for the remaining object (WISEA J220905.75+271143.6 (W2209)), Schneider et al. obtain a much higher temperature. For the last three Y dwarfs the higher gravities or temperature are unlikely, based on age and luminosity arguments. 

We discuss our results in terms of populations, and also discuss individual Y dwarfs of particular interest, in the following sub-sections. Two Y dwarfs without trigonometric parallax measurements are not discussed further: W0304 and WISEA J235402.79$+$024014.1. Two T9 dwarfs appear to have significantly non-solar metallicity and should be followed up: WISE J041358.14$-$475039.3 (metal-rich) and WISEPA J075108.79$-$763449.6 (metal-poor).

\subsection{Likely Young, Metal-Rich, Y dwarfs}

Five Y dwarfs have low tangential velocities of $8 \leq v_{\rm tan}$~kms$^{-1} \leq 40$, appear to be metal-rich and also have an age $\lesssim 3$~Gyr as estimated from  $T_{\rm eff}$ and $\log g$. These are W0350, W0825, W1141, W1206 and W1738. They also appear to be low-mass $\sim$8 Jupiter-mass objects.

\subsection{Likely Solar-Age and Solar-Metallicity Y dwarfs}

Fourteen Y dwarfs have kinematics, metallicities and age, as implied by  $T_{\rm eff}$ and $\log g$, that suggest they are generally solar-like in age and chemistry. These have estimated ages of 3 -- 8 Gyr and masses of 10 -- 20  Jupiter-masses: 
W0146B, W0359, W0410, W0535 (if an equal-mass binary), W0647,  W0713, WISE J073444.02$-$715744.0  (W0734), W1217B, W1405, W1541, W1639, W2056, W2209 and W2220.
All these Y dwarfs have a tangential velocity and estimated age consistent with thin disk membership, although W0713 and W2209 have upper limits on their age of 12 and 15~Gyr respectively.

\subsection{WISEPA J182831.08$+$265037.8} 

The super-luminosity of W1828 in the color-magnitude diagrams does not seem explainable by any other means than binarity. 
The selected model in the case of W1828 being an equal-mass binary implies that this system is relatively young. It would be composed of two $\sim$6 Jupiter-mass objects and have an age $\sim$1.5 Gyr. The $HST$ images imply that the binary separation is $\lesssim 2$~AU (\S 8.2). An age of 1.5~Gyr is notionally at odds with the apparently very metal-poor nature of the system. 

A better near-infrared spectrum, and a mid-infrared spectrum when the {\it James Webb} telescope is on-line should improve our understanding of this Y dwarf. Exploration of a non-identical binary pair solution would also be worthwhile once a better spectrum is available.

\subsection{WD0806$-$661B}

The primary of this binary system, WD0806$-$661A, is a helium-rich DQ-class white dwarf separated from the brown dwarf by 
2500~pc (Luhman, Burgasser \& Bochanski 2011). Hydrogen-deficient post-asymptotic-giant-branch (AGB) stars may evolve into DB (helium-rich white dwarfs)  and then DQ white dwarfs (Althaus et al. 2005; Dufour, Bergeron \& Fontaine 2005). Calculations of the late stages of AGB evolution 
can produce the less common non-DA (non-hydrogen-rich) white dwarfs in about the correct proportion although there are multiple paths that lead to hydrogen deficiency (Lawlor \& MacDonald 2006). One factor in these AGB evolution models is the metallicity of the star, and it is possible that the sub-solar metallicity we find for  the Y dwarf WD0806$-$661B is related to the DQ (i.e. non-DA) nature of the primary.
Table 9 gives the properties of this Y dwarf, using the white dwarf primary to constrain the age of the system to 1.5 -- 2.7~Gyr (Rodriguez et al. 2011).

\subsection{WISE J085510.83$-$071442.5}

Figure 19 shows the photometric data points observed for W0855, as fluxes as a function of wavelength. Table 8 lists 
$YJH$, [3.6], [4.5], W1, W2 and W3 magnitudes for W0855. $YJH$ were derived by us from {\it HST} photometry (\S 6, Table 6). The [3.6] and [4.5] are from LE16, the W1 and W2 magnitudes are from the {\it WISE} catalogue, and W3 was determined here from {\it WISE} images (\S 5.4, Table 3). Also shown are shorter wavelength data points from LE16: LP850 obtained using the {\it HST} Advanced Camera for Surveys (ACS) and the $i$-band upper limit obtained using GMOS at Gemini South.

Model spectra are shown for comparison, all with  $T_{\rm eff}=250$~K. Models with $T_{\rm eff}=225$~K produce too little flux at 5~$\mu$m, and models with  $T_{\rm eff}=275$~K produce too much flux at 5~$\mu$m, by a factor of 1.5. 
The models show that $\sim40$\% of the total flux is emitted through the [4.5] bandpass and  $\sim30$\% through the W3 bandpass. An additional $\sim20$\% is emitted at $19 < \lambda~\mu$m $<28$ (the W4 bandpass), and $\sim5$\% at $\lambda > 30~\mu$m. Less than 1\% of the total flux is emitted at $\lambda < 4~\mu$m.
The effective temperature (or luminosity) is tightly constrained by the mid-infrared flux, and we estimate that for W0855  $T_{\rm eff}=250 \pm 10$~K. This is consistent with previous studies (Luhman 2014,  Beamin et
al. 2014, Leggett et al. 2015, Schneider et al. 2016, Zapatero Osorio et al. 2016).

We compared several models to the spectral energy distribution. T15 models were calculated with $\log g =$ 3.5, 3.8, 4.0, 4.3 and 4.5.   T15 models with $T_{\rm eff}=250$~K and non-solar metallicities of [m/H] $= -0.2$ and $+0.2$ were also calculated. Finally, T15 models were calculated with  $\log K_{\rm zz} = 6$, 8 and 9, as
Figure 7 suggests that W0855 may be undergoing more chemical mixing than the warmer Y dwarfs, and Jupiter's 
atmosphere has been modelled with a vertical diffusion coefficient of $\log K_{\rm zz} = 8$
(Wang et al. 2015). The top panel of Figure 19 demonstrates the effect of varying $K_{\rm zz}$, and the central panel shows models with different gravities.

We also compared the observations to M14 cloud-free and partly-cloudy models that are in chemical equilibrium. These are shown in the bottom panel of Figure 19.
Cloud-free models with solar metallicity and $\log g=$ 3.5 and 4.0 were calculated, as well as a $\log g=$ 4.0 solar metallicity model with thin clouds decks (parameterized by $f_{\rm sed} = 7$) covering 50\% of the surface.  
The models are updated versions of the models published in Morley et al. (2014). 
The new models include updates to both chemistry and opacities, which will be described in detail in an upcoming paper (Marley et al. in prep.).  Briefly, the opacities are as described in Freedman et al. (2014) with the exception of the CH$_4$ and alkali opacities; CH$_4$ line lists have been updated using Yurchenko \& Tennyson (2014) and the alkali line lists have been updated to use the results from Allard, Allard \& Kielkopf (2005). Chemical equilibrium calculations are based on previous thermochemical models (Lodders \& Fegley 2002; Visscher, Lodders \& Fegley 2006; Visscher 2012), and have been revised and extended to include a range of metallicities.

Up to this point in our analysis we have neglected the $\lambda < 0.9~\mu$m region. Given the importance of W0855, and the availability of shorter wavelength data, Figure 19 includes observations and model spectra and photometry at  $0.8 \lesssim \lambda~\mu$m $\lesssim 0.9$. The models are about an order of magnitude too bright at $\lambda < 0.9~\mu$m, with the T15 models more discrepant than the modified M14 models. At the temperatures and pressures that are likely in the W0855 photosphere, H$_2$S is a significant opacity source at $\lambda \lesssim 0.9~\mu$m (M14, their Figures 7 and 8). Possibly the strength of this opacity is dependent on the treatment of the condensation of sulfides at warmer temperatures. This issue will be explored in future work.

All the models show the discrepancy with observations at the [3.6] bandpass noted previously in \S 8.2. 
At this temperature all the models also appear too faint at $H$ by about a factor of two. Increasing the mixing coefficient from $\log K_{\rm zz} = 6$ to 9 improves the agreement at $YJH$ by 10 -- 15\%, and at [4.5] and W3 by 5\%. The addition of water clouds improves the agreement at $H$ but makes $Y$ and $J$ too bright by about an order of magnitude. While the addition of water clouds and associated brightening at $Y$ and $J$ may improve the model fits for the 
$T_{\rm eff} \approx 350$~K Y dwarfs (Figures 17 and 18), at 250~K the addition of water clouds (as currently modelled) does not improve the fit in the near-infrared.  We note that condensation of NH$_3$ is not expected until lower temperatures of $\sim$200~K are reached. 
Esplin et al. (2016) have detected variability at [3.6] and [4.5] for W0855, at the $\sim4$\% (peak-to-peak) level. Similar variability is seen in two Y dwarfs that are too warm for water clouds although they may have low-lying sulphide clouds (Cushing et al. 2016, Leggett et al. 2016b). Skemer et al. (2016) present a $5~\mu$m spectrum for W0855 which suggests that water clouds are present. No analysis yet has robustly confirmed the presence of clouds in the W0855 atmosphere (Esplin et al. 2016) and new models are needed which better reproduce the SED of W0855 before their presence or absence can be confirmed.

About 70\% of the total flux from W0855 emerges through the [4.5] and W3 filters and it is important therefore that the models reproduce the observed [4.5] and W3 magnitudes. However, Figure 12 suggests that there is a systematic offset between modelled and observed values of [4.5] $-$ W3. The analysis presented here has shown good agreement between observations and models at [4.5]  (Figures 7, 9, 14), which suggests that the calculated values of W3 may be $\sim 0.5$ magnitudes too faint. The uncertainty in the measured [4.5] and W3 magnitudes for W0855 are 0.04 and 0.30 magnitudes respectively. We restrict models of the W0855 energy distribution to  those where 
the difference $\delta = M({\rm model}) - M({\rm observed})$ magnitude is such that $-0.25 \leq \delta([4.5]) \leq +0.25$ and   $-0.3 \leq \delta(W3) \leq +0.6$. Table 10 lists the M14 and T15 models considered here which satisfy those criteria. 

We find that current models imply that the 250~K Y dwarf W0855 is undergoing vigorous mixing, has a metallicity within $\sim$0.2 dex of solar, has little or no cloud cover, and has a range in surface gravity of $3.5 \lesssim \log g \lesssim 4.3$. The Saumon \& Marley (2008) evolutionary models then give a mass range of 1.5 -- 8 Jupiter masses, and an age range of 0.3 -- 6 Gyr (Table 9). The relatively high tangential velocity  of W0855 of $86 \pm 3$ km s$^{-1}$ suggests the higher age (and higher mass) may be more likely.

\section{Conclusion}

We present new Gemini near-infrared spectroscopy for three Y dwarfs,
near-infrared photometry for two Y dwarfs, and $5~\mu$m photometry for four late-T and two Y dwarfs. We also present new near- and mid-infrared photometry for 19 T6.5 and later-type brown dwarfs, including 8 Y dwarfs, using archived images. We have determined improved astrometry for four Y dwarfs, also by using archived images. Combining the new photometry with data taken from the literature allows us to transform $CH_4$(short) and WFC3 photometry on to MKO $YJH$. We give  a newly homogenized photometric data set for the known Y dwarfs (Table 8) which enables better comparisons to models as well as the identification of trends and outliers. 

Using MKO-system color-magnitude diagrams and the new 
parallaxes, we find that two of the Y dwarfs are likely to be binaries 
composed of similar-mass objects: W0535 and W1828 (Figures 9, 14). WFC3 and Gemini adaptive optics images of W0535 from Opitz et al. (2016) do not resolve W0535. WFC3 images of W1828 show marginal elongation and ellipticity of $16 
\pm 8$\%. The separation of the putative binaries are $\lesssim 2$~AU 
for W1828 and $< 1$~AU for W0535.

The models show that the $J -$ [4.5]:[3.6] $-$ [4.5] 
and the $J -$ [4.5]:$M_{[4.5]}$ diagrams can be used to estimate
metallicity (Figures 8, 9). 
We refine our atmospheric parameter estimates by comparing near-infrared spectra for 20 of the Y dwarfs
to synthetic spectra generated by cloud-free non-equilibrium chemistry models (Figures 15 -- 18).
We find that all the known Y dwarfs have metallicities within 
0.3~dex of solar, except for W0350 which has
[m/H]$\sim +0.4$~dex and  W1828 which has
[m/H]$\sim -0.5$~dex. All the known Y dwarfs with measured parallaxes are 
within 20~pc of the Sun, and therefore solar-like metallicities are expected. 

Assuming W1828 is an equal-mass binary, we derive a low gravity for the 
pair, which translates into a relatively young age of $\approx 1.5$~Gyr. 
Notionally, this is inconsistent with the degree of metal paucity that 
we find, as the thin disk generally has $-0.3 \lesssim$ [Fe/H] $\lesssim +0.3$ (e.g. Bensby, Feltzing \& Oey 2014). 
An improved near-infrared spectrum is needed for this source, 
preferably taken close in time to photometry, as the current spectrum 
and photometry are discrepant and variability needs to be excluded.

The atmospheric parameters determined by fitting the near-infrared spectra are consistent with the values estimated photometrically. The 
synthetic spectra generated by T15 non-equilibrium chemistry cloud-free models reproduce observations well for the 
 warmer half of the sample with $425 \leq T_{\rm eff}$~K $\leq 450$.
For the cooler Y dwarfs with  $325 \leq T_{\rm eff}$~K $\leq 375$ the models seem consistently faint at $Y$. A comparison of models to a pseudo-spectrum of the 250~K W0855 shows that models with patchy clouds are brighter at $Y$ than cloud-free models, and the discrepancy seen  in the $1~\mu$m flux of Y dwarfs with $T_{\rm eff} \approx$ 350~K may be due to the onset of water clouds. However the cloudy models produce too much flux at $0.8 < \lambda~\mu$m $<1.3$ for the cooler W0855. It is unclear if there is missing opacity at lower temperatures, or if the atmosphere of this cold object is cloud-free. All the Y dwarf atmospheres appear to be turbulent, with vertical mixing leading to non-equilibrium chemistry.

We determine masses and ages for   22 Y dwarfs  from evolutionary 
models, based on our temperature and gravity estimates (Table 9). Approximately 90\% of the sample has an estimated age of 2 to 6~Gyr, i.e. thin-disk-like as would be expected for a local sample. W1141 appears younger, with age $\sim$0.6~Gyr, and W0713 and W2209 appear older, with ages 7 and 8.5~Gyr  respectively. About 70\% of the sample has a mass of 10 -- 15 Jupiter-masses. W0350, WD0806$-$661B, W0825, W0855, W1141 and W1828 (if an equal-pair binary) have masses of 3 -- 8  Jupiter-masses. W0713 appears to be a 20  Jupiter-mass Y dwarf. 
A larger sample is needed to constrain the shape of the mass function and the low-mass limit for star-like brown dwarf formation.
We may not find more Y dwarfs however, unless or until a more sensitive version of {\it WISE} is flown.

\clearpage

\appendix

\section{Astrometric Measurements of WISE J014656.66$+$423410.0AB,\\ WISE J053516.80$-$750024.9, WISEPC J121756.91$+$162640.2AB and \\ WISEPC J140518.40$+$553421.}

Tables 11 -- 14 give the astrometric measurements of the four Y dwarfs or dwarf systems for which we present new trigonometric parallaxes in \S 7:  WISE J014656.66$+$423410.0AB,
WISE J053516.80$-$750024.9, WISEPC J121756.91$+$162640.2AB and
WISEPC J140518.40$+$553421. 

\section{On-Line Data Table}

Table 15 gives the spectral types, distance moduli and photometric data used in this work, together with data sources.
The Table consists of 97 brown dwarfs or brown dwarf systems with spectral types T6 and later.

\acknowledgments

Some of the data presented in this paper were obtained from the Mikulski Archive for Space Telescopes (MAST). STScI is operated by the Association of Universities for Research in Astronomy, Inc., under NASA contract NAS5-26555. Support for MAST for non-HST data is provided by the NASA Office of Space Science via grant NNX09AF08G and by other grants and contracts.

Based on data obtained from the ESO Science Archive Facility.

This research has made use of the NASA/ IPAC Infrared Science Archive, which is operated by the Jet Propulsion Laboratory, California Institute of Technology, under contract with the National Aeronautics and Space Administration.

This publication makes use of data products from the Wide-field Infrared Survey Explorer, which is a joint project of the University of California, Los Angeles, and the Jet Propulsion Laboratory/California Institute of Technology, funded by the National Aeronautics and Space Administration. This research has made use of the NASA/ IPAC Infrared Science Archive, which is operated by the Jet Propulsion Laboratory, California Institute of Technology, under contract with the National Aeronautics and Space Administration.

The Center for Exoplanets and Habitable Worlds is supported by the
Pennsylvania State University, the Eberly College of Science, and the
Pennsylvania Space Grant Consortium.

Based in part on observations obtained at the Gemini Observatory, which is operated by the Association of Universities for
Research in Astronomy, Inc., under a cooperative agreement with the NSF on behalf of the Gemini partnership: the National Science Foundation (United States), the National Research Council (Canada), CONICYT (Chile), 
    Minist\'{e}rio da Ci\^{e}ncia, Tecnologia e Inova\c{c}\~{a}o (Brazil)
    and Ministerio de Ciencia, Tecnolog\'{i}a e Innovaci\'{o}n Productiva
    (Argentina).
S. L.'s research is supported by Gemini Observatory.

%\clearpage

\begin{figure}
\includegraphics[angle=0,width=0.85\textwidth]{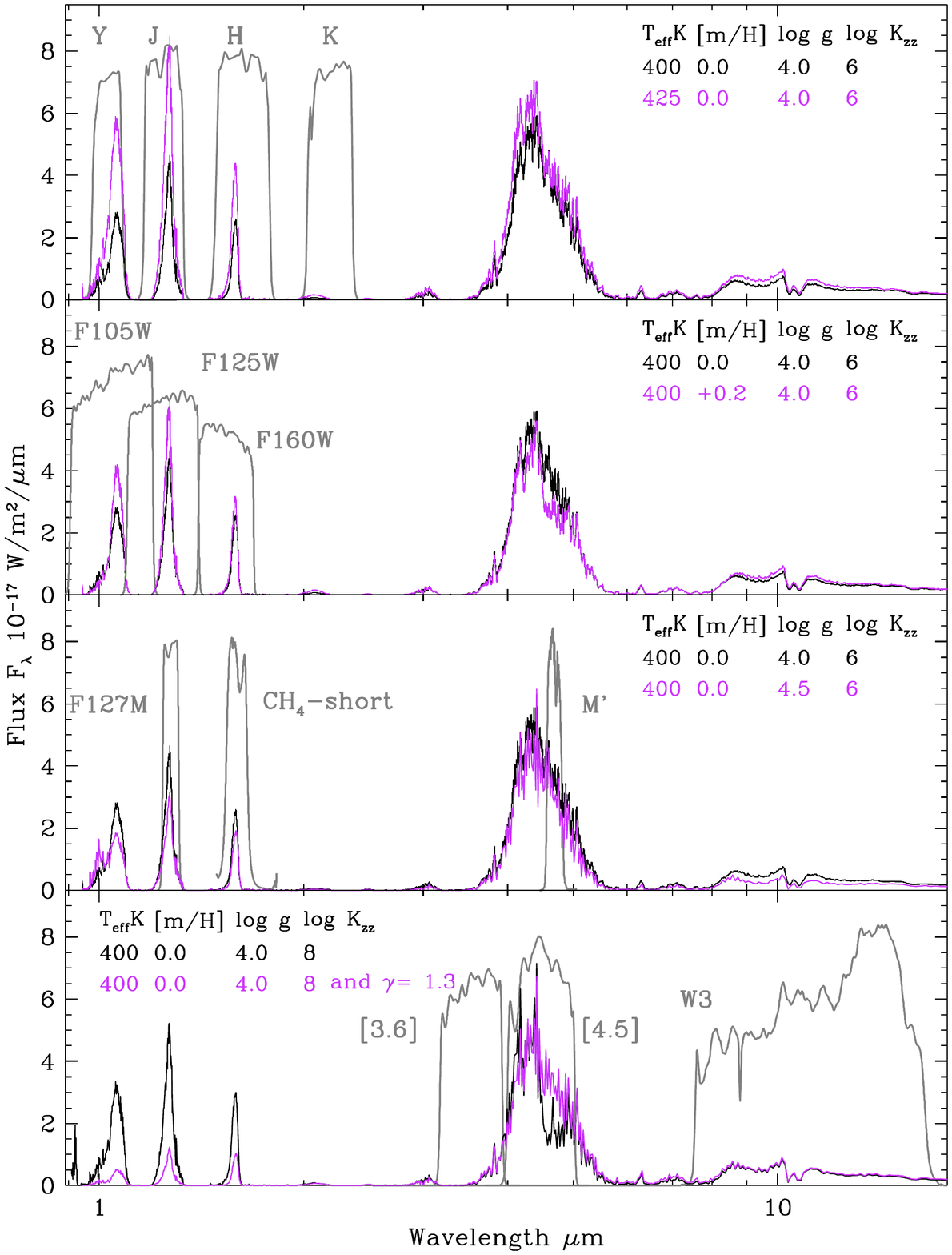}
\vskip -0.6in
\caption{Spectral energy distributions of Y dwarfs with $T_{\rm eff}$=400~K at a distance of 10~pc for $0.9 \leq \lambda~\mu$m $\leq 17.8$, generated by Tremblin et al. (2015) models. In the top panel spectra are shown for different  $T_{\rm eff}$, in the next panel [m/H] is varied, in the next $\log g$ is varied, and the bottom panel demonstrates the effect of changing the diffusion coefficient $K_{\rm zz}$ and the adiabatic index $\gamma$ (see text). The MKO $YJHKM^{\prime}$ and $CH_4$, the {\em HST} WFC3 F105W, F125W, F127M and F160W, the {\em WISE} W3 and the {\em Spitzer} [3.6] and [4.5] bandpasses are also shown.
\label{fig1}}
\end{figure}

\clearpage

\begin{figure}
\includegraphics[angle=-90,width=1.0\textwidth]{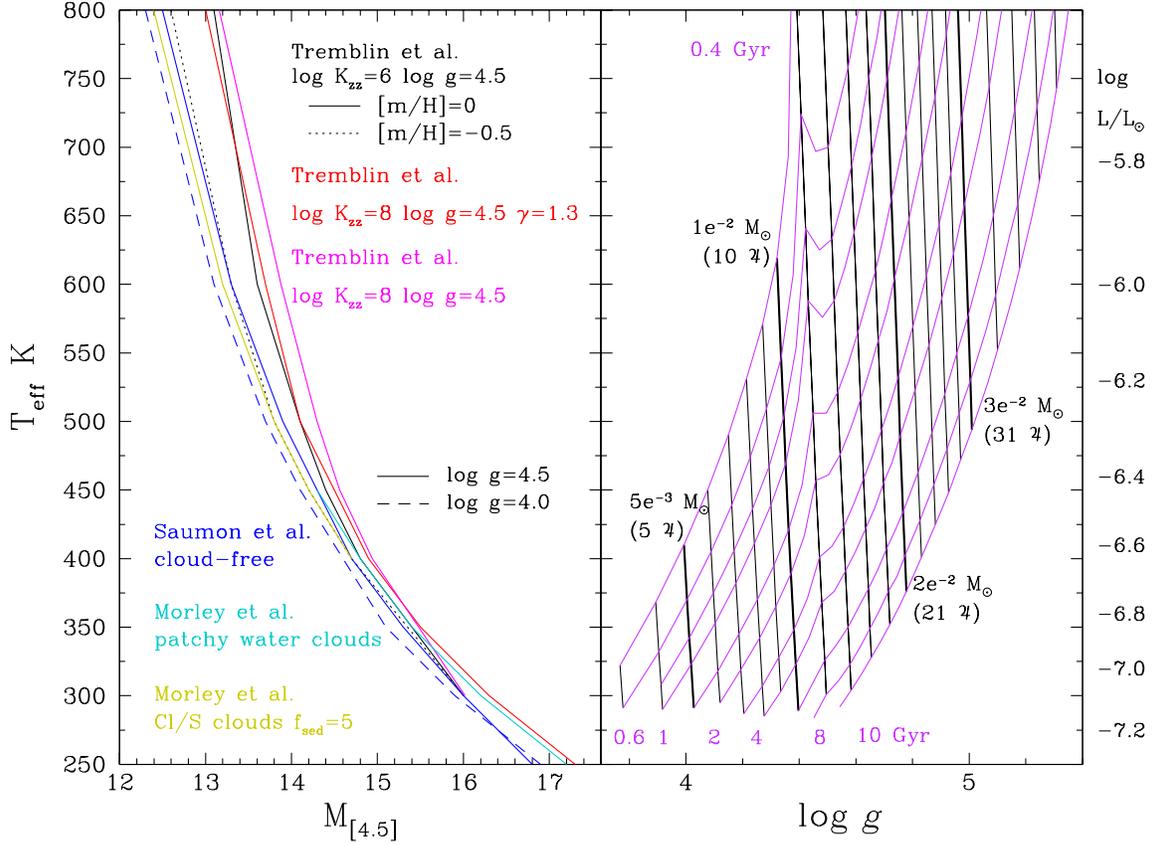}
\caption{Effective temperature $T_{\rm eff}$~K is plotted on the $y$-axis with  
absolute [4.5] magnitude on the $x$-axis in the left panel and $\log g$ in the right panel. The right axis gives
$\log L/L_{\odot}$ values.
Sequences in the left panel are from various model atmospheres as indicated in the legend. Sequences in the right panel are taken from Saumon \& Marley (2008) evolutionary models. In the right panel violet lines are isochrones for ages as indicated along the bottom, and the almost vertical black sequences are lines of constant mass. From left to right sequences for 0.003, 0.004, 0.005, 0.006, 0.007, 0.008, 0.009, 0.010, 0.012, 0.014, 0.016, 0.018, 0.020, 0.022, 0.024, 0.026, 0.028, 0.030, 0.035, 0.040, 0.045 and 0.050 solar mass are shown. One solar mass $=$ 1047 Jupiter masses. 
\label{fig2}}
\end{figure}

\clearpage

\begin{figure}
\includegraphics[angle=0,width=0.9\textwidth]{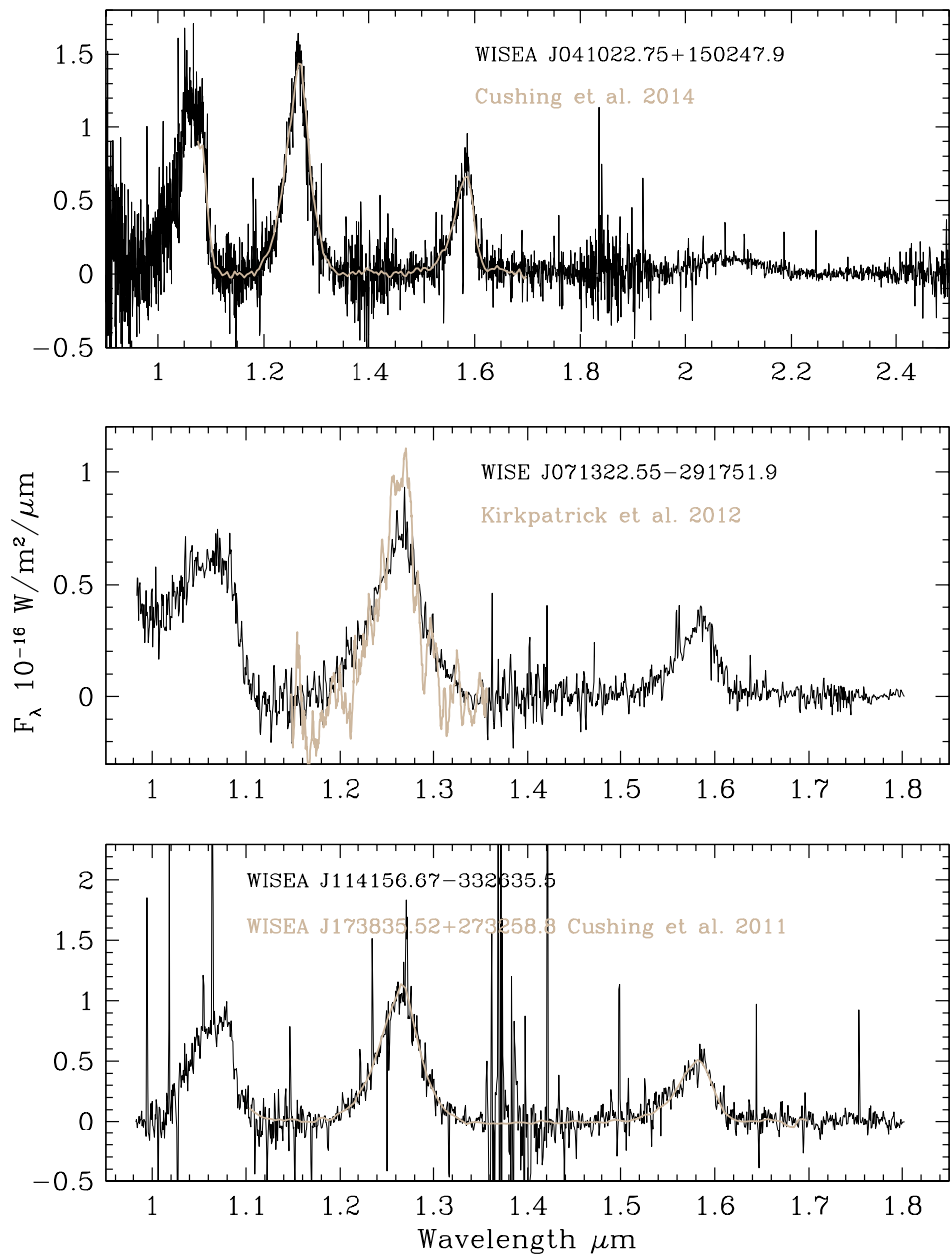}
\vskip -0.5in
\caption{Black lines in the top, middle and bottom panels are 
unsmoothed spectra of WISEA J041022.75$+$150247.9, WISE J071322.55$-$291751.9 and WISEA  J114156.67$-$332635, respectively, determined here. Previously published spectra of the first two objects are shown as tan lines. 
In the bottom panel the tan line is the spectrum of the Y0 standard WISEA J173835.52$+$273258.8, which is very similar in shape to WISE 1141.
\label{fig3}}
\end{figure}

\clearpage

\begin{figure}
\includegraphics[angle=0,width=0.48\textwidth]{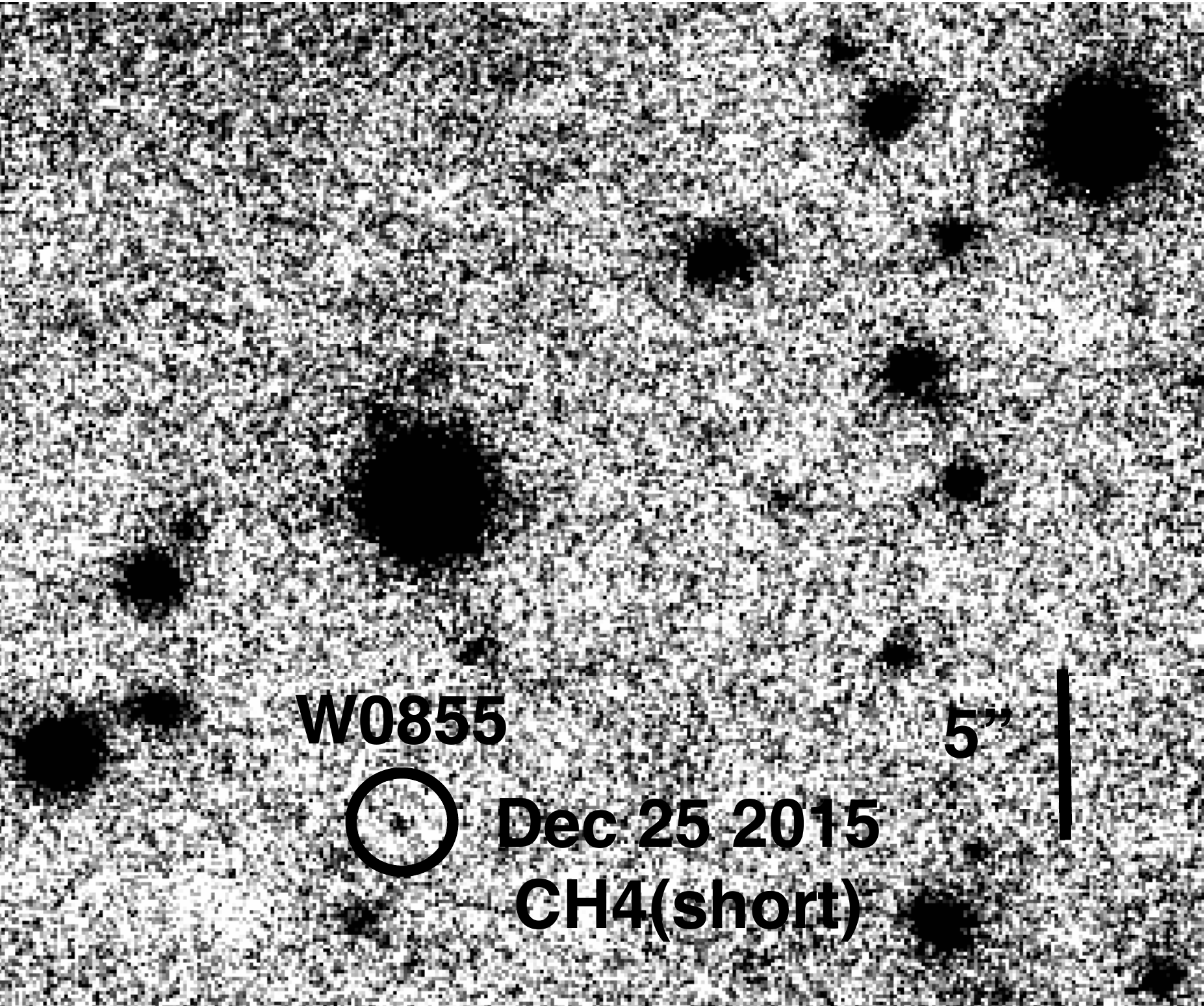}
\includegraphics[angle=0,width=0.48\textwidth]{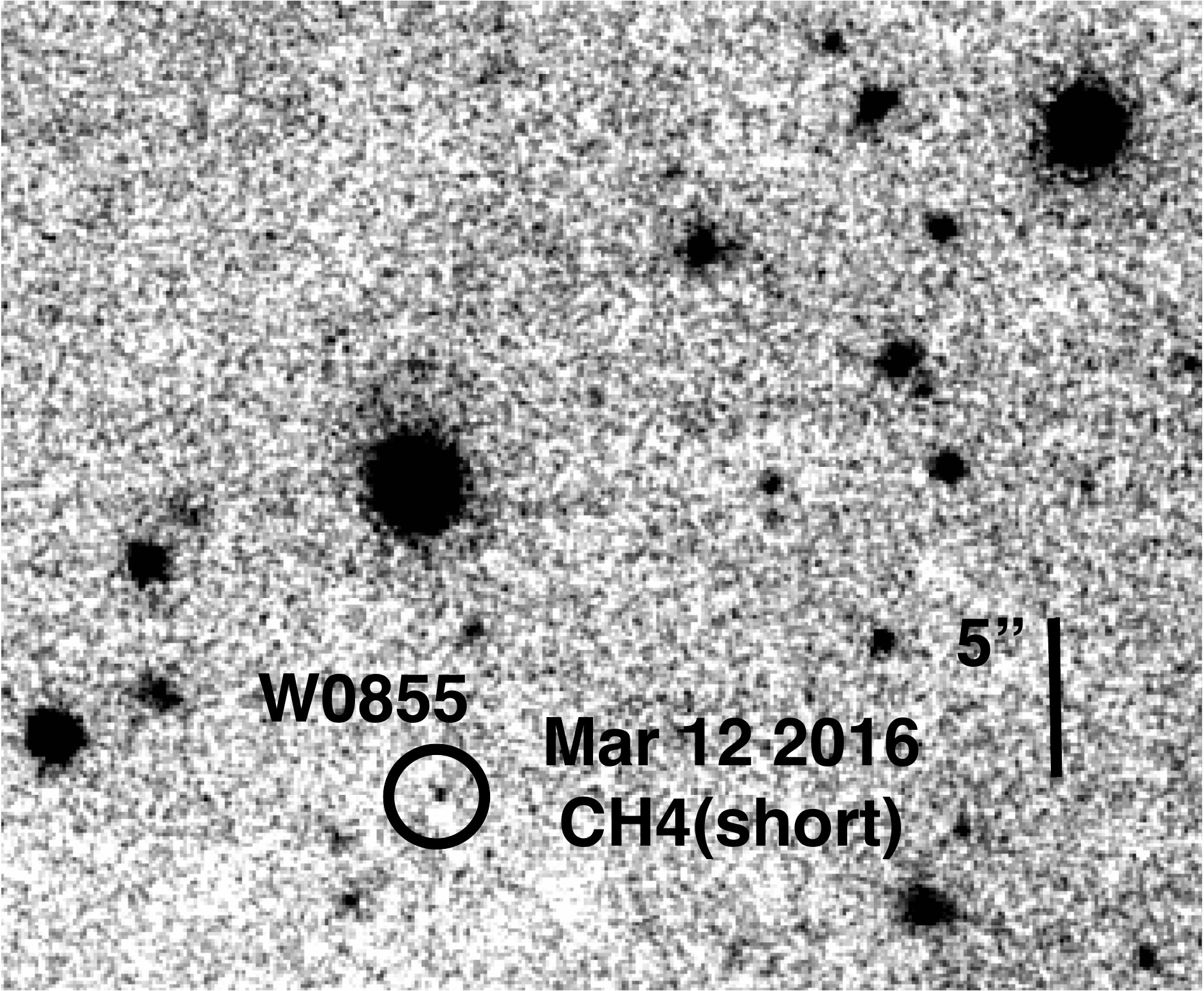}\\
\includegraphics[angle=0,width=0.48\textwidth]{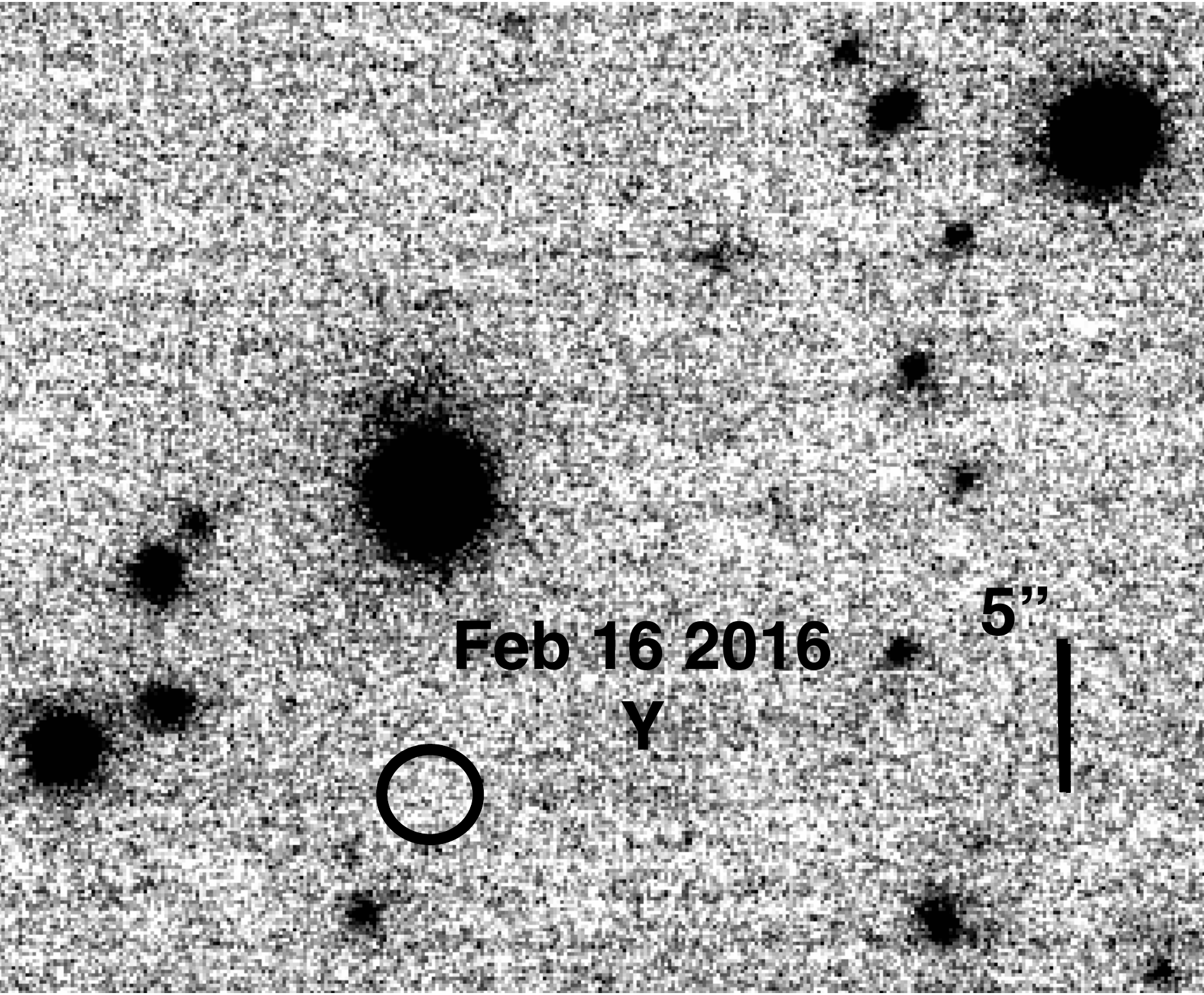}
\includegraphics[angle=0,width=0.48\textwidth]{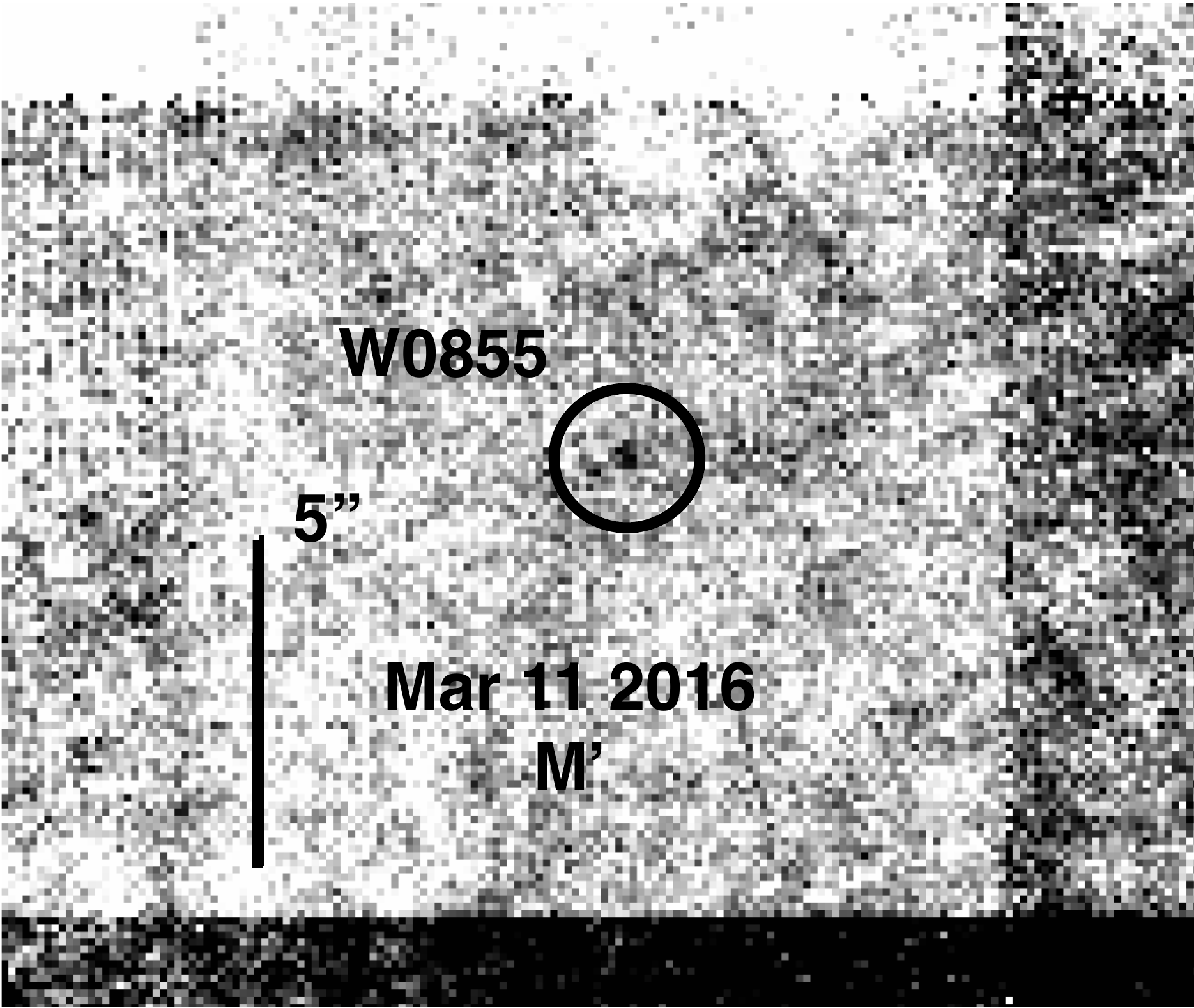}
\caption{NIRI images of WISE J085510.83$-$071442.5. North is up and East to the left, the scale is indicated by the $5''$ vertical bar. The source is circled in the CH$_4$(short) and $M^{\prime}$ images. The source is not detected in $Y$, the circle indicates the expected location.
\label{fig4}}
\end{figure}

\clearpage

\begin{figure}
    \includegraphics[angle=0,width=.85\textwidth]{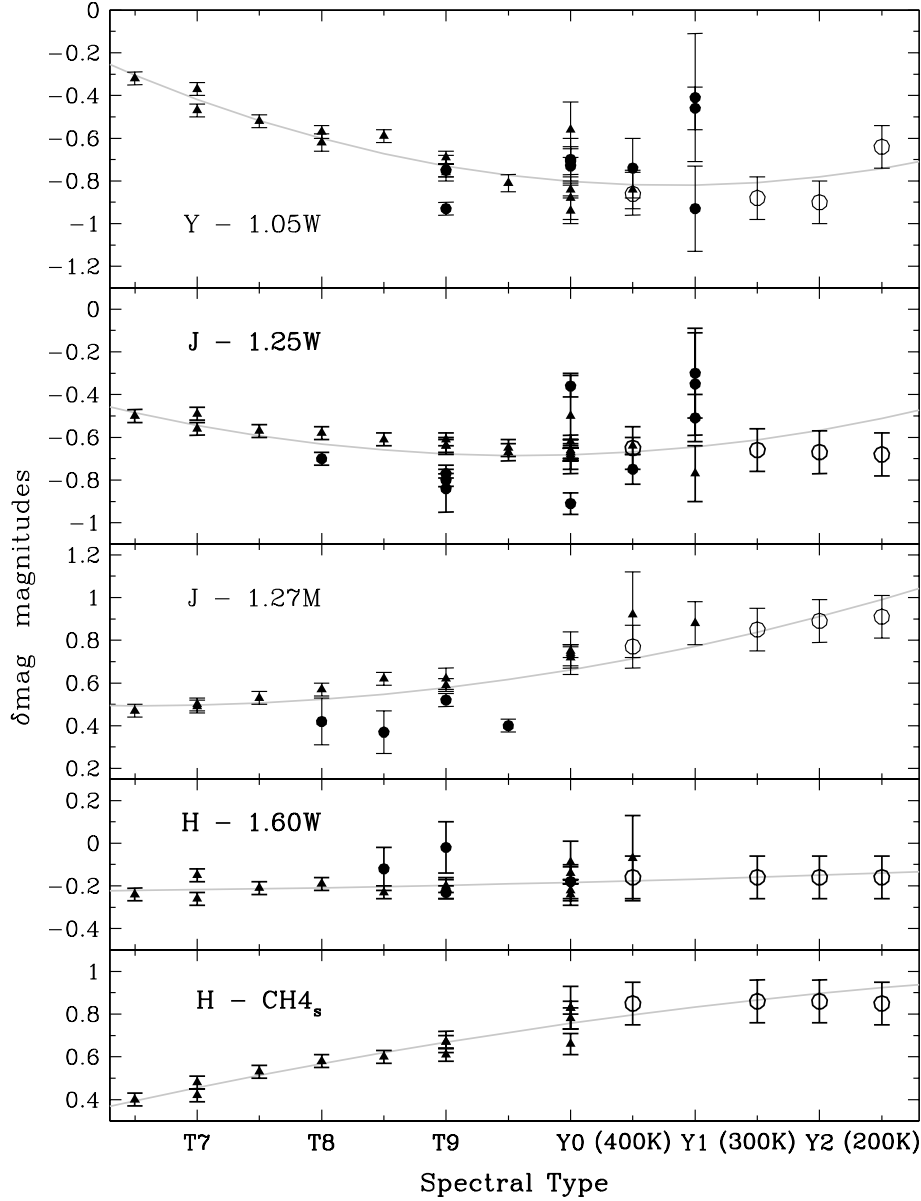}
\vskip -0.4in
\caption{{\em HST}, $CH_4$(short) and MKO near-infrared colors as a function of spectral type. 
WISEPA J182831.08$+$265037.8  has been assigned a type of Y1. Filled circles are measured values, triangles are synthesized from spectra. Open circles are calculated from Tremblin et al. (2015) models with $\log g = 4.5$ and $\log K_{\rm zz} = 6.0$. The model effective temperature is given along the $x$ axis, and we adopt   400~K $=$ Y0.5, 300~K $=$ Y1.5, 250~K $=$ Y2 and 200~K $=$ Y2.5.  The parameters of the weighted quadratic fits are given in Table 5.
\label{fig5}}
\end{figure}

\clearpage

\begin{figure}
\includegraphics[angle=-90,width=1.0\textwidth]{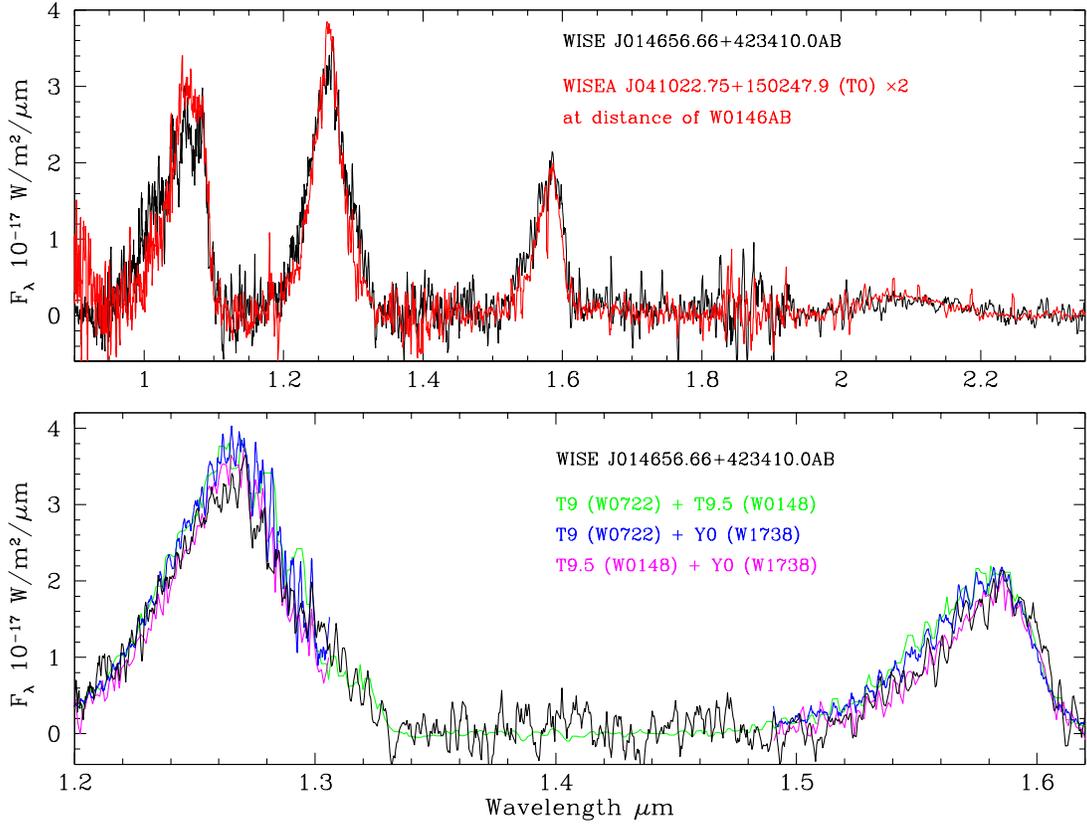}
%\vskip -0.4in
\caption{ The observed spectrum of WISE J014656.66$+$423410.0AB (black line; Dupuy, Liu \& Leggett 2015) is compared to composite spectra of very-late T and early Y dwarfs. In the upper panel the instrinsic $J$-band brightness of the template spectra are retained. In the lower panel the $J$-band brightness is not retained, however the spectra that form the composite are scaled so that $\delta J$ equals that of the components of the W0146 binary (Dupuy, Liu \& Leggett 2015). The upper panel demonstrates that the near-infrared combined-light spectrum of the binary is similar to that of a pair of Y0 dwarfs. The lower panel shows that the binary is likely composed of a T9.5 primary and a Y0  secondary. We adopt a spectral type of T9.5 for W0146AB and W0146A, and a type of Y0 for W0146B.  
\label{fig6}}
\end{figure}

\clearpage

\begin{figure}
    \includegraphics[angle=0,width=.85\textwidth]{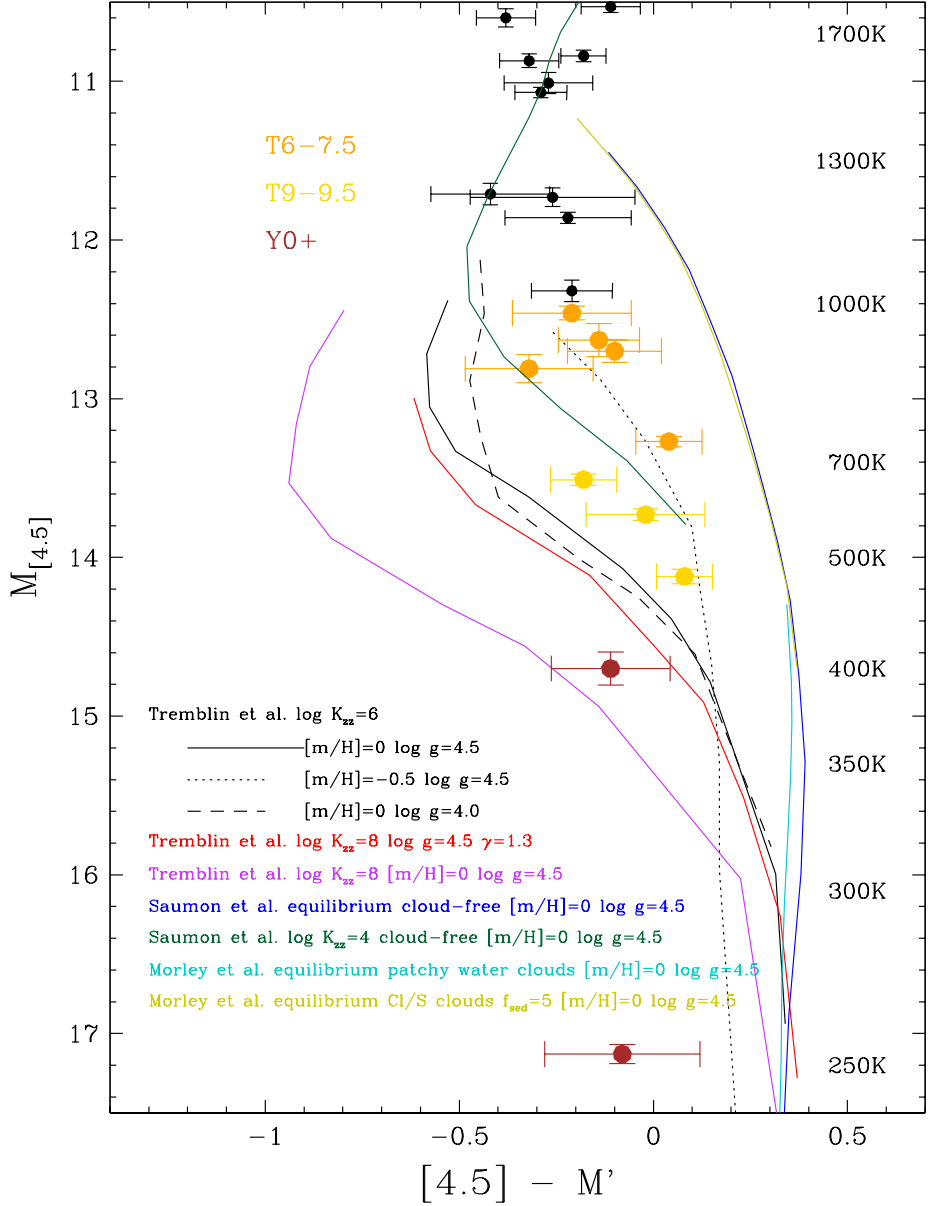}
\vskip -0.2in
\caption{Absolute [4.5] as a function of the color [4.5]$-M^{\prime}$. Filled circles are observational data, with colors indicating late-type T and Y dwarfs, as shown by the legend. Sequences are calculated by the models described in the legend (see also \S 3). Model $T_{\rm eff}$ values corresponding to $M_{[4.5]}$ are shown along the right axis.   Absorption by CO at $4.7~\mu$m, caused by vertical transport of gas, increases $M^{\prime}$.
 \label{fig7}}
\end{figure}

\clearpage

\begin{figure}
    \includegraphics[angle=0,width=.85\textwidth]{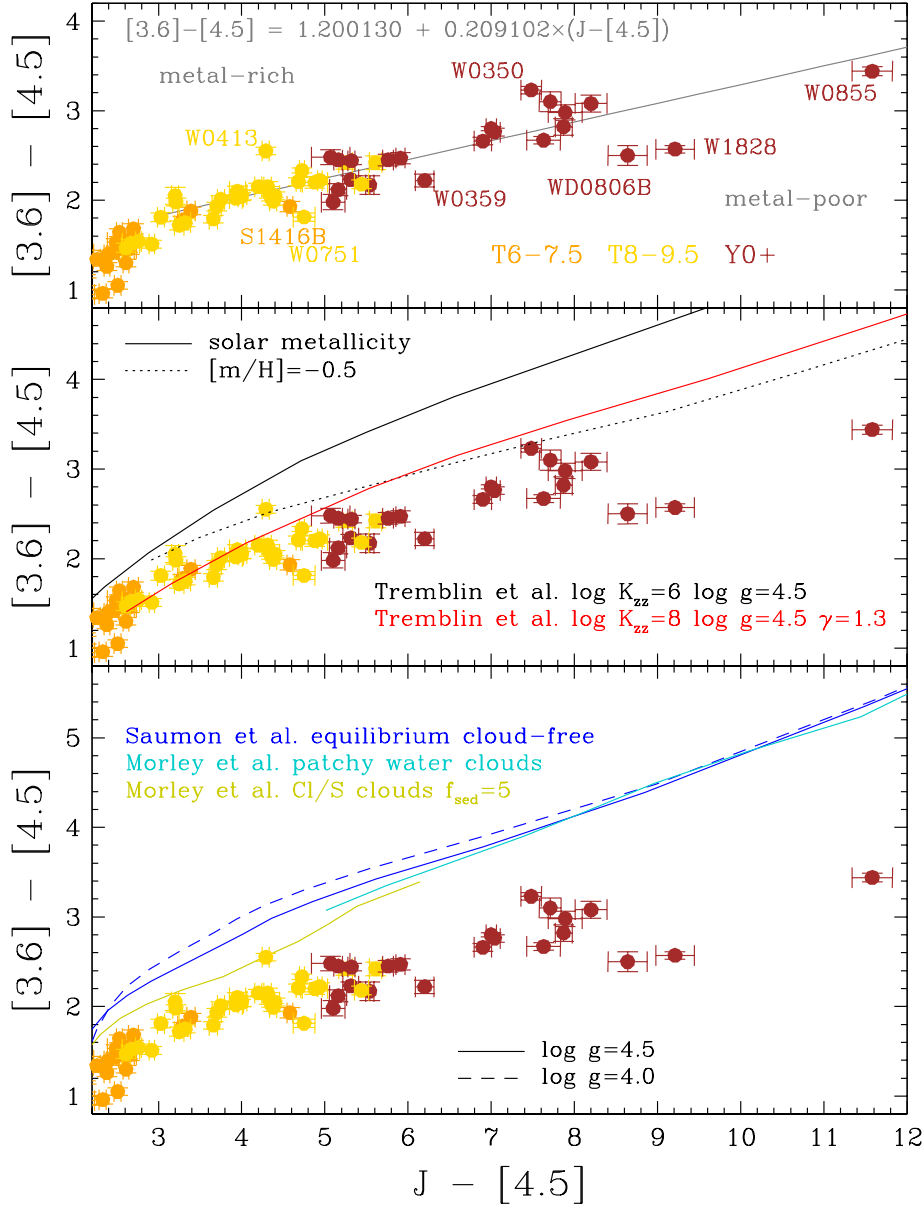}
\vskip -0.4in
\caption{The color [3.6]$-$[4.5] as a function of $J-$[4.5]. Filled circles are observational data, with colors indicating spectral type as given in the legend. Outliers are identified by name in the top panel. In the top panel the grey line is a linear fit to data with $J-[4.5]>3$  magnitudes, excluding sources that deviate by $> 2\sigma$. The middle panel explores changing metallicity and the adiabatic gradient via T15 models, and the bottom panel explores the impact of gravity and clouds via S12, M12 and M14 models (see legends).
 \label{fig8}}
\end{figure}

\clearpage

\begin{figure}
    \includegraphics[angle=0,width=.85\textwidth]{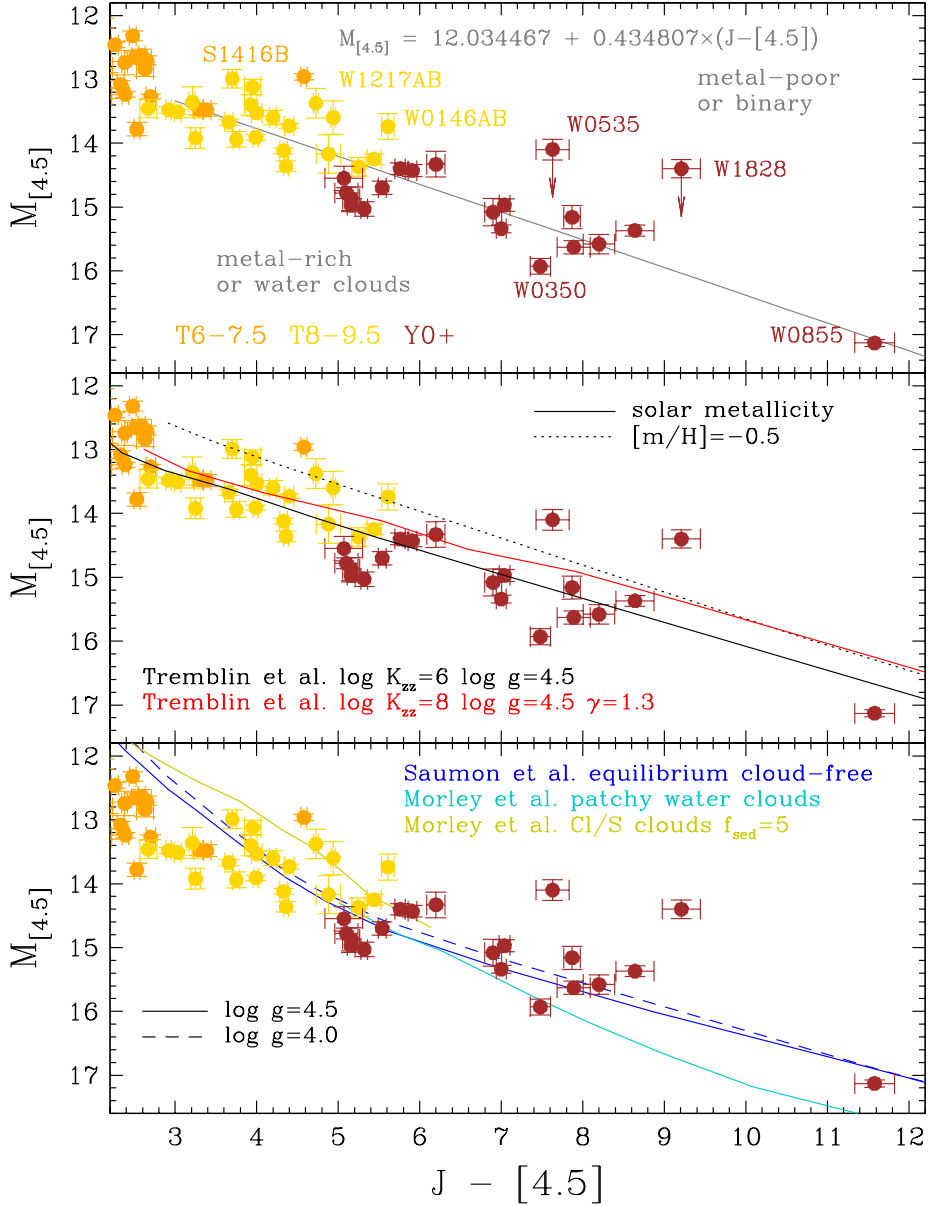}
\vskip -0.2in
\caption{Absolute [4.5] as a function of  the color $J -$ [4.5]. Symbols and lines are as in Figure 8. W0535 and W1828 appear to be multiple sources and the downward arrows indicate their location if the sources are a pair of identical Y dwarfs. Note that, for brown dwarfs, lines of constant gravity are close to iso-mass sequences (Figure 2, right panel). For Y dwarfs $M_{[4.5]}$ correlates with  $T_{\rm eff}$ and for a given $M_{[4.5]}$ lower gravity implies a lower mass and younger brown dwarf, and vice versa (Figure 2). 
 \label{fig9}}
\end{figure}

\clearpage

\begin{figure}
    \includegraphics[angle=0,width=.95\textwidth]{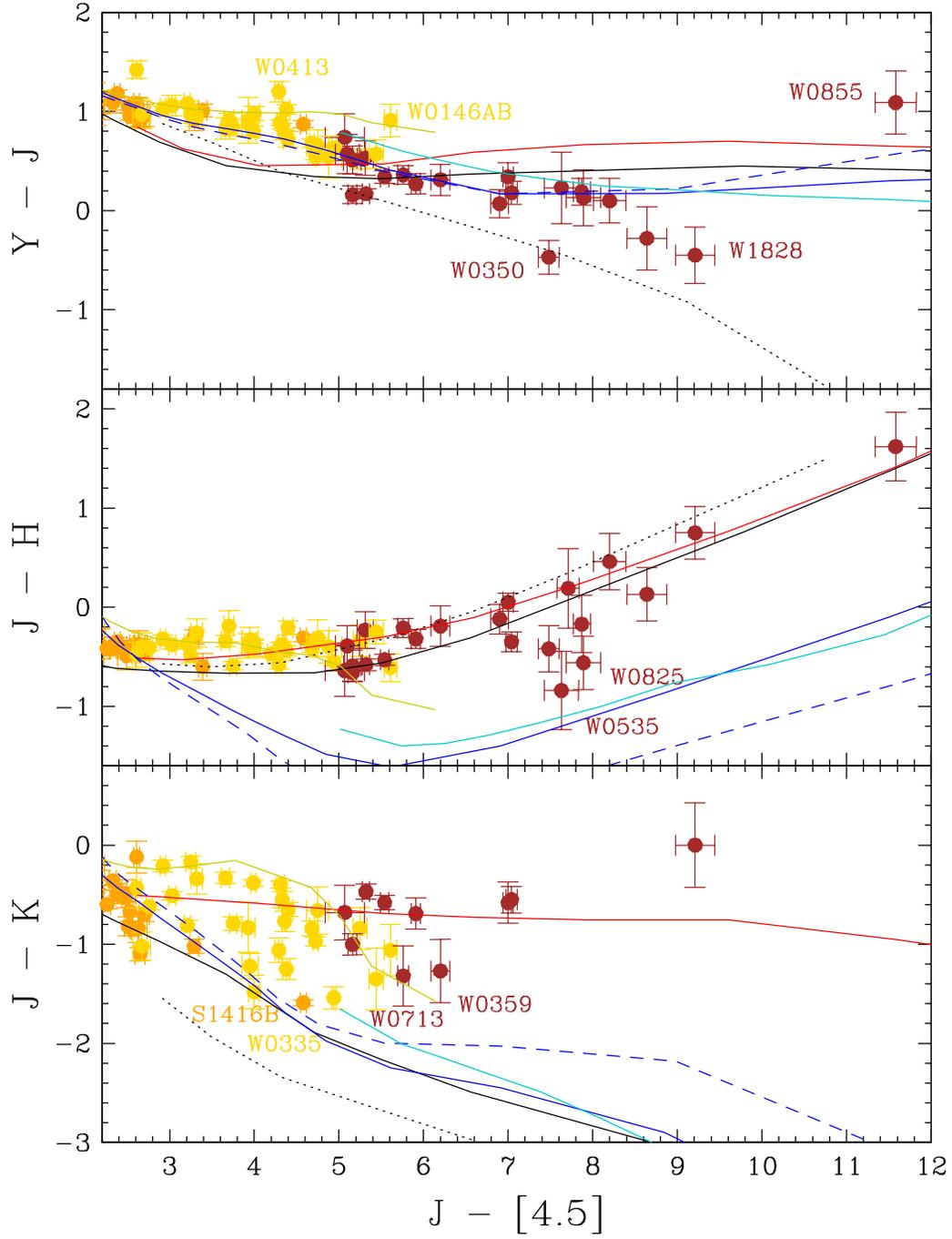}
\caption{Near-infrared colors as a function of  $J -$ [4.5].  Symbols and lines are as in Figure 8.
 \label{fig10}}
\end{figure}

\clearpage

\begin{figure}
    \includegraphics[angle=-90,width=1.0\textwidth]{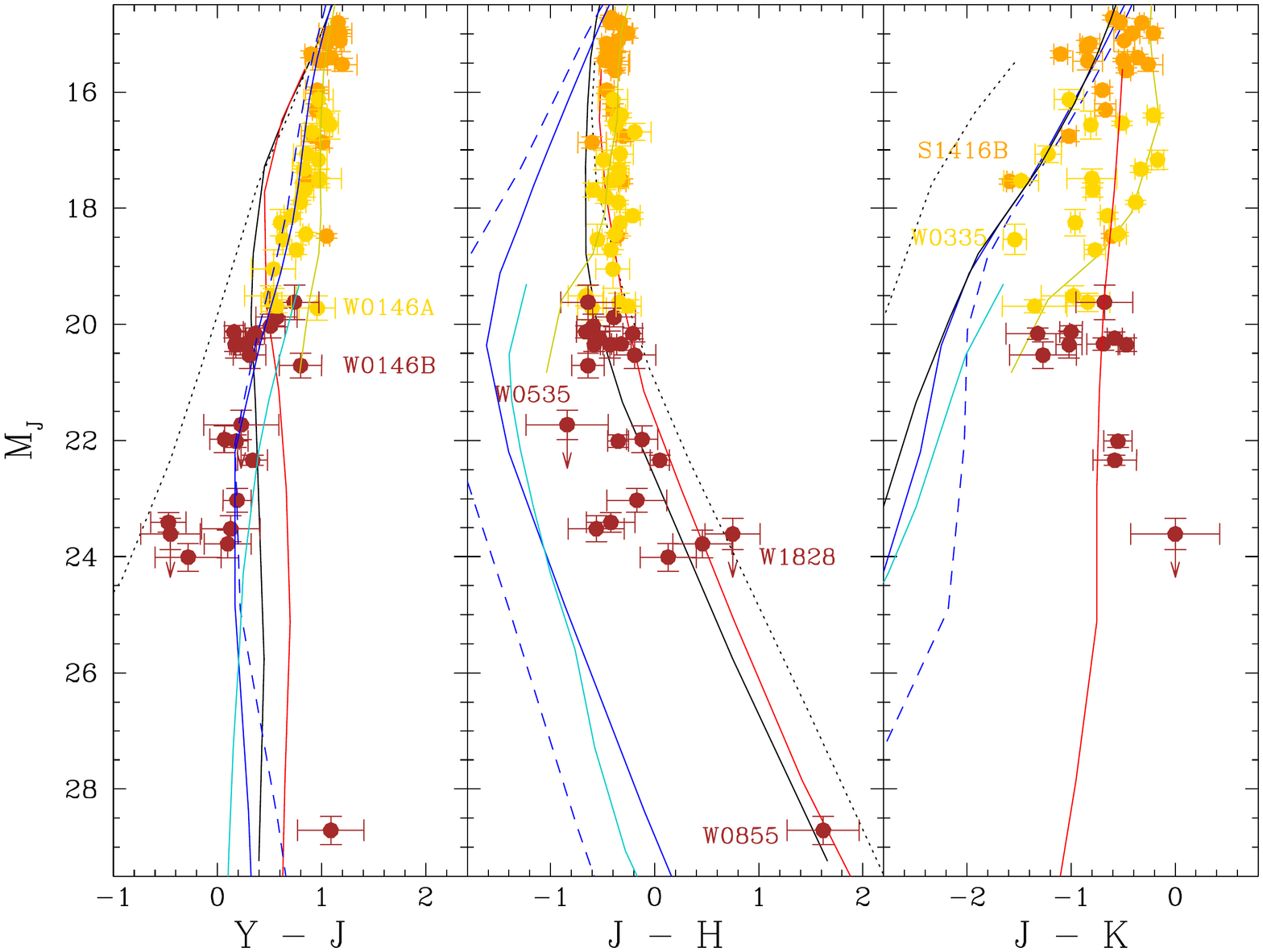}
\caption{Absolute $J$ magnitude as a function of near-infrared colors.  Symbols and lines are as in Figure 8. W0535 and W1828 appear to be multiple sources and the downward arrows indicate their location if the sources are a pair of identical Y dwarfs.
 \label{fig11}}
\end{figure}

\clearpage

\begin{figure}
    \includegraphics[angle=-90,width=1.0\textwidth]{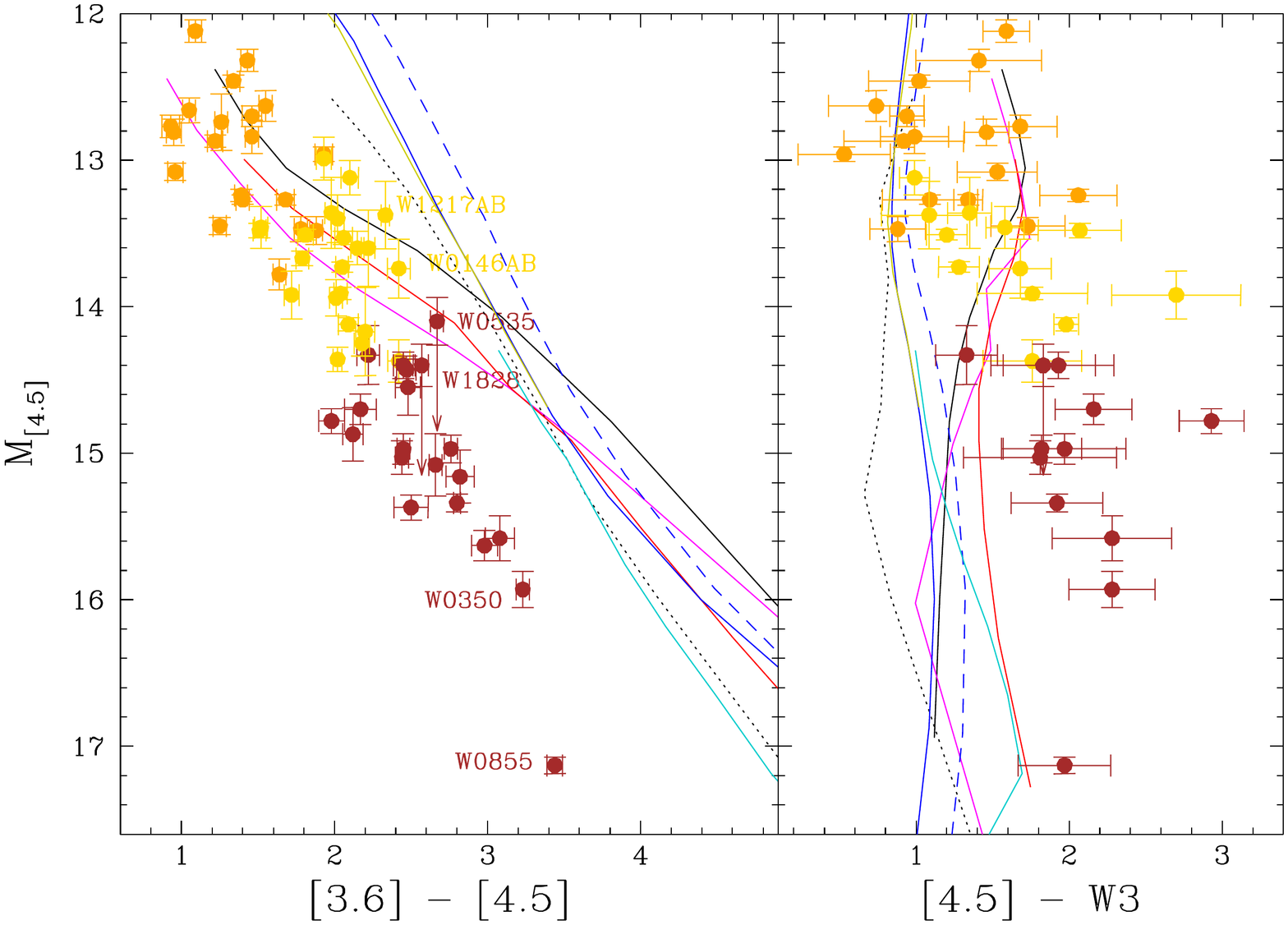}
\caption{Absolute [4.5] as a function of mid-infrared colors. Symbols and lines are as in Figure 8. W0535 and W1828 appear to be multiple sources and the downward arrows indicate their location if the sources are a pair of identical Y dwarfs.
 \label{fig12}}
\end{figure}

\clearpage

\begin{figure}
    \includegraphics[angle=0,width=.85\textwidth]{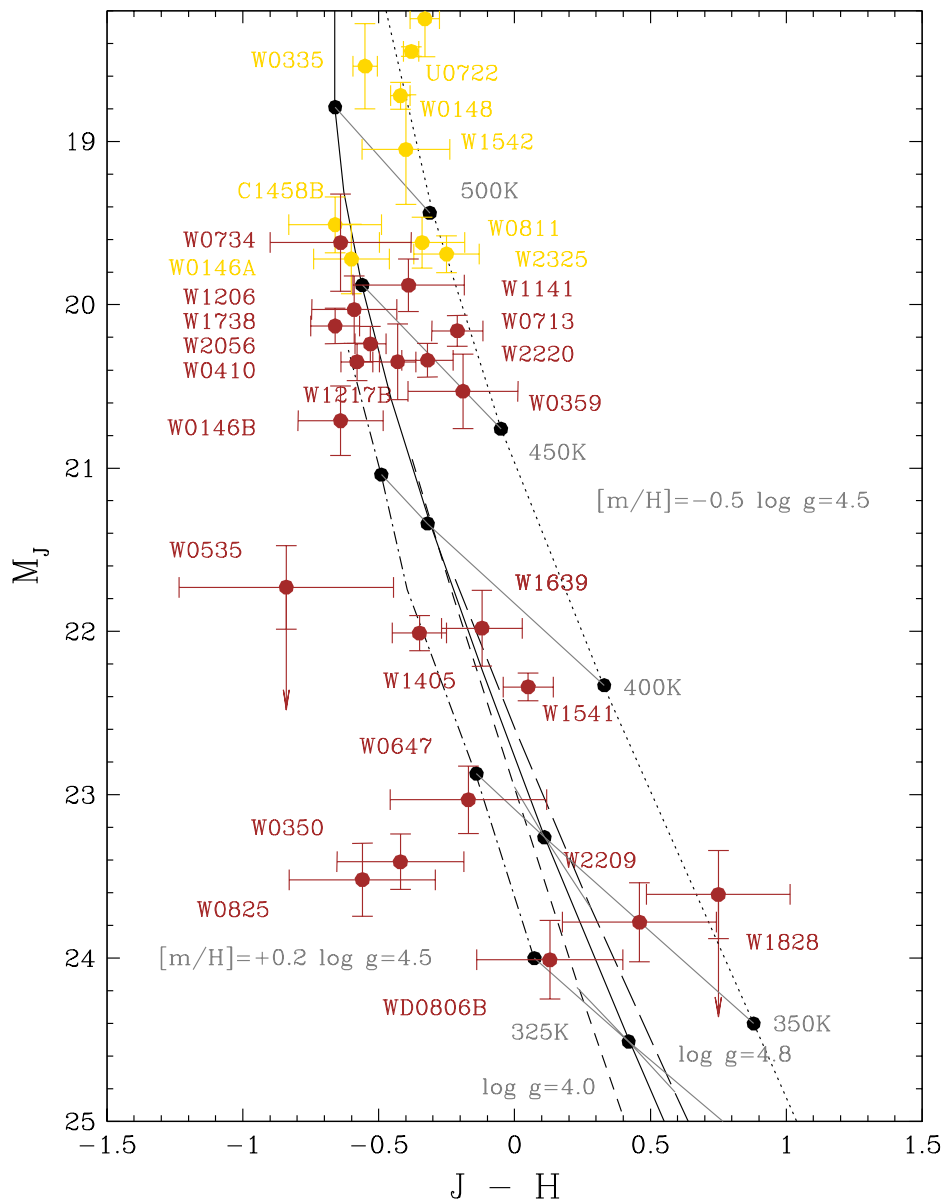}
\vskip -0.4in
\caption{Absolute $J$ as a function of $J-H$ for T9 and Y dwarfs. T15 non-equilibrium sequences are shown for solar metallicity models with $\log g=4.5$ (solid line), $\log g=4.0$ (short dash line) and  $\log g=4.8$ (long dash line). Non-solar metallicity sequences with  $\log g=4.5$ are also shown:  [m/H] $=-0.5$ (dotted line) and [m/H] $=+0.2$ (dash-dot line).
Values of $T_{\rm eff}$ are indicated along the sequences. W0535 and W1828 appear to be multiple sources and the downward arrows indicate their location if they are a pair of identical Y dwarfs.
 \label{fig13}}
\end{figure}

\clearpage

\begin{figure}
    \includegraphics[angle=-90,width=1.0\textwidth]{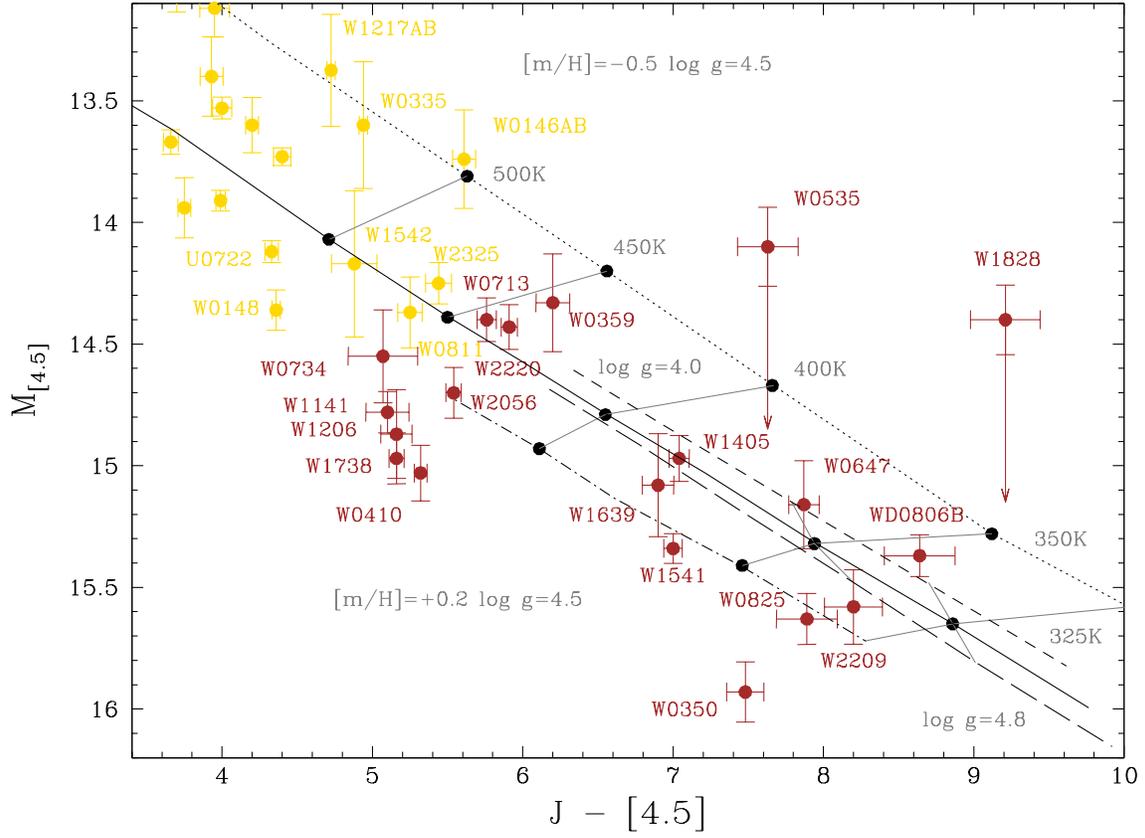}
\caption{Absolute [4.5] as a function of $J-$[4.5] for T9 and Y dwarfs. Lines are as in Figure 13. W0535 and W1828 appear to be multiple sources and the downward arrows indicate their location if they are a pair of identical Y dwarfs.
 \label{fig14}}
\end{figure}

\clearpage

\begin{figure}
\includegraphics[angle=0,width=0.95\textwidth]{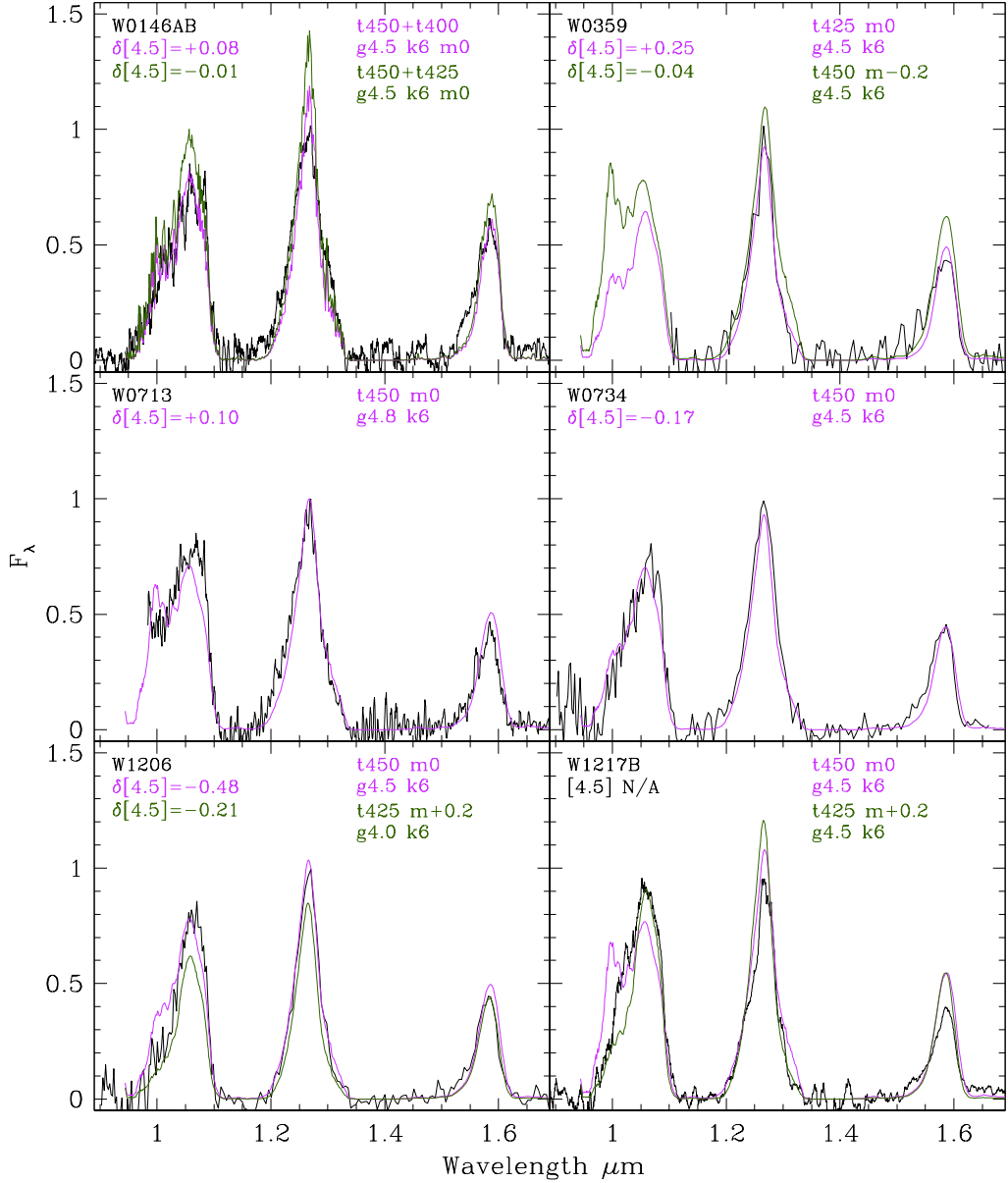}
\vskip -1.2in
\caption{Tremblin et al. (2015) cloud-free non-equilibrium model fits (violet  and dark green lines) to observed near-infrared spectra (black lines), with  $425 \leq T_{\rm eff}$~K $\leq 450$. The legends give name, model parameters and 
$\delta [4.5] = M_{[4.5]}({\rm model}) - M_{[4.5]}({\rm observed})$. There is no resolved [4.5] photometry for the W1217 binary system. Observed spectra are flux calibrated by photometry; the models are flux calibrated by the object's distance and the evolutionary radius for $T_{\rm eff}$ and $\log g$ (Saumon \& Marley 2008). The spectra are  
normalized so the $J$-band peak is $\sim 1.0$. The uncertainty in the flux calibration is 10 -- 20\% (Table 8).
\label{fig15}}
\end{figure}

\clearpage

\begin{figure}
\includegraphics[angle=0,width=0.95\textwidth]{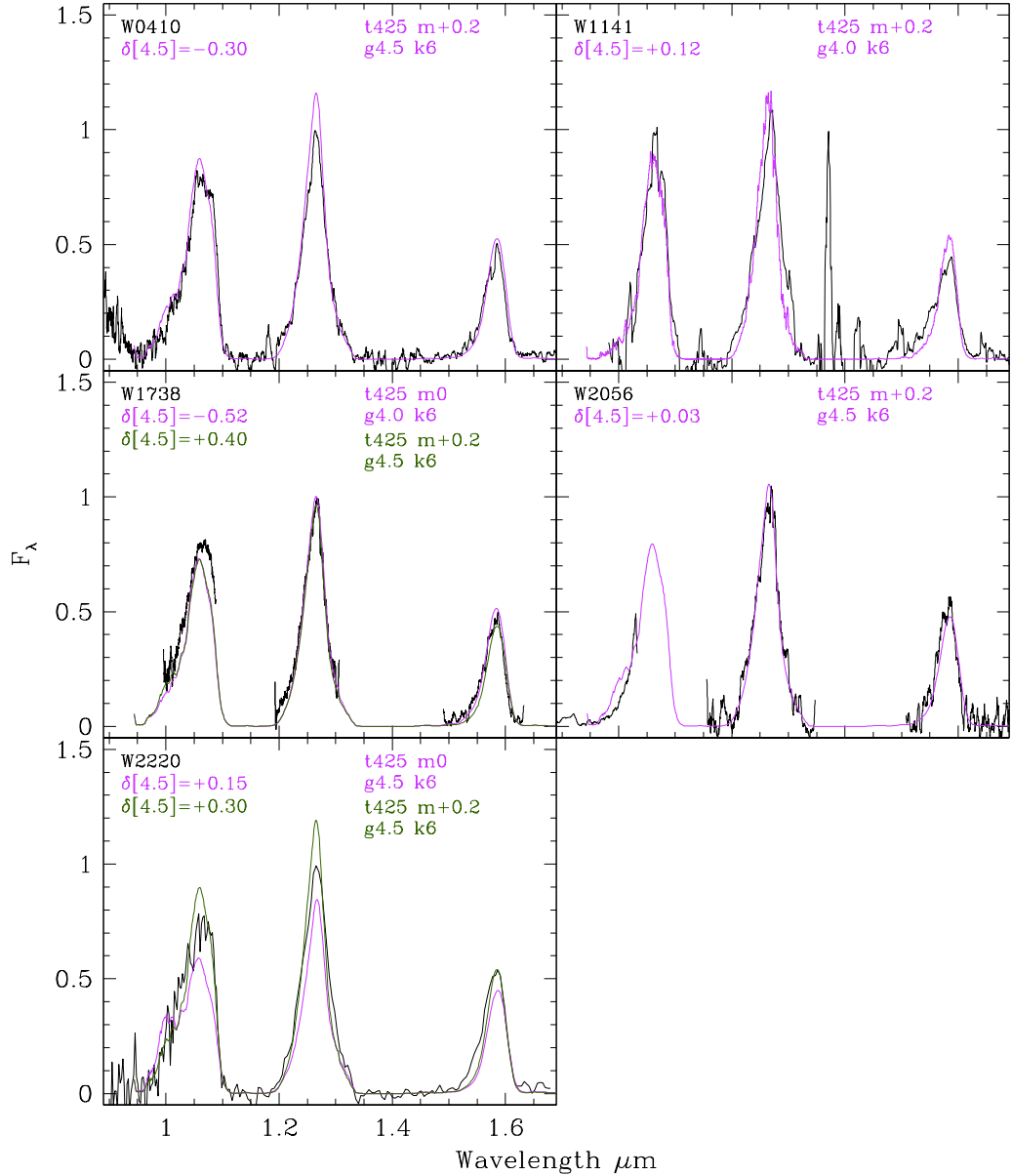}
\vskip -1.1in
\caption{Tremblin et al. (2015) cloud-free non-equilibrium model fits (violet and dark green lines) to 
observed Y dwarf near-infrared spectra (black lines), for which we determine  $T_{\rm eff} = 425$~K. Lines are as in Figure 15. 
\label{fig16}}
\end{figure}

\clearpage

\begin{figure}
\includegraphics[angle=0,width=0.95\textwidth]{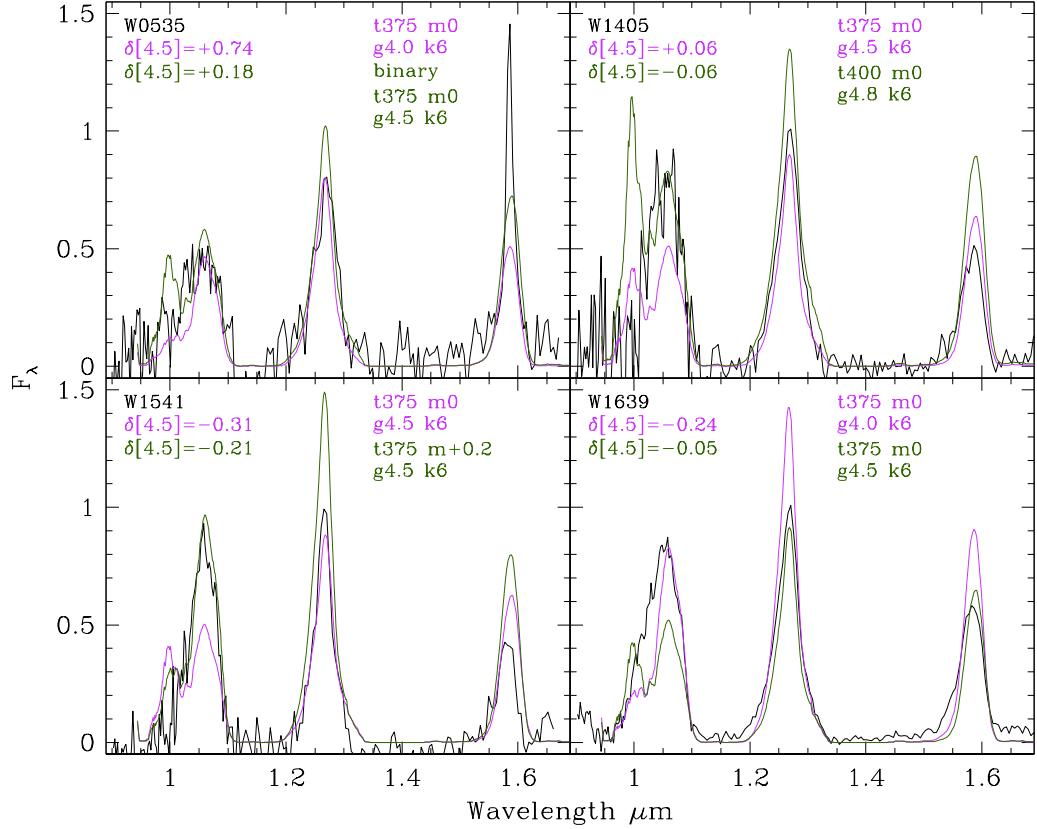}
\vskip -3in
\caption{Tremblin et al. (2015) cloud-free non-equilibrium model fits (violet and dark green lines) to 
observed Y dwarf near-infrared spectra (black lines), for which we determine  $375 \leq T_{\rm eff}$~K $\leq 400$. Lines are as in Figure 15. For W0535 the binary solution is preferred due to the better agreement with $M_{[4.5]}$. For this Y dwarf the $H$-band flux calibration is inconsistent with $Y$ and $J$ and we suggest that there are spurious bright data points in the $H$-region of the spectrum. 
\label{fig17}}
\end{figure}

\clearpage

\begin{figure}
\includegraphics[angle=0,width=0.95\textwidth]{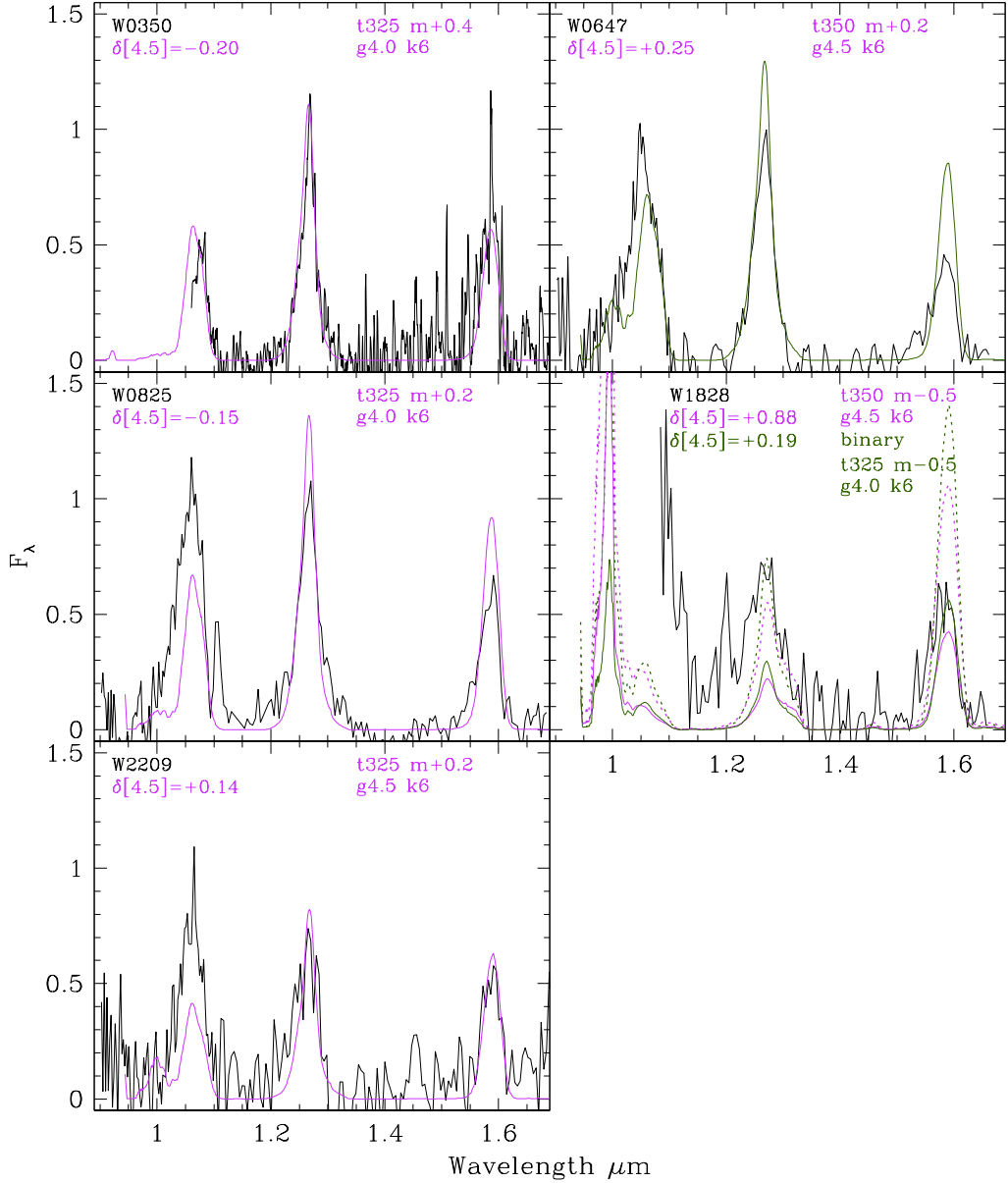}
\vskip -1.2in
\caption{
Tremblin et al. (2015) cloud-free non-equilibrium model fits (violet and dark green lines) to 
observed Y dwarf near-infrared spectra (black lines), with  $325 \leq T_{\rm eff}$~K $\leq 350$. Lines are as in Figure 15. For W1828 the $J$- and $H$-band flux calibrations are inconsistent. The solid violet and dark green lines show the fits assuming that the $H$-band region of the spectrum and the $H$ photometry is correct, and the dotted line shows the same  models if the $J$-band spectrum and photometry are correct. It appears that the observed spectrum may include a spurious signal at $\lambda < 1.35~\mu$m.
\label{fig18}}
\end{figure}

\clearpage

\begin{figure}
\includegraphics[angle=0,width=0.8\textwidth]{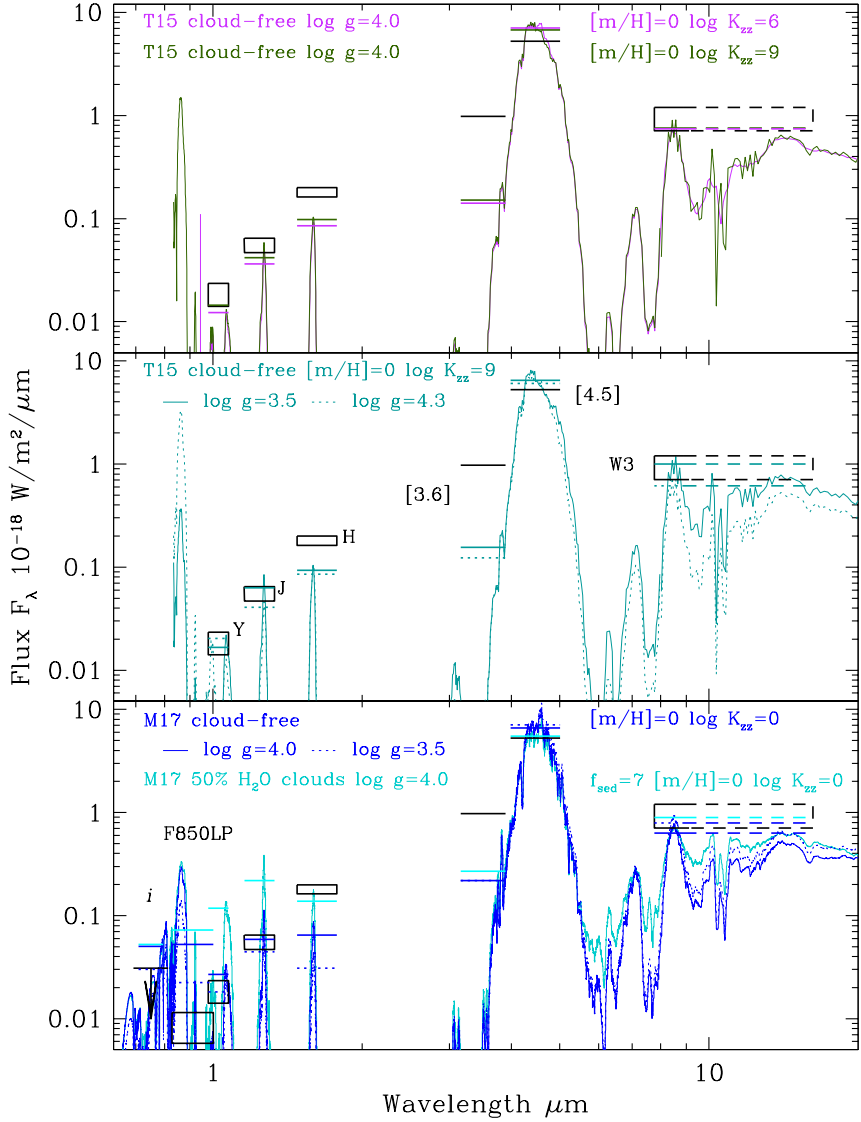}
\vskip -0.5in
\caption{The observed photometry for W0855, if at 10~pc, is shown as black boxes. 
The box height indicates the flux uncertainty, and the box width the filter band pass. The dashed extension for the W3 filter indicates that the flux is expected primarily at the blue end. 
The uncertainties in [3.6] and [4.5] flux are small and these appear as black lines. 
In the bottom panel, the $i_{\rm AB}$ flux is an upper limit only. 
Colored lines are $T_{\rm eff} = 250$~K model spectra and photometry as described in the legends. The model fluxes have been scaled to 10~pc and the evolutionary value of the Y dwarf radius. The top panel compares T15 models that differ in  $K_{\rm zz}$, the middle panel compares T15 models that differ in $g$, and the bottom panel compares updated M14 cloudy and cloud-free equilibrium models. See text and Table 10.
\label{fig19}}
\end{figure}

\clearpage

\begin{deluxetable}{lcrrrr}
%\tabletypesize{\scriptsize}
\tabletypesize{\footnotesize}
%\tablecolumns{9}
\tablewidth{0pt}
%\rotate
\tablecaption{New Gemini NIRI Photometry}
\tablehead{ 
\colhead{Name} & \colhead{Spectral} & \colhead{$Y$\tablenotemark{a}} & \colhead{$CH_4$(short)} &
\colhead{$H$} & \colhead{$M^{\prime}$} \\
\colhead{} & \colhead{Type} & \colhead{(err)} & \colhead{(err)} &
\colhead{(err)} & \colhead{(err)} \\
\colhead{} & \colhead{} & \multicolumn{4}{c}{magnitudes} \\
}
\startdata
CFBDS J005910.90$-$011401.3 & T8.5 & \nodata & \nodata & \nodata & 13.68(0.15) \\
2MASSI J0415195$-$093506 &  T8 & \nodata & \nodata & \nodata & 12.47(0.08) \\
WISEA J064723.24$-$623235.4 & Y1  & \nodata & \nodata &  23.11(0.27) & \nodata \\
UGPS J072227.51$-$054031.2 & T9 & \nodata & \nodata & \nodata & 12.11(0.06)\\
2MASSI J0727182$+$171001 & T7 & \nodata & \nodata & \nodata & 12.97(0.08)\\
WISE J085510.83$-$071442.5 & $>$Y2 & $>$24.5\tablenotemark{b} &  23.38(0.20) & \nodata  & 13.95(0.20)  \\
WISEPC J205628.90$+$145953.3 & Y0 & \nodata & \nodata & \nodata & 14.00(0.15) \\
\enddata
\tablenotetext{a}{The NIRI $Y$ magnitudes have been put on the MKO $Y$ system as described in the text.}
\tablenotetext{b}{$3 \sigma$ detection limit.}
\end{deluxetable}

\begin{deluxetable}{lcrrrrrrc}
\tabletypesize{\scriptsize}
%\tablecolumns{9}
\tablewidth{0pt}
%\rotate
\tablecaption{New Near-Infrared Photometry from WFC3 and HAWK-I Archives}
\tablehead{ 
\colhead{Name(RA)}  & \colhead{Spectral} & \colhead{F105W}  & \colhead{F125W} & \colhead{F127M} & \colhead{F160W}
 & \colhead{$J$}  & \colhead{$H$} & WFC-3 PI,ID  \\
\colhead{(Declination)}  & \colhead{Type} & \colhead{(err)}  & \colhead{(err)} & \colhead{(err)} & \colhead{(err)}
 & \colhead{(err)}  & \colhead{(err)}  & HAWK-I PI,ID \\
\colhead{} & \colhead{} & \multicolumn{6}{c}{magnitudes} & \colhead{} \\
}
\startdata
WISEA J014807.34 & T9.5 & \nodata &  \nodata & 18.52  &  \nodata & 18.92 & \nodata  & Biller 12873 \\
 $-$720258.7     &      &         &          & (0.02) &          & (0.02) &       & Forveille 091.C-0543(D)  \\
WISEA J032504.52 & T8 &  \nodata &  \nodata & \nodata &  \nodata & 18.90 & \nodata   & \\
$-$504403.0      &    &           &         &         &           & (0.03) &  & Cushing 089.C-0042(A)  \\
WISE J041358.14 & T9 &  \nodata & 20.45 & \nodata & 20.22 & 19.61 & 20.20 & Cushing 12970, Gelino 12972 \\
$-$475039.3     &    &          & (0.11) &        & (0.12) & (0.02) & (0.03) & Cushing 089.C-0042(A)  \\
UGPS J072227.51 & T9 & 18.12 & 17.32 & 16.00 & 17.13 & \nodata    & \nodata  & Liu 12504 \\
$-$054031.2    &     & (0.03) & (0.03) & (0.03) & (0.03) &        &          &        \\
WD 0806           & Y1 &  25.97 & 25.92  & 24.80  & 25.46 & \nodata   &  \nodata  & Gelino 13428 \\
$-$661B           &    & (0.20) & (0.30) & (0.30) & (0.10) &          &           &          \\
WISEPC J104245.23 & T8.5 & \nodata & \nodata & 18.37 & 19.33  & 18.74  & 19.21 & Biller 12873, Gelino 12972
\\
$-$384238.3       &      &         &        & (0.10) & (0.10) & (0.03) & (0.02) & Cushing 089.C-0042(A)  \\
ULAS J123828.51 & T8 & \nodata & \nodata & 18.24 & \nodata   &  \nodata  &  \nodata  & Burgasser 11666 \\
$+$095351.3     &     &        &         & (0.11) &          &           &           &            \\ 
\enddata
\end{deluxetable}

\begin{deluxetable}{lcrrrrc}
\tabletypesize{\scriptsize}
%\tablecolumns{9}
\tablewidth{0pt}
%\rotate
\tablecaption{New Mid-Infrared Photometry from IRAC and {\it WISE} Archives}
\tablehead{ 
\colhead{Name}  & \colhead{Spectral} & \colhead{[3.6](err)} & \colhead{[4.5](err)} & \colhead{W1(err)} & \colhead{W3(err)} & ITAC PI,ID \\
\colhead{} & \colhead{Type} & \multicolumn{4}{c}{magnitudes} & \colhead{} \\
}
\startdata
2MASS J00345157$+$0523050\tablenotemark{a}  & T6.5  & 14.07(0.03) & 12.52(0.03) &\nodata   & \nodata & 
Fazio 30179\\
CFBDS J005910.90$-$011401.3 & T8.5 &  \nodata & \nodata & \nodata   & 12.38(0.13) & \\
WISE J014656.66$+$423410.0AB  & T9.5  & \nodata & \nodata & \nodata   &  13.4(0.2) & \\
WISEA J014807.34$-$720258.7 & T9.5 & 16.58(0.02) & 14.56(0.02)  &\nodata   & \nodata & Kirkpatrick 70062 \\
WISE J030449.03$-$270508.3 &  Y0 &  17.71(0.03) & 15.48(0.03)  & \nodata   & \nodata & Pinfield 10135 \\
WISE J035934.06$-$540154.6 &  Y0 &   \nodata & \nodata &  19.1(0.2) & 14.0(0.2) &  \\
WISE J085510.83$-$071442.5 & $>$Y2  & \nodata & \nodata & \nodata   & 11.9(0.3) & \\
ULAS J090116.23$-$030635.0  & T7.5 &  16.38(0.03) & 14.50(0.03) & \nodata   & \nodata & Kirkpatrick 80109\\
WISEA  J114156.67$-$332635.5 & Y0 & 16.64(0.08) & 14.66(0.03) & \nodata   & \nodata & Kirkpatrick 80109\\
ULAS J150457.66$+$053800.8 &  T6.5  &  15.13(0.03) & 14.08(0.03) & \nodata   & \nodata & Kirkpatrick 80109\\
WISEPA J154151.66$-$225025.2  & Y0.5 &   \nodata & \nodata & \nodata   & 12.2(0.3) & \\
WISEA J163940.84$-$684739.4 & Y0pec & 16.23(0.03) & 13.57(0.03) & \nodata   & \nodata  & Kirkpatrick 80109 \\
WISEPA J182831.08$+$265037.8  & $>$Y1 &   \nodata & \nodata &  17.34(0.26)  & \nodata & \\
WISEA J205628.88$+$145953.6 & Y0  & 16.06(0.10) & \nodata & \nodata   & \nodata & Dupuy 80233, Cushing 90015
\enddata
\tablenotetext{a}{For 2MASS J00345157$+$0523050 we also measured [5.8]$= 13.13 \pm 0.03$ and [8.0]$= 12.41 \pm 0.03$ magnitudes.}
\end{deluxetable}

\begin{deluxetable}{lcrrrrrr}
\tabletypesize{\scriptsize}
%\tablecolumns{9}
\tablewidth{0pt}
%\rotate
\tablecaption{Synthesized Near-Infrared Colors from Spectra}
\tablehead{ 
\colhead{Name}  & \colhead{Spectral}  & \colhead{$J -$ F127M} & \colhead{$H -$ F160W} & \colhead{$H -$ CH$_4$} & \colhead{$Y - J$} &  \colhead{$J - H$} &  \colhead{$J - K$}\\
\colhead{}  & \colhead{Type}  & \colhead{(err)} & \colhead{(err)} & \colhead{(err)} & \colhead{(err)} & \colhead{(err)}\\
\colhead{} & \colhead{} & \multicolumn{6}{c}{magnitudes}  \\
}
\startdata
ULAS J003402.77$-$005206.7\tablenotemark{a} & T8.5 &  0.62(0.03)  & 0.60(0.03)  & $-$0.23(0.03) & \nodata & \nodata  & \nodata\\
WISEA J014807.34$-$720258.7 & T9.5 & \nodata  &    \nodata         &   \nodata       &   0.76(0.05)  & $-$0.42(0.03) & $-$0.76(0.10)\\
WISE J030449.03$-$270508.3 &  Y0 & \nodata  &    \nodata         &   \nodata       & 0.53(0.15) & \nodata & \nodata\\
WISEA J033515.07$+$431044.7 & T9 &  \nodata  & \nodata & 0.67(0.05) & \nodata & \nodata & \nodata\\
WISEA J035000.31$-$565830.5 & Y1 & 0.88(0.10) & \nodata  & \nodata  & \nodata & \nodata& \nodata\\
WISEA J041022.75$+$150247.9 & Y0 &  0.74(0.10) & $-$0.09(0.10)  & 0.83(0.10) & \nodata & \nodata & \nodata\\
2MASSI J0415195$-$093506 & T8 &  0.57(0.03)  &  $-$0.19(0.03)  & 0.58(0.03)  & \nodata & \nodata & \nodata \\
UGPS J072227.51$-$054031.2 & T9 &  0.65(0.03) &  $-$0.16(0.03) & 0.67(0.03) & \nodata & \nodata & \nodata\\
WISEA  J114156.67$-$332635.5 & Y0 &  \nodata  &    \nodata         &   \nodata       &   0.57(0.03)  & $-$0.39(0.05)& \nodata \\
WISEPC J121756.91$+$162640.2A & T9 &  0.59(0.03)  & $-$0.20(0.03)  & 0.61(0.03)  & \nodata & \nodata & \nodata\\
WISEPC J121756.91$+$162640.2B & Y0 &   0.75(0.03) &  $-$0.14(0.03) & 0.83(0.03) & \nodata & \nodata & \nodata\\
WISEP J142320.86$+$011638.1\tablenotemark{b} & T8pec &  \nodata  & \nodata  &   \nodata  &  \nodata & $-$0.41(0.06) & $-$1.48(0.15)\\ 
WISE J154151.65$-$225024.9 & Y0.5 &   0.92(0.20) & $-$0.07(0.20) & \nodata & \nodata & \nodata & \nodata\\
WISEA J173835.52$+$273258.8 & Y0 &  0.75(0.03) & $-$0.14(0.03)  & 0.83(0.03)  & \nodata & \nodata & \nodata\\
WISEA J205628.88$+$145953.6 & Y0 &    0.72(0.05) & $-$0.22(0.05) & 0.78(0.05)  & \nodata & \nodata & \nodata\\
WISEA J232519.55$-$410535.1\tablenotemark{c} & T9pec &  \nodata  &  \nodata  &   \nodata  &   0.57(0.10)  & $-$0.25(0.06) & $-$1.35(0.30)\\
\enddata
\tablenotetext{a}{For ULAS J003402.77$-$005206.7 we also synthesized $Y -$ F105W $= -0.59\pm0.03$ and $J -$ F125W $= -0.61\pm0.03$ magnitudes.}
\tablenotetext{b}{WISEP J142320.86$+$011638.1 is also known as BD $+01^{\circ}2920$B (Pinfield et al. 2012).} 
\tablenotetext{c}{For WISEA J232519.55$-$410535.1   Data Release 4 of the VISTA VHS survey  (McMahon et al. 2013) gives $J = 19.53\pm0.08$.}
\end{deluxetable}

\begin{deluxetable}{lrrr}
\tablewidth{0pt}
\tablecaption{Transformations  Between {\em HST}, $CH_4$(short) and MKO Near-Infrared Colors}
\tablehead{ 
\colhead{Color} & \colhead{a0} & \colhead{a1}  & \colhead{a2}  \\
}
\startdata
$Y -$ F105W  & 2.42719& $-$0.601128 & 0.027812 \\
$J -$ F125W  & 1.228187 & $-$0.398120 & 0.020710  \\
$J -$ F127M  & 1.092170  & $-$0.1183161 &  0.014005      \\
$H -$ F160W  & $-$0.21659 & $-$0.008060 & 0.001134  \\
$H -$ CH$_4$ & $-$0.679045  & 0.204865 &  $-$0.006130 \\
\enddata
\tablecomments{The transformation is applied as $$ color = a0 + (a1 \times Type) + (a2 \times Type^2)   $$ where $Type$ runs from  7.5 to 11, corresponding to spectral types T7.5 to Y1. The estimated uncertainty in color is 0.10 magnitudes (see text and Figure 5).}
\end{deluxetable}

\begin{deluxetable}{lcrrrrrrrr}
\tabletypesize{\scriptsize}
\tablewidth{0pt}
%\rotate
\tablecaption{New Estimated MKO-System $Y$, $J$, $H$ Photometry for Y Dwarfs}
\tablehead{ 
\colhead{Name}  & \colhead{Spectral} & \colhead{$Y$} & \colhead{$Y$($YJ$)} & \colhead{$J$} & \colhead{$J$} & \colhead{$J$($YJH$)}
& \colhead{$H$} & \colhead{$H$} & \colhead{$H$($JH$)}\\
\colhead{} &  \colhead{Type} & \colhead{(F105W)} & \colhead{(spectrum)} &
\colhead{(F125W)} & \colhead{(F127M)} & \colhead{(spectrum)}
& \colhead{(F160W)} & \colhead{(CH$_4$)} & \colhead{(spectrum)}\\
\colhead{} &  \colhead{} & \colhead{(err)} & \colhead{(err)} & \colhead{(err)}& \colhead{(err)}& \colhead{(err)} & \colhead{(err)}& \colhead{(err)} \\
\colhead{} & \colhead{} & \multicolumn{8}{c}{magnitudes}  \\
}
\startdata
WD 0806              & Y1   &  25.14\tablenotemark{a} & \nodata & 25.27\tablenotemark{b} & 25.57\tablenotemark{b} & \nodata & 25.29 & \nodata  & \nodata \\
$-$661B &  & (22) &  & (32) & (32) &  & (14) &  &  \\
WISEA J082507.37\tablenotemark{c} & Y0.5  &   22.66 & \nodata & 22.53 & \nodata & \nodata & \nodata & \nodata  & 23.09 \\
$+$280548.2&  & (11) &  & (10) &  &  &  &  & (18)  \\
WISE J085510.83 & $>$Y2\tablenotemark{d} &  26.54\tablenotemark{e} & \nodata & 25.84\tablenotemark{f} & 25.37\tablenotemark{f} & \nodata & 23.71 & 24.19\tablenotemark{g} & \nodata\\
$-$071442.5 &  & (21) &  & (29) & (13) &  & (10) & (17) &  \\
WISEA  J114156.67\tablenotemark{h}  & Y0 & \nodata & 20.33 &  \nodata &  \nodata &  \nodata &  \nodata &  \nodata & 20.15 \\
$-$332635.5&  &  & (14)  &  &  &  &  &  & (15) \\
WISEA J120604.25\tablenotemark{c} & Y0 & 20.89 & \nodata & 20.38 & \nodata & \nodata  & \nodata & \nodata & 20.97 \\
$+$840110.5&  & (10) &  & (10) &  &  &  &  & (12) \\
WISEA J163940.84\tablenotemark{c} & Y0 &  20.54 & \nodata & 20.47 &  \nodata & \nodata  & \nodata &  \nodata & 20.59   \\
$-$684739.4&  & (10) &  & (10) &  &  &  &  & (11) \\
WISEA J220905.75\tablenotemark{c} & Y0 & 23.04 & \nodata &  \nodata     &  \nodata    & 22.94 & \nodata & \nodata &  22.48\\
$+$271143.6&  & (12) &  &  &  & (19) &  &  & (21)  \\
WISEA J235402.79\tablenotemark{c} & Y1 & \nodata     & \nodata & 22.72 &  \nodata    & \nodata  & \nodata & \nodata &  22.53 \\
$+$024014.1 &  &  &  & (13)  &  &  &  &  & (38) \\
\enddata
\tablecomments{Photometric error is in centimag.}
\tablenotetext{a}{Consistent with Luhman et al. (2014) $Y >$ 23.2 magnitudes.}
\tablenotetext{b}{Previous estimate  $J = 25.00 \pm 0.10$ magnitudes from  F110W $= 25.70 \pm 0.08$ magnitudes (Luhman et al. 2014). The transformation  between {\it HST} and MKO colors is better determined here, and  adopt the weighted average of $J$(F125W) and $J$(F127M)  (Table 8).}
\tablenotetext{c}{Photometry is based on Schneider et al. (2015) measurements of {\it HST} magnitudes and their synthetic colors from {\it HST} spectra.}
\tablenotetext{d}{A spectral type of Y2 is used for the estimation of MKO photometry from the {\em HST} and $CH_4$ photometry.} 
\tablenotetext{e}{Consistent with our measurement of $Y > $24.5 magnitudes, and also Beamin et al. (2014) $Y > $24.4 magnitudes.}
\tablenotetext{f}{Consistent with the faint limit of Faherty et al. (2014) $J = 25.0^{+0.53}_{-0.35}$ magnitudes. We adopt the weighted average of $J$(F125W) and $J$(F127M) (Table 8).}
\tablenotetext{g}{Based on the average of $CH_4$(short) $= 23.38 \pm 0.20$ magnitudes (this work) and $CH_4$(short) $= 23.2 \pm 0.2$ magnitudes (Luhman \& Esplin 2016).  We adopt the weighted average of $H$(F160W) and $H$(CH$_4$)  (Table 8).}
\tablenotetext{h}{Photometry is based on the $J$ measurement given in Tinney et al. (2014) and the synthetic colors measured here (Table 4).}
\end{deluxetable}

\begin{deluxetable}{lrrr}
\tablewidth{0pt}
\tablecaption{New Astrometry}
\tablehead{ 
\colhead{Name} & \colhead{Parallax} & \colhead{$\mu_{\alpha}$cos$\delta$}  & \colhead{$\mu_{\delta}$} \\
\colhead{} & \colhead{''(err)} & \colhead{''yr$^{-1}$(err)}  & \colhead{''yr$^{-1}$(err)}  \\
}
\startdata
WISE J014656.66$+$423410.0AB\tablenotemark{a} & 0.054(0.005) & $-$0.455(0.004) & $-$0.024(0.004) \\
WISEA J035000.31$-$565830.5\tablenotemark{b} &  0.184(0.010) & $-$0.206(0.007) & $-$0.578(0.008) \\ 
WISE J053516.80$-$750024.9\tablenotemark{c}     & 0.070(0.005) &  $-$0.127(0.004) &  0.013(0.004) \\
WISEA J082507.37$+$280548.2\tablenotemark{b} & 0.158(0.007) &  $-$0.066(0.008) & $-$0.247(0.010) \\
WISEA J120604.25$+$840110.5\tablenotemark{b} & 0.085(0.007) &  $-$0.585(0.004) & $-$0.253(0.005) \\
WISEPC J121756.91$+$162640.2AB\tablenotemark{d}  &  0.113(0.012) & 0.760(0.011) & $-$1.278(0.010) \\
WISEPC J140518.40$+$553421.5\tablenotemark{e}  & 0.155(0.006) & $-$2.334(0.005) &  0.232(0.005) \\ 
\enddata
\tablenotetext{a}{Astrometric measurements are given in Table 11.}
\tablenotetext{b}{Parallax from  Luhman \& Esplin 2016.}
\tablenotetext{c}{Astrometric measurements are given in Table 12.}
\tablenotetext{d}{Astrometric measurements are given in Table 13.}
\tablenotetext{e}{Astrometric measurements are given in Table 14.}

\end{deluxetable}

\begin{deluxetable}{rcrrrrrcc}
\tabletypesize{\scriptsize}
%\tabletypesize{\tiny}
\tablewidth{0pt}
%\rotate
\tablecaption{Y Dwarf Data Set}
\tablehead{ 
\colhead{Name(RA)}  & \colhead{Spectral} & \colhead{M $-$ m(err)} & \colhead{$Y$(err)}  & \colhead{$J$(err)}  & \colhead{$H$(err)}  & \colhead{$K$(err)}  & \colhead{Ref($\pi$)}   & \colhead{Ref($YJHK$)} \\ 
\colhead{(Declination)}  & \colhead{Type} & \colhead{[3.6](err)}  & \colhead{[4.5](err)}   & \colhead{W1(err)}   & \colhead{W2(err)}    & \colhead{W3(err)}  & \colhead{Ref([3.6][4.5])}   & \colhead{Ref(Discovery)} \\ 
\colhead{} & \colhead{} & \multicolumn{5}{c}{magnitudes} & \colhead{} & \colhead{} \\
}
\startdata
WISE J014656.66\tablenotemark{a} & Y0  &  $-$1.34(20) & 22.85(19) & 22.05(7) & 22.69(14) & \nodata & Table 7 & D15 \\
$+$423410.0B &  & \nodata    & \nodata     & \nodata     & \nodata     & \nodata  & & K12,D15 \\
WISE J030449.03 &  Y0 & \nodata &  21.32(17) & 20.79(09) & 21.02(16) & \nodata  &   & P14, Table 4 \\
$-$270508.3 &         & 17.71(03) & 15.48(03) & \nodata  & 15.59(09) & \nodata & Table 3 & P14\\
WISEA J035000.31 & Y1 &   1.32(12) &  21.62(12) & 22.09(10) & 22.51(20) & \nodata  & LE16 & L15 \\
$-$565830.5 &         & 17.84(03) & 14.61(03) & \nodata & 14.75(04) & 12.33(28) & L15 & K12\\
WISE J035934.06 &  Y0 &  $-$1.00(20) &  21.84(11) & 21.53(11) & 21.72(17) & 22.80(30)  & T14 & L15 \\
$-$540154.6 &         & 17.55(07) & 15.33(02) & 19.10(20) & 15.38(05) & 14.00(20) & K12 & K12 \\
WISEA J041022.75 & Y0  &   0.91(11) &  19.61(04) & 19.44(03) & 20.02(05) & 19.91(07) & B14 & L13 \\
$+$150247.9  &         & 16.56(03) & 14.12(03) & \nodata & 14.11(05) & 12.31(50) & L13 & C11 \\
WISE J053516.80  &   $\geq$Y1 & $-$0.77(16) &  22.73(30) & 22.50(20) & 23.34(34) &  \nodata & Table 7 & L15 \\
$-$750024.9  &   &  17.54(03) & 14.87(03) & 17.94(14) & 14.90(05) & \nodata & L15 & K12 \\
WISE J064723.23& Y1 &  0.09(18) &   23.13(09) & 22.94(10) & 23.11(27) & \nodata & T14 & L15,Table 1 \\
$-$623235.5 &       & 17.89(09) & 15.07(02) & \nodata & 15.22(05) & \nodata &   K13 & K13 \\
WISE J071322.55 & Y0 &   0.18(08) &  20.34(08) & 19.98(05) & 20.19(08) & 21.30(30) & B14 & L15 \\
$-$291751.9 &         & 16.67(05) & 14.22(04) & \nodata & 14.46(05) & 12.29(36) & B14 & K12 \\
WISE J073444.02   &   Y0 &   $-$0.66(19) &  21.02(05) & 20.05(05) & 20.92(12) & 20.96(15) & B14 &  L15 \\
$-$715744.0     &        & 17.69(08) & 15.21(02) & 18.75(28) & 15.19(05) &  \nodata  & K12  & K12 \\
WD 0806         & Y1  & $-$1.41(07) &  25.14(22) & 25.42(23) & 25.29(14) & \nodata & S09 & Table 6 \\
$-$661B              &   & 19.28(10) & 16.78(05) &  \nodata &  16.88(05) & \nodata & L15 & LBB11 \\
WISEA J082507.37  & Y0.5  & 0.99(10)   & 22.66(11) & 22.53(10) & 23.09(18) & \nodata & LE16 & Table 6 \\
$+$280548.2  &        & 17.62(08) & 14.64(03) &  \nodata & 14.58(06) & \nodata & S15 & S15 \\
WISE J085510.83\tablenotemark{b} & $>$Y2 & 3.26(04) & 26.54(21) & 25.45(24) & 23.83(24) & \nodata  &  LE16 &  Table 6 \\
$-$071442.5  &   & 17.31(03) & 13.87(04) & 17.82(33) & 14.02(05) & 11.90(30) & LE16 & L14\\
WISEA  J114156.67 & Y0 & 0.12(08) & 20.33(14) &  19.76(14) & 20.15(15)  & \nodata & T14 & T14,Table 6 \\
$-$332635.5 &       & 16.64(08) &  14.66(03) &  17.08(12) & 14.61(06) & 11.73(21) & Table 3 & T14 \\
WISEA J120604.25 & Y0 & $-$0.35(18) & 20.89(10) & 20.38(10) &  20.97(12) & \nodata  &  LE16 & Table 6 \\
$+$840110.5 &        & 17.34(06) & 15.22(03) & \nodata & 15.06(06) & \nodata &  S15 & S15 \\
WISEPC J121756.91\tablenotemark{a} & Y0 &  0.27(23) & 20.26(03) & 20.08(03) & 20.51(06) & 21.10(12) & Table 7 & L12 \\
$+$162640.2B & &  \nodata    & \nodata     & \nodata     & \nodata     & \nodata  & & K12,L12 \\
WISEPC J140518.40 & Y0.5 &  0.95(09) & 21.24(10) & 21.06(06) & 21.41(08) & 21.61(12) & Table 7 & L13 \\
$+$553421.5 &           & 16.78(03) & 14.02(03) & 18.77(40) & 14.10(04) & 12.20(26) & L13 & C11 \\
WISE J154151.65 & Y1  &   1.22(06) &   21.46(13) & 21.12(06) & 21.07(07) & 21.70(20)  & B14 & L13,L15 \\
$-$225024.9 &         & 16.92(02) & 14.12(02) & \nodata & 14.25(06) & 12.20(30) & L13 & C11 \\
WISEA J163940.84 & Y0pec  &  1.51(21) & 20.54(10) & 20.47(10) & 20.59(11) & \nodata & T14 & Table 6 \\
$-$684739.4 &                 & 16.23(03) & 13.57(03) & 17.27(19) & 13.54(06) & \nodata & Table 3 & T12 \\
WISEA J173835.52& Y0 &   0.55(10) &  19.74(08) & 19.58(04) & 20.24(08) & 20.58(10)  & B14 & L16b \\
$+$273258.8 &        & 16.87(03) & 14.42(03) & 17.71(16) & 14.50(04) & 12.45(40) & L16b & C11 \\
WISEPA J182831.08 & $\geq$Y2 &  0.13(14) & 23.03(17) & 23.48(23) & 22.73(13) & 23.48(36)& B14 & L13,L15 \\
$+$265037.8       &          &  16.84(03) & 14.27(03) & 17.34(26) & 14.35(05) & 12.44(34) & L13 & C11 \\
WISEA J205628.88\tablenotemark{b} & Y0  &  0.81(10) &   19.77(05) & 19.43(04) & 19.96(04) & 20.01(06)  & B14 & L13 \\
$+$145953.6 &  & 16.06(10) & 13.89(03) & 16.48(08) & 13.84(04) & 11.73(25) & Table 3,L13 &  C11 \\
WISEA J220905.75 & Y0 & 0.84(15) &  23.04(12) & 22.94(19) & 22.48(21) & \nodata  & B14 & Table 6 \\
$+$271143.6 &     & 17.82(09) & 14.74(03) & \nodata & 14.77(06) & 12.46(39) & C14  & C14 \\
WISE J222055.31 & Y0  & $-$0.30(09) & 20.91(09) & 20.64(05) & 20.96(08) & 21.33(15)  & B14 &  L15\\
$-$362817.4 &  & 17.20(06) & 14.73(02) & \nodata & 14.71(06) & \nodata & K12  & K12 \\
WISEA J235402.79 & Y1 & \nodata &  \nodata & 22.72(13) & 22.53(38)   & \nodata &  & Table 6 \\
$+$024014.1 &    & 18.11(11) & 15.01(02) & \nodata & 15.01(09) & \nodata & S15 & S15 \\
\enddata
\tablecomments{Photometric error is in centimag. W1, W2 and W3 photometry is from the ALLWISE catalog. References: Beichman et al. 2014; Cushing et al. 2011, 2014; Dupuy, Liu \& Leggett 2015; Kirkpatrick et al. 2012, 2013; Leggett et al. 2013, 2015, 2016b; Liu et al. 2012; Luhman, Burgasser \& Bochanski 2011; Luhman 2014; Luhman \& Esplin 2016; Pinfield et al. 2014; Schneider et al. 2015; Subasavage et al. 2009; Tinney et al. 2012, 2014.}
\tablenotetext{a}{WISE J014656.66$+$423410.0B, and 
WISEPC J121756.91$+$162640.2B, are in close binary systems. Mid-infrared photometry is not available for the individual components of the systems.}
\tablenotetext{b}{We have also measured for WISE J085510.83$-$071442.5 $M^{\prime} =  13.95\pm0.20$ magnitudes and for WISEA J205628.88$+$145953.6 $M^{\prime} =  14.00\pm0.15$ magnitudes .}
\end{deluxetable}

\begin{deluxetable}{lccccccc}
\tabletypesize{\scriptsize}
%\tabletypesize{\tiny}
\tablewidth{0pt}
%\rotate
\tablecaption{Y Dwarf Estimated Properties}
\tablehead{ 
\colhead{Name}  & \colhead{Type} & \colhead{v$_{\rm tan}$}  &  \colhead{[m/H]}  &
\colhead{$T_{\rm eff}$} & \colhead{log $g$} &  \colhead{Mass} &  \colhead{Age} \\
\colhead{}  & \colhead{}  & \colhead{km s$^{-1}$}    &   \colhead{}  &
\colhead{K} & \colhead{cm s$^{-2}$} & \colhead{Jupiter}  & \colhead{Gyr} \\
}
\startdata
WISE J014656.66$+$423410.0B & Y0 & 40(4) & $\sim$ 0 & 400 -- 430 & 4.25 -- 4.75  & 8 -- 20 & 0.8 -- 7  \\ 
WISE J030449.03$-$270508.3\tablenotemark{a} &  Y0 & \nodata & $\lesssim 0$ & 450 -- 500 & \nodata & \nodata & \nodata\\
WISEA J035000.31$-$565830.5 & Y1 & 16(1) &   $0.25$ -- $+0.55$ & 310 -- 340 & 3.75 -- 4.25 & 3 -- 8  & 0.3 -- 3 \\
WISE J035934.06$-$540154.6 &  Y0 & 57(18) & $-0.25$ -- $+0.05$ & 420 -- 450 & 4.25 -- 4.75 & 8 -- 20 & 0.7 -- 6 \\
WISEA J041022.75$+$150247.9 & Y0 & 72(4)  & $0$ -- $+0.3$ & 410 -- 440 & 4.25 -- 4.75 & 8 -- 20 & 0.8 -- 6 \\
WISE J053516.80$-$750024.9\tablenotemark{b}     &   $\geq$Y1 & 9(1) &  $-0.15$ -- $+0.15$ & 360 -- 390 & 4.25 -- 4.75 & 8 -- 20 & 1 -- 10 \\
WISE J064723.23$-$623235.5 & Y1 &  19(7) & $-0.05$ -- $+0.25$  & 320 -- 350 & 4.0 -- 4.5 &  5 -- 13  & 0.8 -- 6  \\
WISE J071322.55$-$291751.9 & Y0 &  26(3)  &  $-0.15$ -- $+0.15$ & 435 -- 465 & 4.5 -- 5.0 & 13 -- 29 & 2 -- 12 \\
WISE J073444.02$-$715744.0     &   Y0 &   37(11)  &  $-0.15$ -- $+0.15$ & 435 -- 465 & 4.25 -- 4.75 & 9 -- 20 & 0.8 -- 5 \\
WD 0806$-$661B\tablenotemark{c}              & Y1  & 41(1)  & $< 0$  & 325 -- 350 & 4.2 -- 4.3   & 7 -- 9  &  1.5 -- 2.7\\
WISEA J082507.37$+$280548.2 & Y0.5  &  8(1)  &  $0$ -- $+0.3$  & 310 -- 340 & 3.75 -- 4.25 & 3 -- 8 & 0.3 -- 3 \\
WISE J085510.83$-$071442.5\tablenotemark{d}  & $>$Y2 & 86(3)  & $-0.2$ -- $+0.2$ & 240 -- 260 & 3.5 -- 4.3  & 1.5 -- 8 & 0.3 -- 6 \\
WISEA  J114156.67$-$332635.5 & Y0 & 41(8) & $0$ -- $+0.3$ & 410 -- 440 & 3.75 -- 4.25 & 3 -- 8 & 0.1 -- 1.0 \\
WISEA J120604.25$+$840110.5 & Y0 &  36(3) & $-0.05$ -- $+0.25$   & 420 -- 450 & 4.0 -- 4.5 & 6 -- 14 & 0.4 -- 3 \\ 
WISEPC J121756.91$+$162640.2B & Y0 & 62(6)  & $-0.05$ -- $+0.25$  &  420 -- 450 & 4.25 -- 4.75 & 8 -- 20 & 0.7 -- 6 \\
WISEPC J140518.40$+$553421.5\tablenotemark{d} & Y0.5 & 72(3)  & $-0.15$ -- $+0.15$ & 370 -- 400 & 4.3 -- 4.8 & 9 --21 & 1.5 -- 10\\
WISE J154151.65$-$225024.9 & Y1  & 23(1)  &   $-0.05$ -- $+0.25$  & 360 -- 390  &  4.25 -- 4.75 & 8 -- 20 & 1 -- 10\\
WISEA J163940.84$-$684739.4 & Y0pec  & 74(9)  & $-0.15$ -- $+0.15$ & 360 -- 390  & 4.0 -- 4.5 & 5 -- 14 & 0.5 -- 5 \\
WISEA J173835.52$+$273258.8\tablenotemark{d} & Y0 & 17(1)  &  $-0.05$ -- $+0.25$  & 410 -- 440 & 4.0 -- 4.5 & 5 -- 14 & 0.3 -- 3 \\
WISEPA J182831.08$+$265037.8\tablenotemark{b}  & $\geq$Y2 & 46(3)  & $-0.6$ -- $-0.3$ & 310 -- 340  & 3.75 -- 4.25  & 3 -- 8 &  0.3 -- 3 \\
WISEA J205628.88$+$145953.6 & Y0 & 33(2)  & $0$ -- $+0.3$ & 410 -- 440 & 4.25 -- 4.75 & 8 -- 20 & 0.8 -- 6 \\
WISEA J220905.75$+$271143.6 & Y0 & 59(4) &   $0$ -- $+0.3$ & 310 -- 340 & 4.25 -- 4.75 & 8 -- 19 & 2 -- 15 \\
WISE J222055.31$-$362817.4 & Y0  & 10(1) & $-0.05$ -- $+0.25$ & 410 -- 440 & 4.25 -- 4.75 & 8 -- 20 & 0.8 -- 6 \\
WISEA J235402.79$+$024014.1\tablenotemark{e} & Y1 & \nodata  & $\sim$ 0 & $\sim$350  & \nodata   &\nodata  & \nodata \\
\enddata
\tablecomments{For Y dwarfs with near-infrared spectroscopy, the estimated uncertainty in $T_{\rm eff}$, $\log g$ and [m/H] is  $\pm 15$~K, $\pm 0.25$ dex and $\pm 0.15$ dex (Figures 15 -- 18, see text).}
\tablenotetext{a}{The color-color plots $J-$[4.5]:[3.6]$-$[4.5] and  $J-$[4.5]:$J-H$ suggest W0304 is slightly metal-poor. Temperature is estimated from $J-$[4.5] color, assuming the metallicity is solar or sub-solar.}
\tablenotetext{b}{W0535 and W1828 appear to be similar-mass binary systems, based on the color-magnitude plots. The estimated properties assume these are same-mass binary systems.}
\tablenotetext{c}{The white dwarf primary constrains the age of the system (Rodriguez et al. 2011). The brown dwarf appears significantly metal-poor in the   $J-$[4.5]:[3.6]$-$[4.5] plot. Temperature is estimated from $M_J$:$J-H$ and $M_{[4.5]}$:$J-$[4.5].  $T_{\rm eff}$ and age together constrain gravity and mass.}
\tablenotetext{d}{Rotation periods based on variability are: for W0855 5 -- 16 hours (Luhman \& Esplin 2016); for W1405 8.5 hours (Cushing et al. 2015); for W1738 6.0 hours (Leggett et al. 2016b). Brown dwarfs are expected to spin up with time such that brown dwarfs younger than around 10 Myr would have a period of 14 -- 48 hours (Bouvier et al. 2014).} 
\tablenotetext{e}{The color-color plots $J-$[4.5]:[3.6]$-$[4.5] and 
$J-$[4.5]:$J-H$ suggest W2354 has solar metallicity. Temperature is 
estimated from $J-$[4.5] color, assuming solar metallicity.}
\end{deluxetable}

\begin{deluxetable}{crrccrrrrr}
\tabletypesize{\scriptsize}
%\tabletypesize{\tiny}
\tablewidth{0pt}
%\rotate
\tablecaption{WISE J085510.83$-$071442.5 $T_{\rm eff} =$ 250~K Model Photometric Comparison}
\tablehead{ 
\colhead{Family\tablenotemark{a}}  & \colhead{$\log g$} & \colhead{[m/H]}  &  \colhead{Cloud Cover\tablenotemark{b}}  &
\colhead{$\log K_{\rm zz}$\tablenotemark{c}} & \colhead{$\delta Y$} &  \colhead{$\delta J$} &  \colhead{$\delta H$} 
  & \colhead{$\delta$[4.5]}  & \colhead{$\delta$W3}  \\
}
\startdata
M14 & 4.0 & 0.0    & nc                       & 0 & $-$0.35 & $-$0.16 & $+$1.11 & $-$0.25 & $+$0.45 \\
M14 & 4.0 & 0.0    & $f_{\rm sed}=7$ $h=50$\% & 0 & $-$1.94 & $-$1.58 & $+$0.29 & $-$0.05 & $+$0.07 \\
T15 & 4.3 & 0.0    & nc                       & 9 & $-$0.09 & $+$0.36 & $+$0.62 & $-$0.18 & $+$0.49 \\
T15 & 3.8 & $+$0.2 & nc                       & 9 & $-$0.32 & $-$0.61 & $+$0.49 & $-$0.15 & $-$0.05 \\
T15 & 3.5 & 0.0    & nc                       & 9 & $+$0.14 & $-$0.12 & $+$0.72 & $-$0.25 & $-$0.04 \\
\enddata
\tablecomments{$\delta  = M({\rm model}) - M({\rm observed})$.}
\tablenotetext{a}{Models are from Tremblin et al. 2015 or updated Morley et al. 2014.}
\tablenotetext{b}{Models are cloud-free (nc) or have thin cloud decks with half the surface covered  ($f_{\rm sed}=7$ $h=50$\%).}
\tablenotetext{c}{The parameter $K_{\rm zz}$ is the diffusion coefficient, $\log K_{\rm zz} = 0$ implies that the models exclude mixing and are in chemical equilibrium.} 
\end{deluxetable}

\begin{deluxetable}{lllll}
\tablewidth{0pt}
%\tabletypesize{\scriptsize}
\tabletypesize{\footnotesize}
\tablecaption{IRAC Astrometry of WISE J014656.66+423410.0AB}
\tablehead{
\colhead{$\alpha$ (J2000)} & \colhead{$\sigma_\alpha$} & \colhead{$\delta$ (J2000)} & \colhead{$\sigma_\delta$} & \colhead{MJD} \\
\colhead{(\arcdeg)} & \colhead{(\arcsec)} & \colhead{(\arcdeg)} & \colhead{(\arcsec)}
%\cline{2-4}
}
\startdata
 26.7360260 & 0.017 & 42.5694192 & 0.017 & 55656.09 \\
 26.7358510 & 0.017 & 42.5694213 & 0.017 & 55993.05 \\
 26.7357920 & 0.017 & 42.5694404 & 0.017 & 56215.08 \\
 26.7356892 & 0.017 & 42.5694181 & 0.017 & 56364.26 \\
 26.7356867 & 0.017 & 42.5694095 & 0.017 & 56372.31 \\
 26.7356746 & 0.017 & 42.5694131 & 0.017 & 56388.82 \\
 26.7356809 & 0.017 & 42.5694104 & 0.017 & 56393.13 \\
 26.7356172 & 0.017 & 42.5694212 & 0.017 & 56579.19 \\
 26.7356091 & 0.017 & 42.5694272 & 0.017 & 56592.47 \\
 26.7356057 & 0.017 & 42.5694355 & 0.017 & 56602.48 \\
 26.7356033 & 0.017 & 42.5694283 & 0.017 & 56616.07 \\
 26.7355123 & 0.017 & 42.5694088 & 0.017 & 56737.38 \\
 26.7355131 & 0.022 & 42.5694102 & 0.022 & 56742.07 \\
 26.7355103 & 0.017 & 42.5694020 & 0.017 & 56750.46 \\
 26.7355038 & 0.017 & 42.5694024 & 0.017 & 56758.35 \\
 26.7354998 & 0.017 & 42.5693956 & 0.017 & 56768.07 \\
 26.7354985 & 0.022 & 42.5694047 & 0.022 & 56772.30 \\
 26.7354517 & 0.022 & 42.5694158 & 0.022 & 56947.21 \\
 26.7354201 & 0.022 & 42.5694134 & 0.022 & 56980.00 \\
 26.7353199 & 0.022 & 42.5693954 & 0.022 & 57145.85 \\
 26.7352655 & 0.022 & 42.5694112 & 0.022 & 57340.06
\enddata
\end{deluxetable}

\begin{deluxetable}{lllll}
\tabletypesize{\footnotesize}
\tablewidth{0pt}
\tablecaption{IRAC Astrometry of WISE J053516.80$-$750024.9}
\tablehead{
\colhead{$\alpha$ (J2000)} & \colhead{$\sigma_\alpha$} & \colhead{$\delta$ (J2000)} & \colhead{$\sigma_\delta$} & \colhead{MJD} \\
\colhead{(\arcdeg)} & \colhead{(\arcsec)} & \colhead{(\arcdeg)} & \colhead{(\arcsec)}
%\cline{2-4}
}
\startdata
  83.8199010  & 0.020 & $-75.0067327$ & 0.020 &  55486.25 \\
  83.8197269  & 0.020 & $-75.0067534$ & 0.020 &  55668.91 \\
  83.8197605  & 0.020 & $-75.0067398$ & 0.020 &  55885.34 \\
  83.8195342 & 0.020 & $ -75.0067408$ & 0.020 &  56036.95 \\
  83.8196093  & 0.020 & $-75.0067301$ & 0.020 &  56264.47 \\
  83.8195703  & 0.020 & $-75.0067413$ & 0.020 &  56317.21 \\
  83.8194065  & 0.020 & $-75.0067341$ & 0.020 &  56421.97 \\
  83.8193712  & 0.020 & $-75.0067281$ & 0.020 &  56450.27 \\
  83.8194404  & 0.020 & $-75.0067059$ & 0.020 &  56545.18 \\
  83.8194623  & 0.022 & $-75.0067298$ & 0.024 &  56626.48 \\
  83.8195022  & 0.020 & $-75.0067309$ & 0.020 &  56641.18 \\
  83.8194624  & 0.020 & $-75.0067359$ & 0.020 &  56673.39 \\
  83.8192796  & 0.020 & $-75.0067386$ & 0.020 &  56777.50 \\
  83.8192769 & 0.022 & $ -75.0067403$ & 0.024 &  56780.25 \\
  83.8192548  & 0.020 & $-75.0067212$ & 0.020 &  56821.73 \\
  83.8192894  & 0.022 & $-75.0067191$ & 0.024 &  56854.02 \\
  83.8192649  & 0.020 & $-75.0067064$ & 0.020 &  56875.58 \\
  83.8193254  & 0.022 & $-75.0067185$ & 0.024 &  56969.02 \\
  83.8191664  & 0.022 & $-75.0067402$ & 0.024 &  57160.80 
\enddata
\end{deluxetable}

\begin{deluxetable}{lllll}
\tabletypesize{\footnotesize}
%\tabletypesize{\scriptsize}
\tablewidth{0pt}
\tablecaption{IRAC Astrometry of WISEPC J121756.91$+$162640.2AB}
\tablehead{
\colhead{$\alpha$ (J2000)} & \colhead{$\sigma_\alpha$} & \colhead{$\delta$ (J2000)} & \colhead{$\sigma_\delta$} & \colhead{MJD} \\
\colhead{(\arcdeg)} & \colhead{(\arcsec)} & \colhead{(\arcdeg)} & \colhead{(\arcsec)}
%\cline{2-4}
}
\startdata
 184.4873562  & 0.022 & 16.4443712  & 0.020 & 55633.00 \\
 184.4875633  & 0.034 & 16.4440433  & 0.037 & 55972.46 \\
 184.4875693  & 0.022 & 16.4440330  & 0.020 & 55972.47 \\
 184.4875799  & 0.034 & 16.4440030  & 0.037 & 56004.42 \\
 184.4876063  & 0.034 & 16.4438993  & 0.037 & 56131.98 \\
 184.4876109  & 0.022 & 16.4439086  & 0.020 & 56136.37 \\
 184.4876155  & 0.034 & 16.4438590  & 0.037 & 56163.50 \\
 184.4880327  & 0.034 & 16.4432928  & 0.037 & 56741.92 \\
 184.4880550  & 0.034 & 16.4431730  & 0.037 & 56890.37 
\enddata
\end{deluxetable}

\begin{deluxetable}{lllll}
\tabletypesize{\footnotesize}
\tablewidth{0pt}
\tablecaption{IRAC Astrometry of WISEPC J140518.40$+$553421.5}
\tablehead{
\colhead{$\alpha$ (J2000)} & \colhead{$\sigma_\alpha$} & \colhead{$\delta$ (J2000)} & \colhead{$\sigma_\delta$} & \colhead{MJD} \\
\colhead{(\arcdeg)} & \colhead{(\arcsec)} & \colhead{(\arcdeg)} & \colhead{(\arcsec)}
%\cline{2-4}
}
\startdata
  211.3259099 & 0.021 & 55.5725996 & 0.020 & 55583.17  \\  
  211.3247196 & 0.031 & 55.5726728 & 0.030 & 55958.16  \\   
  211.3246853 & 0.021 & 55.5726730 & 0.020  & 55978.03  \\ 
  211.3245332 & 0.031 & 55.5727157 & 0.030 & 56026.12    \\ 
  211.3242453 & 0.031 & 55.5727649 & 0.030 & 56098.29    \\ 
  211.3242425 & 0.021 & 55.5727605 & 0.020  & 56100.80  \\ 
  211.3239573 & 0.031 & 55.5727589 & 0.030 & 56169.62    \\ 
  211.3235275 & 0.021 & 55.5727358 & 0.020  & 56344.47   \\  
  211.3233998 & 0.021 & 55.5727742 & 0.020  & 56393.76 \\    
  211.3232349 & 0.021 & 55.5728044 & 0.020  & 56435.74  \\   
  211.3230239 & 0.021 & 55.5728328 & 0.020  & 56483.50    \\  
  211.3228253 & 0.021 & 55.5728286 & 0.020  & 56529.38   \\   
  211.3223622 & 0.021 & 55.5728070 & 0.020  & 56718.77  \\    
  211.3222866 & 0.031 & 55.5728223 & 0.030  & 56741.77 \\     
  211.3222193 & 0.021 & 55.5728423 & 0.020  & 56768.01  \\    
  211.3220628 & 0.021 & 55.5728802 & 0.020  & 56810.47   \\   
  211.3218706 & 0.021 & 55.5729028 & 0.020  & 56854.10   \\   
  211.3217215 & 0.021 & 55.5728930 & 0.020  & 56886.75    \\  
  211.3216683 & 0.021 & 55.5728971 & 0.020  & 56902.01    \\  
  211.3212277 & 0.031 & 55.5728659 & 0.020  & 57076.29     
\enddata
\end{deluxetable}

\begin{deluxetable}{llllllllllllllllllll}
\tabletypesize{\footnotesize}
\tablewidth{0pt}
\tablecaption{On-Line Data Table}
\tablehead{}
\startdata
 & & & & & & & & & & & & & & & & & & &\\
\enddata
\tablecomments{An on-line data table giving the distance moduli and photometry used in this work, for T6 and later brown dwarfs, will be made available at the Astrophysical Journal site.}
\end{deluxetable}


\begin{thebibliography}{}
\bibitem[Ackerman \& Marley (2001)]{}Ackerman, A. S. \& Marley, M. S. \ 2001, \apj, 556, 872
\bibitem[Allard, Allard \& Kielkopf (2005]{}Allard, N. F., Allard, F. \& Kielkopf, J. F. \ 2005, \aap, 440, 1195
\bibitem[Althaus et al. (2005)]{} Althaus, L. G., Serenelli, A. M., Panei, J. A., Corsico, A. H., Garcia-Berro, E. \& Scoccola, C. G. \ 2005, \aap, 435, 631
\bibitem[Baraffe et al. (2003)]{}Baraffe, I., Chabrier, G., Barman, T. S., Allard, F. \& Hauschildt, P. H. \ 2003, \aap, 402, 701
\bibitem[Beamin et al. (2014)]{}Beamin. J. C.  et al. \ 2014, \aap, 570, L8
\bibitem[Beichman et al. (2014)]{} Beichman, C., Gelino, C. R., Kirkpatrick, J. D,. Cushing, M. C., Dodson-Robinson, S,, Marley, M. S., Morley, C, V. \& Wright, E. L. \ 2014, \apj, 783, 68
\bibitem[Bensby et al. (2005)]{}Bensby, T., Feltzing, S., Lundstrom, I. \& Ilyin, I. \ 2005, \aap, 433, 185
\bibitem[Bensby, Feltzing \& Oey (20140]{}Bensby, T., Feltzing, S. \& Oey, M. S. \ 2014, \aap, 562, 71
\bibitem[Bouvier et al. (2014)]{}Bouvier, J., Matt, S. P., Mohanty, S., Scholz, A., Stassun, K. G. \& Zanni, C. \ 2014,         
Protostars and Planets VI, Henrik Beuther, Ralf S. Klessen, Cornelis P. Dullemond, and Thomas Henning (eds.), University of Arizona Press, Tucson, 433
\bibitem[Burgasser et al. (2002)]{}Burgasser, A.J., Kirkpatrick, J. D., Brown, M. E. et al. \ 2002, \apj, 564, 421 
\bibitem[Burgasser, Looper \& Rayner (2010)]{}Burgasser, A.J., Looper, D. \& Rayner, J.T. \ 2010, \aj, 139, 2448 
\bibitem[Burrows et al. (1997)]{} Burrows, A., Marley, M. S., Hubbard, W. B., Lunine, J. I., Guillot, T., Saumon, D., Freedman, R., Sudarsky, D. \& Sharp, C. \ 1997, \apj, 491, 856
\bibitem[Burrows, Sudarsky \& Lunine (2003]{}Burrows, A., Sudarsky, D. \& Lunine, J. I. \ 2003, \apj, 596, 587
\bibitem[Covey et al. (2007)]{} Covey K. R., Ivezic, Z., Schlegel, D. et al. \ 2007, \aj, 134, 2398
\bibitem[Cushing et al. (2011)]{ys} Cushing, M. C., Kirkpatrick, J. D., Gelino, C. R. et al. \ 2011, \apj, 743, 50
\bibitem[Cushing et al. (2014)]{}Cushing, M. C., Kirkpatrick, J. D., Gelino, C. R., Mace, G. N., Skrutskie, M. F. \& Gould, A. \ 2014, \aj, 147, 113
\bibitem[Cushing et al. (2016)]{}Cushing, M. C., Hardegree-Ullman K. K., Trucks J. L. et al \ 2016, \apj, 823, 152
\bibitem[Davenport et al. (2014)]{} Davenport J. R. A., Ivezic, Z., Becker, A. C., Ruan, J. J., Hunt-Walker, N. M., Covey, K. R., Lewis, A. R., AlSayyad, Y. \& Anderson, L. M. \ 2014, \ mnras, 440, 3430  
\bibitem[Delorme et al. (2008)]{}Delorme, P., Delfosse, X., Albert, L. et al. \ 2008, \aap, 482, 961
\bibitem[Dufour, Bergeron \& Fontaine (2005)]{} Dufour, P., Bergeron, P. \& Fontaine, G. \ 2005, \apj, 627, 404
\bibitem[Dupuy \& Kraus (2013)]{}Dupuy, T. J. \& Kraus, A. L. \ 2013, {\it Science}, 341, 1492 
\bibitem[Dupuy \& Liu (2012)]{}Dupuy, T. J. \& Liu, M. C. \ 2012, \apjs, 201, 19
\bibitem[Dupuy, Liu \& Leggett (2015)]{}Dupuy, T. J., Liu, M. C. \& Leggett, S. K. \ 2015, \apj, 803, 102
\bibitem[Eikenberry et al. (2004)]{}Eikenberry, R., Elston, R., Guzman, R. et al. \ 2004, SPIE, 5492, 1196 
\bibitem[Elias et al. (2006)]{}Elias, J. H,  Joyce, R. R., Liang, M., Muller, G. P., Hileman, E. A., \& George, J. R., in Ground-based and Airborne Instrumentation for Astronomy, eds. I. S. McLean and I. Masanori \ 2006, SPIE, 6269, 138
\bibitem[Esplin \& Luhman (2016)]{}Esplin, T. L. \& Luhman, K. L. 2016, \aj, 151, 9
\bibitem[Esplin et al. (2016)]{}Esplin, T. L., Luhman, K. L., Cushing, M. C., Hardegree-Ullman, K. K., Trucks, J. L., Burgasser, A. J. \& Schneider, A. C. \ 2016, \apj, 832, 58
\bibitem[Faherty et al. (2014)]{} Faherty, J. K.; Beletsky, Y., Burgasser, A. J., Tinney, C., Osip, D. J., Filippazzo, J. C. \& Simcoe, R. A. \ 2014, \apj, 790, 90
\bibitem[Freedman et al. (2014)]{}Freedman, R. S., Lustig-Yaeger, J., Fortney, J. J., Lupu, R. E., Marley, M. S. \& Lodders, K. \ 2014, \apjs, 214, 25
\bibitem[Freytag et al. (2010)]{}Freytag, B., Allard, F., Ludwig, H.-G., Homeier, D. \& Steffen, M. \ 2010, \aap, 513, 19
\bibitem[Geballe et al. (2009)]{}Geballe, T. R., Saumon, D., Golimowski, D. A., Leggett, S. K., Marley, M. S. \& Noll, K. S. \ 2009, \apj, 695, 844
\bibitem[Golimowski et al. (2004)]{}Golimowski, D. A., Leggett, S. K., Marley, M. S. et al. \ 2004, \aj, 127, 3516
%\bibitem[Heber  (2016)]{}Heber, U. \ 2016, \pasp, 128, 2001
\bibitem[Helling et al. (2001)]{} Helling, Ch., Oevermann, M., Luttke, M. J. H., Klein, R. \& Sedlmayr, E. \ 2001, \aap, 376, 194
\bibitem[Hodapp et al. (2003)]{NIRI} Hodapp, K. W., Jensen, J. B., Irwin, E. M.  et al.\ 2003, \pasp, 115, 1388
\bibitem[Hubeny \& Burrows (2007)]Hubeny, I. \& Burrows, A. \ 2007, \apj, 669, 1248 
\bibitem[Kirkpatrick et al. (2011)]{} Kirkpatrick, J. D., Cushing, M. C., Gelino, C. R., Griffith, R. L., Skrutskie, M. F., Marsh, K. A., Wright, E. L.,  Mainzer, A., Eisenhardt, P. R. \& McLean , I. S. \ 2011, \apjs, 197, 19
\bibitem[Kirkpatrick et al. (2012)]{} Kirkpatrick, J. D., Gelino, C. R., Cushing, M. C. et al. \ 2012, \apj, 753, 156
\bibitem[Kirkpatrick et al. (2013)]{} Kirkpatrick, J. D., Cushing, M. C., Gelino, C. R., Beichman, C. A., Tinney, C. G., Faherty, J. K., Schneider, A. \& Mace, G. N. \ 2013, \apj, 776, 128
\bibitem[Knapp et al. (2004)]{} Knapp, G. R., Leggett, S. K., Fan, X. et al. \ 2004, \aj, 127, 3553
\bibitem[Lawlor \& MacDonald (2006)]{}Lawlor, T. M. \& MacDonald, J. \ 2006, \mnras, 371, 263
\bibitem[Leggett et al. (2003)]{} Leggett, S. K., Hawarden, T. G., Currie, M. J., Adamson, A. J., Carroll, T. C., Kerr, T. H., Kuhn, O. P., Seigar, M. S., Varricatt, W. P. \& Wold, T. \ 2003, \mnras, 345, 144
\bibitem[Leggett et al. (2006)]{} Leggett, S. K., Currie, M. J., Varricatt, W. P. et al. \ 2006, \mnras, 378, 781
\bibitem[Leggett et al. (2007)]{} Leggett, S. K., Saumon, D., Marley, M. S., Geballe, T. R., Golimowski, D. A., Stephens, D. \& Fan, X. \ 2007, \apj, 655, 1079
\bibitem[Leggett et al. (2009)]{} Leggett, S. K., Cushing, M. C., Saumon, D. et al. \ 2009, \apj, 695, 1517
\bibitem[Leggett et al. (2012)]{} Leggett, S. K., Saumon, D., Marley, M. S. et al. \ 2012, \apj, 748, 74
\bibitem[Leggett et al. (2014)]{} Leggett, S. K., Liu, M. C., Dupuy, T. J., Morley, C. V., Marley, M. S. \& Saumon, D.
 \ 2014, \apj, 780, 62
\bibitem[Leggett et al. (2015)]{} Leggett, S. K., Morley, C. V., Marley, M. S. \& Saumon, D. \ 2015, \apj, 799, 37
\bibitem[Leggett et al. (2016)]{} Leggett, S. K., Tremblin, P., Saumon, D., Marley, M. S., Morley, C. V., Amundsen, D. S., Baraffe, I. \& Chabrier, G. \ 2016, \apj, 824, 2 (16a) 
\bibitem[Leggett et al. (2016b)]{} Leggett, S. K., Cushing, M. C., Hardegree-Ullman, K. K., Trucks, J. L., Marley, M. S. et al. \ 2016, \apj, 830, 141  (16b)
\bibitem[Liu, Leggett \& Chiu (2007)]{}Liu, M. C., Leggett, S. K. \& Chiu, K. \ 2007, \apj, 660, 1507
\bibitem[Liu et al. (2012)]{}Liu, M. C., Dupuy, T. J., Bowler, B. P., Leggett, S. K. \& Best, W. M. J. \ 2012, \apj, 758, 57
\bibitem[Lodders \& Fegley (2002)]{}Lodders, K. \& Fegley, B. \ 2002, \icarus, 155, 393
\bibitem[Lucas et al. (2010)]{}Lucas, P. W., Tinney, C. G., Burningham, B.  et al. \ 2010, \mnras, 408, L56
\bibitem[Luhman, Burgasser \& Bochanski (2011)]{} Luhman, K. L., Burgasser, A. J. \& Bochanski, J. J. \ 2011,  \apj, 730, L9 
\bibitem[Luhman (2014)]{}Luhman, K. L.  \ 2014, \apj, 786, L18
\bibitem[Luhman \& Esplin (2014)]{}Luhman, K. L. \& Esplin, T. L. \ 2014, \apj, 796, 6
\bibitem[Luhman et al. (2014)]{}Luhman, K. L., Morley, C. V., Burgasser, A. J., Esplin, T. L. \& Bochanski, J. J. \ 2014, \apj, 794, 16
\bibitem[Luhman \& Esplin (2016)]{}Luhman, K. L.\& Esplin, T. L. \ 2016, \aj, 152, 78 
\bibitem[Marley, Saumon \& Goldblatt (2010)]{}Marley, M, S., Saumon, D. \& Goldblatt, C. \ 2010, \apj, 723, L11
\bibitem[Marley et al. (2012)]{}Marley, M. S., Saumon, D., Cushing, M., Ackerman, A. S., Fortney, J. J. \& Freedman, R. \ 2012, \apj, 754, 135
\bibitem[Marsh et al. (2013)]{}Marsh K. A., Wright, E. L., Kirkpatrick, J. D., Gelino, C. R., Cushing, M. C., Griffith, R. L., Skrutskie, M. F. \& Eisenhardt, P. R. \ 2013, \apj, 762, 119
\bibitem[McMahon et al. (2013)]{}McMahon, R. G., Banerji, M., Gonzalez, E., Koposov, S. E., Bejar, V. J., Lodieu, N. \& Rebolo, R. \ 2013, Msngr, 154, 35
\bibitem[Morley et al. (2012)]{}Morley, C. V., Fortney, J. J., Marley, M. S., Visscher, C., Saumon, D. \& Leggett, S. K. \ 2012, \apj, 756, 172 (M12)
\bibitem[Morley et al. (2014)]{}Morley, C. V., Marley, M. S.,Fortney, J. J., Lupu, R., Saumon, D., Greene, T. \& Lodders, K. \ 2014,  \apj, 787, 78 (M14)
\bibitem[Noll, Geballe \& Marley (1997)]{} Noll, K. S., Geballe, T. R. \& Marley, M. S. \ 1997, \apj, 489, L87
\bibitem[Opitz et al. (2016)]{}Opitz, D., Tinney, C. G., Faherty, J. K., Sweet, S., Gelino, C. R. \& Kirkpatrick, J. D. \ 2016, \apj, 819, 17 
\bibitem[Pickles (1998)]{}Pickles, A. J. \ 1998, \pasp, 110,863
\bibitem[Pinfield et al. (2012)]{}Pinfield, D. J., Burningham, B., Lodieu, N. et al. \ 2012, \mnras, 422, 1922 
\bibitem[Pinfield et al. (2014)]{}Pinfield, D. J., Gromadzki, M., Leggett, S. K. et al. \ 2014, \mnras, 444, 1931
\bibitem[Radigan et al. (2012)]{}Radigan, J., Jayawardhana, R., Lafreniere, D., Artigau, E., Marley, M. S. \& Saumon, D. \ 2012, \apj, 750, 105
\bibitem[Rayner et al. (2009)]{}Rayner, J. T., Cushing, M. C. \& Vacca, W. D. \ 2009, \apjs,   185, 289 
\bibitem[Rodriguez et al. (2011)]{}Rodriguez, D. R., Zuckerman, B., Melis, C. \& Song, I. \ 2011, \apj, 732, L29
\bibitem[Saumon et al. (2000)]{} Saumon, D., Geballe, T. R., Leggett, S. K., Marley, M. S., Freedman, R. S., Lodders, K., Fegley, B., Jr. \& Sengupta, S. K. \ 2000, \apj, 541, 374
\bibitem[Saumon et al. (2006)]{} Saumon, D., Marley, M. S., Cushing, M. C., Leggett, S. K., Roellig, T. L., Lodders, K. \& Freedman, R. S. \ 2006, \apj, 647, 552
\bibitem[Saumon et al. (2007)]{} Saumon, D.,  Marley, M. S., Leggett, S. K., Geballe, T. R., Stephens, D., Golimowski, D. A., Cushing, M. C., Fan, X., Rayner, J. T., Lodders, K. \& Freedman, R. S. \ 2007, \apj, 656, 1136
\bibitem[Saumon \& Marley (2008)]{}Saumon, D. \& Marley, M. S. \ 2008, \apj, 689, 1327
\bibitem[Saumon et al. (2012)]{}Saumon D., Marley, M. S., Abel, M., Frommhold, L. \& Freedman, R. S. \ 2012,  \apj, 750, 74 (S12)
\bibitem[Schneider et al. (2015)]{}Schneider A. C.,
Cushing, M. C., Kirkpatrick, J. D., Gelino, C. R., Mace, G. N., Wright, E. L., Eisenhardt, P. R., Skrutskie, M. F., Griffith, R. L. \& Marsh, K. A. \ 2015, \apj, 804, 92
\bibitem[Schneider et al. (2016)]{}Schneider A. C., Cushing, M. C., Kirkpatrick, D. J. \& Gelino, C. R. \ 2016, \apjl, 823, 35
\bibitem[Skemer et al. (2016)]{}Skemer A. J., Morley C. V., Allers K. N. et al. \ 2016, \apj, 826, L17
\bibitem[Skrzypek, Warren \& Faherty (2016)]{} Skrzypek, N., Warren, S. J. \& Faherty, J. K. \ 2016, \aap, 589, 49
%\bibitem[Smart et al. (2016)]{}Smart, R. et al. \ 2016, \mnras, submitted
\bibitem[Stephens et al. (2009)]{}Stephens D. C., Leggett, S. K., Cushing, M. C., Marley, M. S., Saumon, D., Geballe, T. R., Golimowski, D. A., Fan, X. \& Noll, K. S. \ 2009, \apj, 702, 154
\bibitem[Subasavage et al. (2009)]{}Subasavage, J. P., Jao, W. -C., Henry, T. J., Bergeron, P., Difour, P., Ianna, P. A., Costa, E. \& Mendez, R. A. \ 2009, \aj, 137, 4547
\bibitem[Tinney et al. (2012)]{}Tinney C. G., Faherty, J. K., Kirkpatrick, J. D., Wright, E. L., Gelino, C. R.; Cushing, M. C.; Griffith, R. L. \& Salter, G. \ 2012, \apj, 759, 60
\bibitem[Tinney et al. (2014)]{}Tinney, C. G., Faherty, J. K., Kirkpatrick, J. D., Cushing, M., Morley, C. V. \& Wright, E. L. \ 2014, \apj, 796, 39
\bibitem[Tremblin et al. (2015)]{}Tremblin, P., Amundsen, D. S., Mourier, P., Baraffe, I., Chabrier, G., Drummond, B., Homeier, D. \& Venot, O. \ 2015, \apj, 804, L17 (T15)
\bibitem[Tremblin et al. (2016)]{}Tremblin, P., Amundsen, D. S., Chabrier, G., Baraffe, I.,  Drummond, B., Hinkley, S., Mourier, P. \&  Venot, O. \ 2016, \apj, 817, L19 
\bibitem[Tokunaga\& Vacca (2005)]{MKO3} Tokunaga, A. T. \& Vacca, W. D. \ 2005, \pasp, 117, 421
\bibitem[Tsuji et al. (1996)]{} Tsuji, T., Ohnaka, K., Aoki, W. \& Nakajima, T. \ 1996, \aap, 308
\bibitem[Visscher, Lodders \& Fegley (2006)]{}Visscher, C., Lodders, K. \& Fegley, B. \ 2006, \apj, 648, 1181
\bibitem[Visscher \& Moses (2011)]{} Visscher, C. \& Moses, J. I. \ 2011, \apj, 738, 72
\bibitem[Visscher (2012)]{}Visscher, C. \ 2012, Comparative Climatology of Terrestrial Planets,LPI Contribution No. 1675
\bibitem[Wang et al. (2015)]{}Wang, D., Gierasch, P. J., Lunine, J. I. \& Mousis, O. \ 2015, \icarus, 250, 154
\bibitem[Warren et al. (2007)]{}Warren, S. J., Mortlock, D. J., Leggett, S. K. et al. \ 2007, \mnras, 381, 1400
\bibitem[Wright et al. (2010)]{WISE} Wright, E. L.,  Eisenhardt, P. R. M., Mainzer, A. K. et al. \ 2010, \aj, 140, 1868
\bibitem[Yurchenko, Barber \& Tennyson (2011)]{}
Yurchenko, S. N., Barber, R. J. \& Tennyson, J. \ 2011, \mnras, 413, 1828
\bibitem[Yurchenko & Tennyson (2014)]{} Yurchenko S. N. \& Tennyson J. \ 2014, \mnras, 440, 1649 
\bibitem[Zahnle \& Marley (2014)]{}Zahnle, K. J. \& Marley, M. S. \ 2014, \apj, 797, 41
\bibitem[Zapatero Osorio et al. (2016)]{}Zapatero Osorio, M. R., Lodieu, N., Béjar, V. J. S., Martin, E. L., Ivanov, V. D., Bayo, A., Boffin, H. M. J., Muzik,, K., Minniti, D. \& Beamín, J. C. \ 2016, \aap, 592, 80
\end{thebibliography}
\end{document}